%% file: main.tex
\documentclass[manuscript,screen,nonacm]{acmart}
\pdfoutput=1
\makeatletter\def\@ACM@checkaffil{}\makeatother% allow authors without full ACM affiliation (no country line)
\acmJournal{IMWUT}
\usepackage{enumitem}
\usepackage{placeins}
\usepackage{float}
\usepackage{tabularx}
\usepackage{wrapfig}
\usepackage{subcaption}
\usepackage{booktabs}
\usepackage{graphicx}

\AtBeginDocument{}

\setcopyright{none}

\title{From Wearable Data to Personalized and Actionable Health Insights}

\author{Esther Brown\textsuperscript{1}, Karis Moon, Victoria Dean, PhD\textsuperscript{2}, Finale Doshi-Velez, PhD\textsuperscript{1}}
\affiliation{%
  \institution{\\[-1pt]\normalfont\textsuperscript{1}Harvard University, Cambridge, MA;\quad \textsuperscript{2}Princeton University, Princeton, NJ}}

\date{}

\newcommand{\figsubcaption}[1]{%
  \par\smallskip
  \captionsetup{width=0.9\linewidth,justification=raggedright}%
  \caption*{\small #1}%
  \captionsetup{width=\linewidth,justification=centering}%
}
\newcommand{\figsubcaptionnospace}[1]{%
  \captionsetup{width=0.9\linewidth,justification=raggedright}%
  \caption*{\small #1}%
  \captionsetup{width=\linewidth,justification=centering}%
}

\makeatletter
\fancypagestyle{standardpagestyle}{%
  \fancyhf{}%
  \fancyhead[LE]{\ACM@linecountL\@shortauthors}
  \fancyhead[RO]{\ACM@linecountR\@shorttitle}
  \fancyhead[LO]{\ACM@linecountL}
  \fancyhead[RE]{\ACM@linecountR}
  \fancyfoot[LE,RO]{\thepage}
  \fancyfoot[LO,RE]{Manuscript submitted to Proc.\ ACM Interact.\ Mob.\ Wearable Ubiquitous Technol.}
}
\makeatother

\begin{document}

\begin{teaserfigure}
  \centering
  \includegraphics[width=0.95\textwidth]{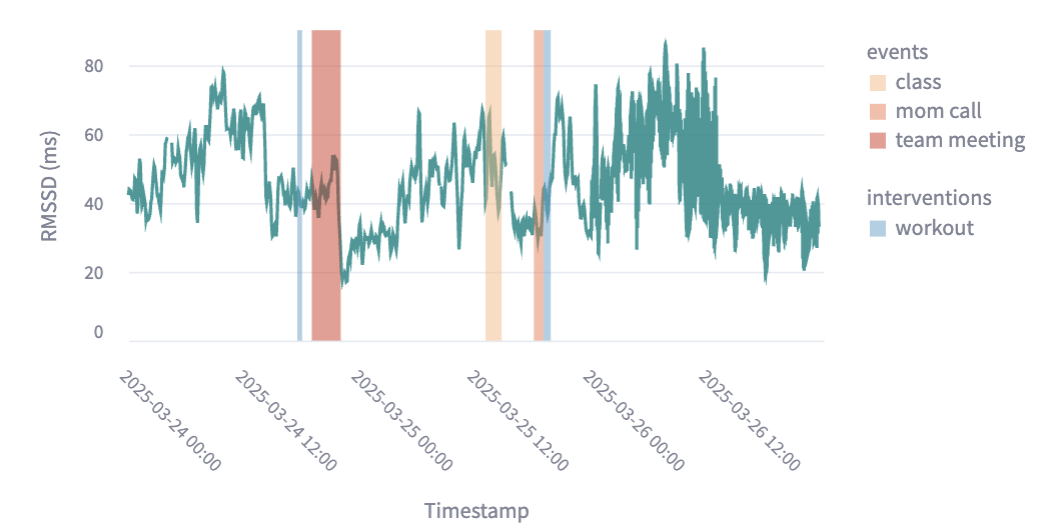}
  \caption{\textbf{Combining wearable data with human reflection:} Our interactive tool enables users to overlay user-tagged life events and interventions on physiological variables such as heart rate variability, shown here.}%Interactive view that lets users overlay activity categories on any physiological variable.}
  %\figsubcaption{This example shows a line plot of Heart Rate Variability (HRV) over a few days for a single participant. The colored horizontal bars represent user-tagged activities, allowing for visual correlation between HRV fluctuations and contextual events.}
  \label{fig:teaser_hrv}
\end{teaserfigure}

\begin{abstract}
{\normalfont\bfseries\large Abstract}\par\medskip
Commercial wearable devices continuously capture rich physiological data (e.g., heart rate, respiration), opening new possibilities for monitoring health conditions, notably around stress. 
Despite their promise, turning raw wearable physiological data streams into visualizations that surface stress-related insights in daily activities, and that ultimately foster reflection, awareness, and better stress management, remains a significant challenge. The data are noisy and context‑dependent: the same spike in heart rate can come from sprinting, a tense presentation, or laughing with friends. %As a result, people struggle to reflect and translate these ambiguous signals into insights that can help them manage their stress.
To address these challenges, we propose a framework that combines user annotations with wearable data to support better stress management.
We introduce a web framework offering interactive visualizations that layer daily activities, stress events, and interventions onto raw physiological streams, enabling users to reflect and identify trends. %allowing for insights around stress. The proof of concept empowers users to personalize their visualizations, track how different events affect their physiology over time, and uncover trends that inform better stress management. 
%Our work makes two contributions: (1) we examine how interactive and personalized visualizations that integrate participant-tagged interventions and contextual events with continuous wearable data empower users to dynamically explore their physiological responses across different contexts; and 
%(2) building on this examination, we consider various visualization techniques and propose a framework that overlays self-reported tags on physiological time-series data, enabling users to analyze the relationships between their daily activities and physiological responses—an approach that encourages reflection and provides deeper insight into when stress manifests and when an intervention is effective in recovery.
In a four-week pilot with seven university graduate and undergraduate student participants who logged 269 events, our tool revealed patterns between different types of interventions and stress: social interaction reduced average heart rate by $4.35$ to $5.0$ beats per minute, deliberate rest reduced average Garmin stress scores by $10.03$ to $13.83$ points, and mindfulness activities \emph{decreased} average HRV by $6.61$ to $13.22$ milliseconds. 
\end{abstract}

%\keywords{<keyword1>, <keyword2>, <keyword3>} % (optional for submission)

\maketitle

\section{Introduction}

Commercial wearable devices, such as Garmin, Apple Watch, Fitbit, and Whoop, generate continuous streams of physiological data (e.g., heart rate (HR), heart‑rate variability (HRV), inter‑beat intervals (BBI), and respiration), offering opportunities for real‑world health monitoring of conditions like stress and sleep. Stress, for example, is a major concern in both general and workplace settings, where it can lead to accidents, absenteeism, and reduced productivity. Recent estimates suggest that stress‑related conditions cost the global economy around \$1 trillion per year in lost productivity \cite{WHO2019MentalHealthProductivity}, with direct healthcare costs in the United States approaching \$190 billion annually \cite{Goh2016WorkplaceStress}.  Although these devices provide continuous tracking beyond occasional clinical assessments, turning raw wearable physiological data streams into visualizations that surface stress-related insights in daily activities - and that ultimately foster reflection, awareness, and better stress management - remains a significant challenge. In particular, inherent noise, context dependency, and inter‑individual variability mean that the same physiological signal (e.g., an elevated heart rate) may result from stress, exercise, or even positive emotional experiences. This ambiguity makes it difficult to reflect on how and when different activities or events can help reduce physiological stress.

To overcome this ambiguity, our work uses \textbf{activity tagging}—the process by which participants annotate their daily lives with events and activities. By pairing these self‑reported tags with their physiological sensor data, we empower users to see how different activities and events in their day are associated with stress‑related physiological changes. In our approach, we define an \emph{activity} as either an intervention (an intentional action aimed at reducing stress) or any other contextual event. These self‑reported tags provide essential contextual information that enables users to gain a deeper understanding of their physiological data by linking sensor measurements to their daily experiences.

Prior work has demonstrated the potential of wearable data for stress monitoring. For instance, Chalmers \textit{et al.}~\cite{chalmers2021stress} explored differences in HR and HRV between resting and stress states in controlled settings, while Dunn \textit{et al.}~\cite{dunn2021wearable} showed that wearable data can reliably predict clinical laboratory outcomes. Similarly, Hickey \textit{et al.}~\cite{hickey2021smart} and Pakhomov \textit{et al.}~\cite{pakhomov2020using} found that metrics such as HRV and heart rate changes correlate with stress, and Smith \textit{et al.}~\cite{smith2020integrating} applied wearable technology to monitor stress in workplace environments. However, these studies primarily focus on stress detection without addressing the dynamic impact of interventions. Moreover, previous work on contextual self-reporting \cite{neupane2024momentary,novak2023empowering} illustrates that user-reported context can significantly enhance the interpretation of wearable signals.

\textbf{In this work we ask:}
\begin{enumerate}
    \item \textbf{RQ1:} When self‑tagged context is integrated with continuous wearable data, what patterns emerge between everyday activities/interventions and stress‑related physiology (HR, HRV, BBI, subjective stress)?
    \item \textbf{RQ2:} Can interactive visualizations that overlay self-reported tags on physiological time-series data empower and influence users’ ability to interpret their data and choose stress‑management strategies?
\end{enumerate}

Based on these questions, we propose the following hypotheses:

\begin{enumerate}
    \item \textbf{H1 (Pattern Discovery):} Integrating self-tagged contextual data with wearable sensor outputs will uncover patterns that correlate specific activities or interventions with improvements in physiological markers related to stress.
    \item \textbf{H2 (Enhanced Self-Management):} When self‑tagged activities are overlaid on their physiological time series, users will find their data easier to interpret and feel empowered to explore their data and adopt more effective stress management strategies.
\end{enumerate}

\textbf{Our Contributions:}
\begin{enumerate}[noitemsep]
    \item We examine how interactive and personalized visualizations that integrate participant-tagged interventions and contextual events with continuous wearable data empower users to dynamically explore their physiological responses across different contexts. 
    \item Building on this examination, we consider various visualization techniques and propose a framework that overlays self-reported tags on physiological time-series data, enabling users to analyze the relationships between their daily activities and physiological responses. Participants found the tool intuitive and reported uncovering actionable insights for managing stress through the tool. One user realized they became more stressed about meetings than necessary, with stress levels gradually decreasing as the meeting progressed. This insight prompted them to try using positive self-talk to reduce anticipatory stress before the scheduled meeting.

\end{enumerate}

In a four-week pilot with seven undergraduate and graduate university participants who logged 269 events, our tool showed stress patterns: social interaction reduced heart rate by $\approx$5 bpm in 30 minutes, deliberate rest reduced Garmin stress scores by $\approx$8 points, and sometimes mindfulness \emph{decreased} HRV.  We propose a framework that combines user annotations with wearable data to support better stress management in social, workplace, and clinical settings.

\section{Related Work}
\label{sec:relatedwork}

\subsection{Systems to Support Self-experimentation}
A growing body of work has focused on building systems that help people use self-tracking data to run N-of-1 experiments in daily life. For example, Daskalova \textit{et al.}~\cite{daskalova2016sleepcoacher} designed \textit{SleepCoacher}, a self-experimentation framework focused on sleep. The SleepCoacher framework integrates mobile sleep sensing with clinician-authored recommendation templates and guides users through short AB-style experiments to evaluate which behaviors improve their own sleep. Daskalova \textit{et al.}~\cite{daskalova2021selfe} also introduced \textit{Self-E}, which guides and generalizes users' self-experimentation beyond a single domain (e.g., sleep) to include nutrition, physical activity, and meditation. In Self-E, the smartphone app scaffolds the full self-experimentation loop: selecting and designing interventions, randomizing conditions, collecting and interpreting results, to lower the barrier for users with little to no personal analytics background. In addition, Karkar \textit{et al.}~\cite{karkar2017tummytrials} introduced \textit{TummyTrials}, a system that supports people with irritable bowel syndrome (IBS) in testing individualized dietary triggers via randomized self-experiments. Related work has broadened the ecosystem: Kravitz et~al.'s Trialist (PREEMPT) randomized clinical trial evaluated smartphone-supported N-of-1 trials for chronic musculoskeletal pain~\cite{kravitz2018preempt}, and platforms such as Taylor \textit{et al.}'s \textit{QuantifyMe}~\cite{taylor2018quantifyme}, Konigorski \textit{et al.}'s \textit{StudyU}~\cite{konigorski2022studyu}, and Wernicke \textit{et al.}'s \textit{StudyMe}~\cite{wernicke2022studyme} facilitate the design and execution of N-of-1 trials. Together, these systems show the feasibility of everyday self-experiments while highlighting open challenges around balancing freedom and guidance, maintaining rigor with low burden, and supporting trustworthy interpretation of individualized results~\cite{daskalova2016sleepcoacher,daskalova2021selfe,karkar2017tummytrials,kravitz2018preempt,taylor2018quantifyme,konigorski2022studyu,wernicke2022studyme}. Our work extends this line of research by empowering people to layer annotations onto their raw physiological data—capturing, for example, a tense team meeting or a supportive chat with friends—and to reflect on the resulting patterns, turning personal data into actionable insights.

% \subsection{Social Context in Physiological Data}

% Prior CSCW work shows that physiological self-tracking can be
% insightful when the data are interpreted \emph{with} people’s input and situated through users’ own reflections and annotations.  
% In clinical settings, Chung \textit{et al.}~\cite{Chung2016CSCW} and
% Mamykina \textit{et al.}~\cite{Mamykina2008CHI} describe how
% patients and clinicians treat logged biometrics as
% \emph{boundary-negotiating artifacts}—reference points for reflection and coordination—to co-construct treatment plans.  
% In workplace contexts, Holten Møller \textit{et al.}~\cite{HoltenMoller2021GROUP} ran reflection workshops in which employees questioned algorithmic stress labels and reconciled the numbers with their lived experience. Together, these studies demonstrate that physiological data from wearable devices can
% support stress monitoring in social settings when user experiences and annotations are incorporated. Our work extends this line of research by empowering people to layer annotations onto their raw physiological data, capturing, for example, a tense team meeting or a supportive chat with friends, and to reflect on the resulting patterns, turning personal data into actionable insights for cooperative sense-making across different social contexts.

\subsection{Wearable Data Analysis and Stress Monitoring}
Wearable devices continuously collect physiological data such as heart rate, heart rate variability (HRV), respiration, and beat-to-beat interval (BBI), enabling the detection of stress-related changes. For example, Chalmers \textit{et al.}~\cite{chalmers2021stress} explored differences in HR and HRV between resting and stress states, although without considering the effect of interventions. Dunn \textit{et al.}~\cite{dunn2021wearable} demonstrated that wearable data can predict clinical laboratory outcomes more accurately than conventional measures, emphasizing the reliability of resting heart rate data. Moreover, Hickey \textit{et al.}~\cite{hickey2021smart} compared various wearable statistics and found that HRV, electrodermal activity, and respiratory rate may serve as more robust stress indicators than heart rate alone. Pakhomov \textit{et al.}~\cite{pakhomov2020using} showed that wearable-detected heart rate changes correlate with stressful events, while Smith \textit{et al.}~\cite{smith2020integrating} applied wearable technology to monitor stress following mindfulness interventions. Collectively, these studies highlight the potential of wearable data for stress monitoring but focus on stress detection rather than the dynamic impact of interventions.

\subsection{Self-Tagging and Context-Based Monitoring}
A growing body of work highlights the importance of integrating user-generated context with sensor data. Neupane \textit{et al.}~\cite{neupane2024momentary} used a mobile app to prompt participants to log stressors throughout the day, while Novak \textit{et al.}~\cite{novak2023empowering} used a combination of wearable and environmental sensors to support participatory research in urban health. Suh \textit{et al.}~\cite{suh2024toward} took an alternative approach by having participants log their feelings multiple times per day and then inviting those who were stressed to participate in interventions. Although these methods demonstrate the value of contextual self-reporting, they largely rely on retrospective analysis rather than integrating real-time annotations with continuous wearable data.

\subsection{Interactive Machine Learning and Visual Analytics}
Interactive machine learning (ML) and visual analytics have proven to be effective in improving the interpretability and performance of the model by incorporating user feedback. Comito \textit{et al.}~\cite{comito2021diagnosis} used ML to identify diagnostic patterns in physiological parameters, while Evin \textit{et al.}~\cite{evin2022personality} correlated physiological data with personality and behavioral metrics. Olyanasab and Annabestani~\cite{Olyanasab2024} reviewed ML applications in wearables, categorizing methods into bioelectrical, bioimpedance/electrochemical, and electromechanical. Tazarv \textit{et al.}~\cite{tazarv2023active} explored active reinforcement learning for personalized stress monitoring, and Xu \textit{et al.}~\cite{xu2014cluster} applied clustering techniques to personalize stress evaluation. Despite these advances, many of these approaches operate as “black boxes” without incorporating user context, thereby limiting their practical relevance.

\subsection{Machine Learning-Driven Personalization in Wearable Health}
Recent work has begun to focus on enhancing personalization in wearable health systems through ML. Olyanasab and Annabestani~\cite{Olyanasab2024} provided a comprehensive review of ML applications for personalized wearable biomedical devices, emphasizing early intervention. Brons \textit{et al.}~\cite{Brons2024} reviewed ML methods for personalizing persuasive strategies in mHealth interventions, while Fang \textit{et al.}~\cite{Fang2024} developed a deep reinforcement learning algorithm for dynamically personalizing exercise goals. In addition, interactive visualization systems such as CarePortal~\cite{Sadhu2023} and RemoteHealthConnect~\cite{Arun2024} have been proposed to present wearable data in clinician- and patient-friendly formats. Abdelaal \textit{et al.}~\cite{Abdelaal2024} have explored explainable AI techniques for making wearable data analytics more interpretable, and Holzinger~\cite{Holzinger2016} highlighted the benefits of human-in-the-loop ML in health informatics. Esna Ashari and Ghasemzadeh~\cite{Ashari2019} introduced an active learning framework tailored to wearable sensor data, while Wu \textit{et al.}~\cite{Wu2023} demonstrated that integrating clinical expertise into reinforcement learning can improve treatment outcomes. Finally, Zhu \textit{et al.}~\cite{Zhu2022} and Lazarou and Exarchos~\cite{Lazarou2024} further advanced multimodal and real-time stress detection models. These studies collectively suggest that integrating ML with wearable data has the potential to deliver more personalized and actionable health insights.

\section{System Design}

\noindent
\subsection{Visualization Tool}

For this work, we built a web‑based visualization tool with the goal of empowering participants by showing how their self‑tagged interventions and events relate to changes in physiological responses.  
This tool directly supports both of our hypotheses, which center on integrating self‑tagged contextual data with wearable sensor output to help participants (1) discover patterns around their stress and (2) gain insight into how certain interventions might be more or less helpful for stress management.\\

\noindent \textbf{Tool Architecture and Workflow}

The visualization tool merges continuous wearable data—such as beat‑to‑beat interval (BBI) and heart‑rate variability (HRV)—with events and interventions tagged by participants. Its main view shows interactive time‑series plots annotated with these tags. Users can zoom in to inspect the pre‑ and post‑intervention time windows and filter by tag type (e.g., interventions vs.\ contextual events) to explore how specific activities impact their physiological response curves.\\

\noindent \textbf{Viewing Annotations}

\begin{figure}[htbp]
    \centering
    \includegraphics[width=0.48\textwidth]{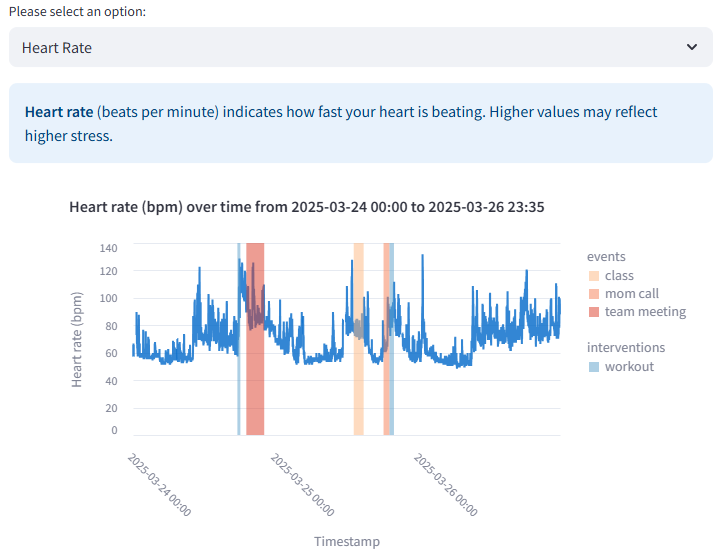}
    \includegraphics[width=0.48\textwidth]{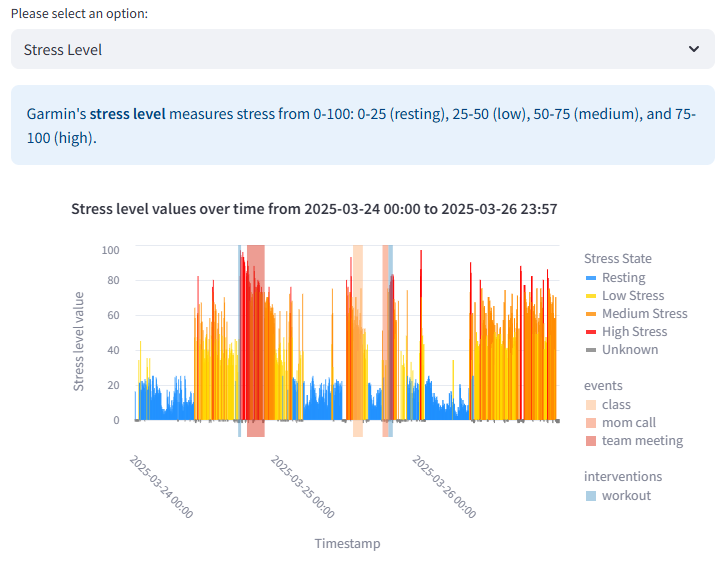}
     \includegraphics[width=0.48\textwidth]{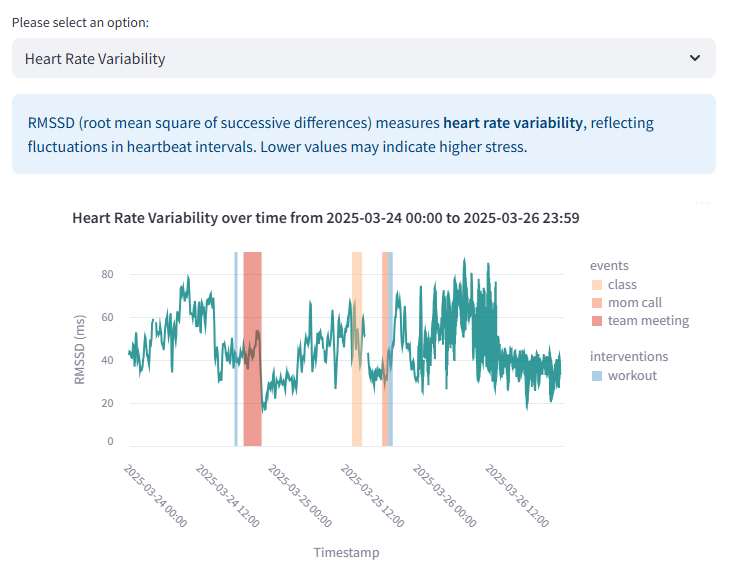}
    \caption{\textbf{View Annotations}: Users can view their data for any of the six variables, with any combination of intervention categories overlaid. Here, a user's data for Heart Rate, Stress Level, and HRV are shown.}
    \label{fig:view_annotations}
\end{figure}

Users can view their annotations overlaid on their data for any of the physiological variables, as shown in Figures~\ref{fig:teaser_hrv} and ~\ref{fig:view_annotations}. Examples of visualizations with the additional physiological variables can be seen in Figure~\ref{fig:more_annotated_visuals} in Appendix A. Users can also toggle events and interventions to view one, both, or none overlaid on their physiological data using a checkbox interface in the tool's sidebar. \\

% \begin{wrapfigure}{l}{0.48\textwidth} 
%   \vspace{-6pt} % (optional) pull figure up a bit
%   \centering
%   \includegraphics[width=\linewidth]{img/show_annotations/hr.png}\\[3pt]
%   \includegraphics[width=\linewidth]{img/show_annotations/stress.png}\\[3pt]
%   \includegraphics[width=\linewidth]{img/show_annotations/hrv.png}
%   \caption{View Annotations}
%   \figsubcaptionnospace{
%     Users can view their data for any of the six variables, with any combination of intervention categories overlaid. Here, a participant’s Heart Rate, Stress Level, and HRV are shown.
%   }
%   \label{fig:view_annotations}
% \end{wrapfigure}

\noindent \textbf{Intervention Comparison} 
\label{sec:Intervention_Comparison}

Users can select the intervention they want to compare, the physiological features they are interested in analyzing, and the amount of time before and after each intervention or event to visualize, as shown in Figure~\ref{fig:intervention_selection} in Appendix A.

\begin{figure}[htbp]
    \centering
    \includegraphics[width=0.48\textwidth]{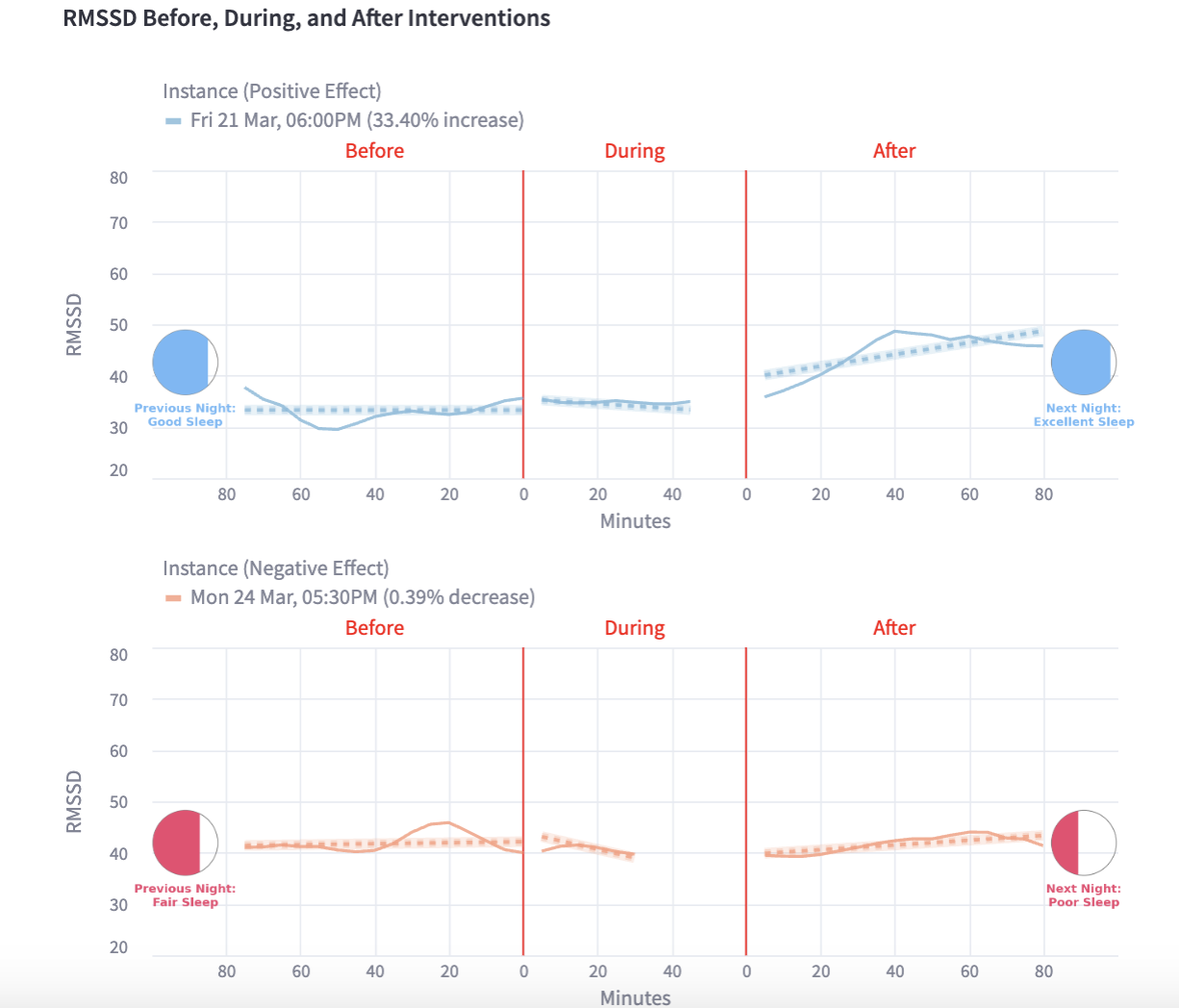}
    
    \caption{Compare Interventions}
\figsubcaption{This visualization displays a user’s data for the selected variable before, during, and after each intervention, together with the sleep quality of the preceding and following nights. Each intervention instance can be shown individually or overlaid with the others (see Figure~\ref{fig:intervention_aggregate} in Appendix~A). The comparison plot also highlights the overall effect across all instances—blue indicates a positive effect and red a negative effect.}
    \label{fig:intervention_visualizations}
\end{figure}

The visualization tool then generates a set of plots (see Figure~\ref{fig:intervention_visualizations}) displaying the participant's physiological sensor data before, during, and after the selected intervention or event. Each plot shows separate instances of the same intervention or event, for example, the physiological response to ``meditation'' on Monday compared to the response to "meditation'' on Tuesday.

Participants can customize these visualizations further by choosing to view each instance of the intervention in separate plots or overlay all instances onto a single plot. When visualizations are overlaid, the physiological data for each instance is aligned at a common starting point (0 minutes), marking the beginning of each intervention or event. Additionally, each intervention or event instance is color-coded to indicate its overall reported effect on stress (blue for positive effects and red for negative effects) for that selected time window. By changing the size of the time window, users can explore the amount of time it takes for an intervention to have a positive effect on their physiological data and when it stops having a positive effect or ``stops working". 

Finally, we also display participants' sleep quality scores (measured by Garmin’s proprietary sleep metric) from the night before each intervention to provide additional context around the intervention’s impact on their stress-related metrics. Showing sleep quality alongside these visualizations helps participants understand why certain instances of an intervention appear more effective than others. For example, even if meditation generally has a positive effect on a user's stress-related physiological markers, it might be less effective following a night of poor sleep, as the user may not have fully recovered overnight.

\noindent \textbf{Aggregate Comparison} 

In addition to the plots comparing the detailed trajectories of the user's physiological responses, we provide a bar plot showing the aggregate effect of each intervention or event. This is reported as the percent change in the sensor data variable before and after the intervention or event. The aggregate plot allows participants to quickly compare the overall effect of many different instances of an intervention or event (e.g. "meditation" on Monday decreased stress by 10\% but "meditation" on Tuesday decreased stress by only 1\%), seen in Figure~\ref{fig:percent_comparison} in Appendix A.

% 5. FEATURE EXTRACTION DESIGN CHOICES AND RATIONALE
\subsection{Feature Extraction Visualizations}
\label{sec:feature_extraction_design}

Within our visualization tool, we provide additional \emph{feature extraction visualizations} designed to help users quickly compare the effects of all their recorded events and interventions at the same time. These visual summaries enable users to rapidly identify when their stress-related physiological metrics show improvement and uncover trends across different types of activities.

\noindent \textbf{Analysis Window Configuration}

In the feature extraction visualizations, users can customize the analysis window, choosing intervals ranging from 5 to 120 minutes before and after each event or intervention, as depicted in Figure~\ref{fig:analysis_options} in Appendix A. This flexibility allows users to examine both short-term and longer-term physiological responses associated with their activities, events, or interventions.

\noindent \textbf{Change Analysis}

\begin{figure}[htbp]
  \centering
  % ---------- left panel ----------
  \begin{minipage}[t]{0.48\textwidth}
    \centering
    \captionsetup{width=\linewidth,justification=centering}
    \includegraphics[width=\linewidth]{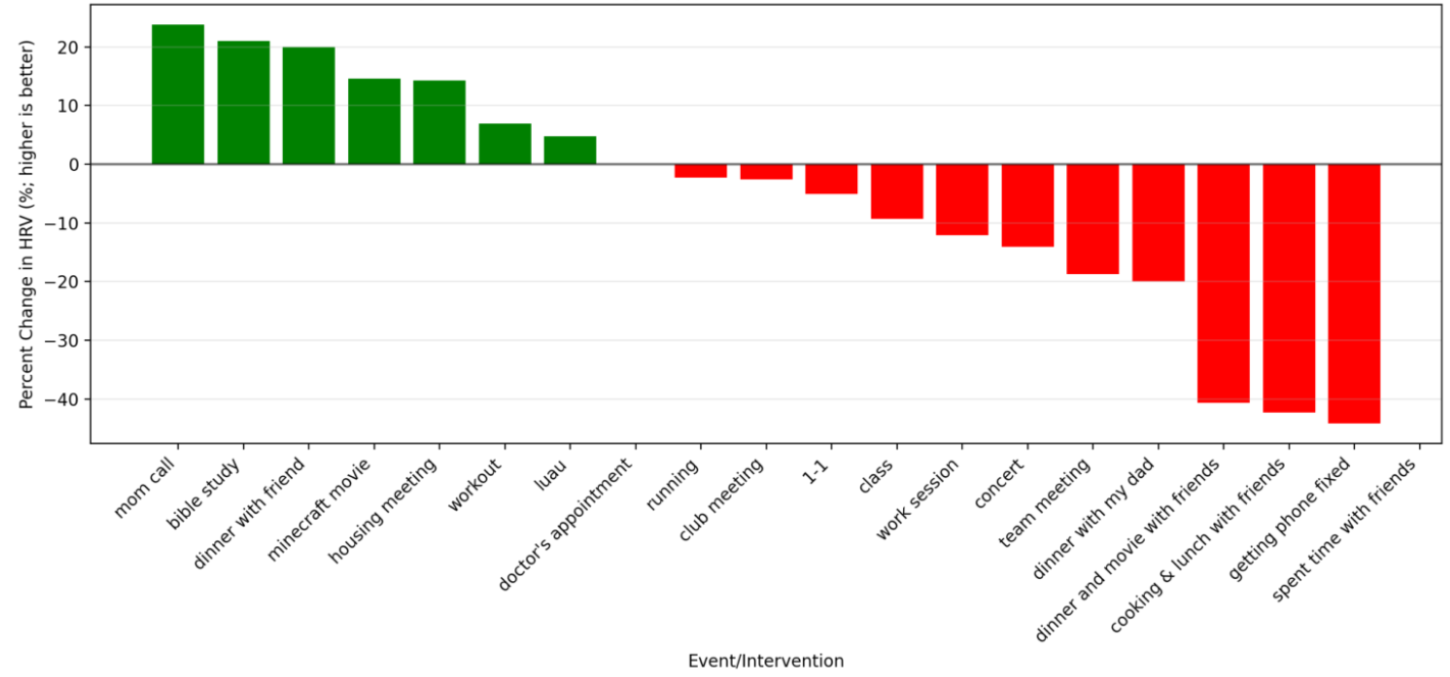}
    \caption{HRV Change Analysis Visualization}
    \figsubcaption{This visualization shows the percentage change
      in HRV before and after each activity for a single user. Green bars indicate improvements, while red bars indicate declines. Activities are sorted from positive to negative, allowing for quick identification of the most or least effective interventions.}
    \label{fig:hrv_change_analysis}
  \end{minipage}
  \hfill
  % ---------- right panel ----------
  \begin{minipage}[t]{0.48\textwidth}
    \centering
    \captionsetup{width=\linewidth}
    \includegraphics[width=\linewidth]{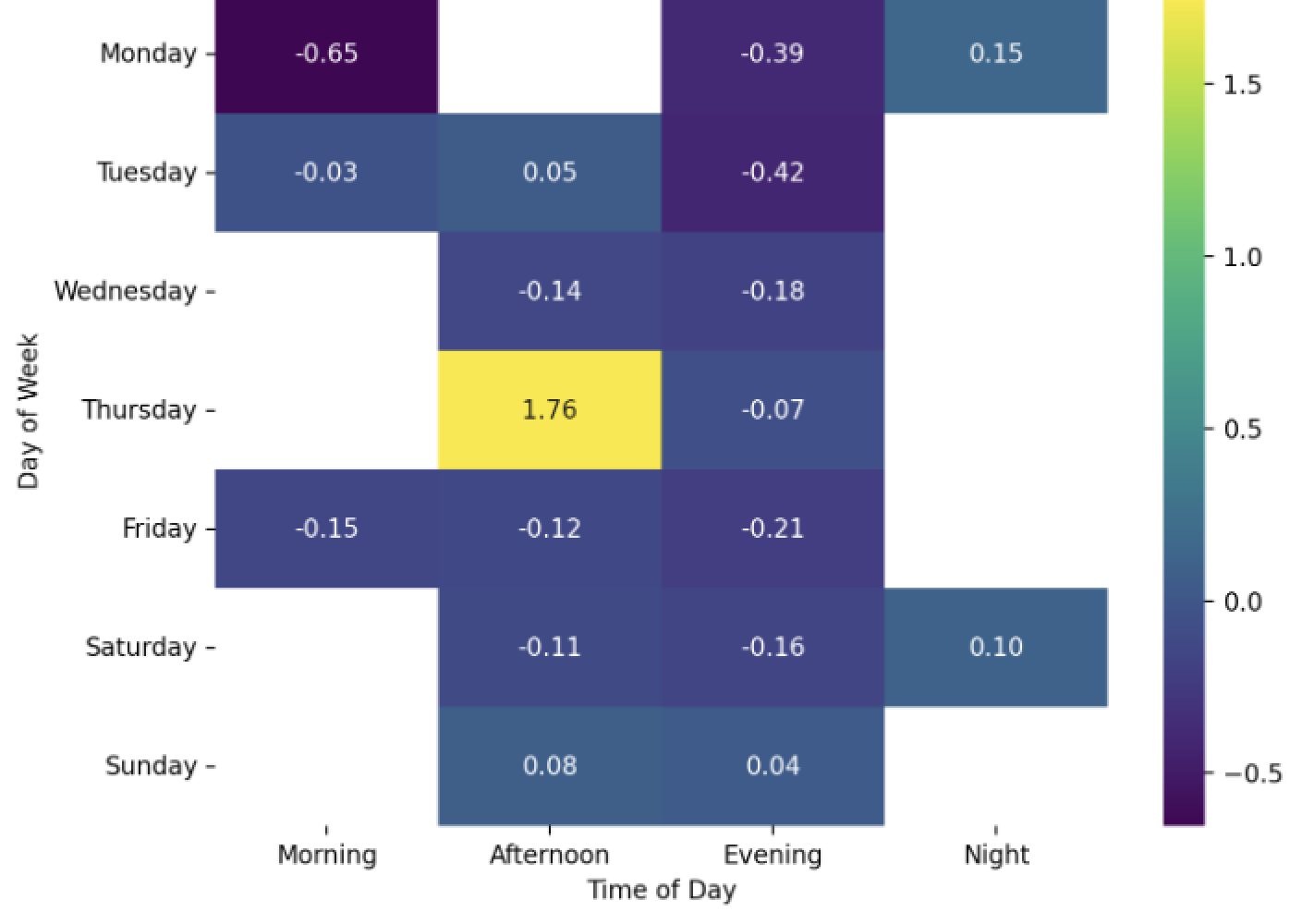}
    \caption{Temporal Heatmap Patterns of Stress Reduction}
    \figsubcaptionnospace{Average standardized stress-reduction scores for one participant, aggregated by day of week and time of day. Lighter cells mark larger reductions (e.g.\ +1.76 on Thursday afternoon); darker cells mark increases (e.g.\ –0.65 on Monday morning). In an exit interview, the participant noted that Thursday afternoons coincide with the end of their weekly meetings, which may explain the pronounced stress reduction at that time.}
    \label{fig:temporal_patterns_heatmap}
  \end{minipage}
\end{figure}

To summarize the physiological impact of various activities and interventions, we include a bar chart visualization showing the average percentage change of physiological variables such as HRV, stress level, and BBI across different event or intervention types. Figure~\ref{fig:hrv_change_analysis} illustrates this visualization specifically for Heart Rate Variability (HRV). In this figure, higher values (represented in green) indicate a more relaxed physiological state and improved recovery. Similar visualizations for BBI and stress can be found in Figure~\ref{fig:bbi_change_analysis} and Figure~\ref{fig:stress_metric_analysis} in Appendix A, respectively.

\noindent \textbf{Temporal Patterns Heatmap}  

Users are provided with a temporal patterns heatmap that visualizes standardized stress reduction scores, organized by day of the week and time of day, as shown in Figure~\ref{fig:temporal_patterns_heatmap}. By aggregating a user’s physiological data across multiple events and interventions, the heatmap helps identify optimal timing for stress management activities. Darker colors (purple/blue) indicate periods of less effective or negative stress reduction, while brighter colors (yellow) reflect higher average stress reduction effectiveness. For example, users might find that they experience greater stress relief on Thursday afternoons and less effective stress reduction on Monday mornings—insights that can prompt reflection on the types of activities or interventions performed during those times and guide more intentional scheduling of future stress-reducing actions.

\noindent \textbf{Continuous Change Analysis}

To help users better understand their recovery trajectories, the tool also plots \emph{average percentage change} in physiological measurements, HRV, stress, and BBI, for the first 60 minutes after the end of an activity. Time \(t = 0\) marks the exact moment the event or intervention ends, so users can compare how they respond to different activities over time. An example plot for a participant in the study is shown in Appendix A (Figure \ref{fig:continuous_hrv_change_analysis}). The purpose of this plot is to help users assess which interventions generally improve recovery and how long their effects typically last. In that figure, each colored line represents one activity type that a user logged (e.g., \textit{Meditation}, \textit{Walk}, \textit{Nap}) and is calculated by averaging all instances of that activity (e.g., meditation sessions on Monday, Tuesday, etc.). By summarizing the typical post-intervention trajectory of each intervention type, users gain an overview of how they respond to different interventions. If users want finer details, they can drill down to individual instances of that intervention type through the intervention-comparison view described in Section \ref{sec:Intervention_Comparison}.

\noindent \textbf{Intervention Category Impact Table} 

To complement the continuous change plots, we include a summary table showing different categories of interventions and their impact on BBI or HRV at 15, 30, and 60-minute intervals, with associated frequency percentages personalized to the interventions logged by that specific user. We chose these three time windows because prior work on wearable stress
dynamics suggests that most autonomic recovery unfolds within the first hour
after a stressor or intervention \cite{Shaffer2017}.  The 15‑minute
column captures immediate effects, the 30‑minute column reflects medium‑term
adaptation, and the 60‑minute column shows whether physiological impacts persist. Figures~\ref{fig:hrv_intervention_category_impact_table} and \ref{fig:bbi_intervention_category_impact_table}, both found in Appendix A, display the table for HRV and BBI, respectively.\\

\section{Methods}
\subsection{Study Design and Data Collection}
\label{sec:study_design}

\textbf{Study Overview:}The IRB-approved study by [anonymous institutions] consisted of two phases. In the first phase, participants wore a Garmin Vívosmart 5 for two weeks to collect raw physiological data while tagging interventions and contextual events. They were given login credentials for our visualization web tool, where they could explore personalized physiological responses to their own annotations (logged interventions and events). In a follow-up phase, participants from the first phase who agreed to continue collected additional physiological data while continuing to tag events and interventions. Participants were then given enhanced visualizations, also personalized to their own data and annotations. The purpose of these additional visualizations was to make insights from their data easier to find. \\

\noindent
In Phase 1, our primary objective was to collect rich, self-tagged data linking wearable sensor streams with daily activities and interventions. This phase forms the core of our study.

\textbf{Participants:} Seven students (graduate and undergraduate) from two academic institutions participated in the study. For demographics, participants’ ages averaged 22.6 years (median 24, SD $\approx$ 2.9); five identified as women and two as men. Participants were recruited through university-wide email lists and campus posters. Based on pre-study survey responses, participants reported a range of stressors (from academic deadlines to poor sleep) and used various strategies to manage stress, including walking, journaling and music. Most of the participants had previous experience with wearable devices or physiological self-tracking. Retention and participation were high, and six of the seven participants continued in the optional second phase of the study. Despite the small size of the cohort, participants logged 269 events and interventions over the course of the study, capturing a diverse set of self-tagged activities.

\textbf{Devices:} Each participant was provided a Garmin Vívosmart 5, which was to be worn continuously (day and night) for a period of two weeks. It was recommended that the device be removed only for charging.

\textbf{Procedure:} Participants completed a pre-study survey that captured their baseline stress behaviors, use of wearable devices, common stressors, and preferred coping strategies. Informed consent was obtained via a required checkbox at the beginning of the survey form. Over the two-week period, participants were asked to tag daily interventions (e.g., meditation, listening to music) and contextual events (e.g., meetings, work sessions). In addition, participants synced their wearable data daily, allowing us to continuously capture physiological data. At the end of the study, exit interviews were conducted to gather anecdotal evidence of participants' experiences, insights gained, and how the tool influenced their stress awareness and self-management strategies.

\textbf{Interventions and Context:} Participants annotated both intentional stress-reduction actions (interventions) and everyday events. They had three main options for tagging: (i) syncing a calendar (we encouraged creating a study-specific calendar to protect privacy); (ii) using an annotation form to log events or interventions as they occurred—capturing start and end times, names, and anticipated effects on stress (positive, negative, or neutral); or (iii) annotating directly in our web tool (see Figures~\ref{fig:google_form} and~\ref{fig:intervention_selection} in the appendix). All annotations were automatically synced with participants’ physiological data and displayed in the visualization interface.

\textbf{Daily Tagging and Syncing:} Participants were expected to log at least two events per day related to stress and sleep management. Tagging and syncing together took approximately 5–10 minutes daily.

\textbf{Compensation:} Each participant received a \$25 gift card as token compensation for their participation.

\noindent
After the first week of Phase One, participants received login credentials for a website where they could visualize their annotations overlaid on their physiological data. Sensor data and logged annotations were automatically synced to their accounts every three days by the research team.

Phase 2 lasted an additional two weeks and began immediately after Phase One for participants who opted to continue. During this phase, participants continued wearing their Garmin devices and tagging events and interventions daily. They were also provided with an expanded set of visualizations (see Section~\ref{sec:feature_extraction_design}) to help them more easily identify when interventions were most effective and how different events influenced their physiological responses.

\subsection{Feature Selection and Processing}
\label{sec:features_design_calculations}

In our analysis, we process several physiological features from Garmin data,
each with its own design choices tailored to capture the underlying signal
while balancing noise. In this subsection, we describe the processing
methodology for each feature, explain our design decisions, and provide the
mathematical formulas.%, particularly for HRV (calculated via RMSSD with BBI).

% 1. FEATURE DESIGN AND PROCESSING CHOICES
\subsubsection{Physiological Features and Data Processing}

\label{sec:feature_design_processing}

We analyze six physiological signals collected from a Garmin
Vívosmart 5 using the Labfront Continuous Data Collection Companion App
\cite{LabfrontHelp}. The physiological features and their temporal resolutions are: beat-to-beat intervals (BBI) at every heartbeat (the device’s finest resolution), heart
rate and step count at 1-minute resolution, Garmin stress score and respiration
rate at 3-minute resolution, and nightly sleep summaries auto-detected by Garmin Connect. All of these features were configured to the most granular sampling settings supported by the device at the time of this study. All timestamps were then converted from UTC to the participant’s local time (America, Eastern Time, Daylight Saving Time-aware). While Garmin does not expose HRV directly on the Vívosmart 5, we compute it from the raw BBI data (as described in Section \ref{sec:calculations_formulas}). We also use Garmin’s 0–100 stress index and map it to four
levels (resting: 0–25; low: 26–50; medium: 51–75; high: 76–100). The 0–100 sleep score was similarly classified into Garmin's tiers: excellent (90–100), good (80–89), fair (60–79), and poor (< 60).

Table~\ref{tab:design_choices} summarizes each variable, its display unit or
scale, and its relevance to stress monitoring and recovery.

\newcolumntype{Y}{>{\raggedright\arraybackslash}X}
\begin{table*}[htbp]
\small
\setlength{\tabcolsep}{6pt}
\renewcommand{\arraystretch}{1.15}
\centering
\begin{tabularx}{\textwidth}{@{}l c Y@{}}
\toprule
\textbf{Physiological Feature} & \textbf{Unit / Scale} & \textbf{Interpretation and Rationale} \\
\midrule
Heart-Rate Variability (HRV) & ms &
HRV derived from beat-to-beat intervals \textbf{BBI} quantifies the variability between consecutive heartbeats and is widely considered one of the best indicators of stress and recovery \cite{Shaffer2017,Kim2018}. High HRV reflects robust autonomic adaptability, whereas low HRV is linked to increased stress and fatigue. Because HRV can be influenced by physical activity and
age, contextual data improve interpretation \cite{Li2023}. \\ \midrule
Stress & 0–100 scale &
Garmin’s proprietary stress score (Stress Score) combines HR and HRV to give an estimate of
stress~\cite{jerath2023future}. Lower values indicate rest; higher values
indicate physiological arousal. Because excitement and exercise can also
elevate the score \cite{booth2022toward}, it should be interpreted in context.
We include it because Garmin users are already familiar with the metric
\cite{Rosenbach2025}. \\ \midrule
Steps Taken & steps at timestamp &
Step count provides general activity context. Including it helps distinguish
whether HR/HRV changes stem from psychological stress or physical exertion,
improving stress-detection accuracy \cite{Smets2018}. \\ \midrule
Heart Rate & beats per minute (bpm) &
Heart rate can be an indicator of autonomic arousal (``fight-or-flight'').
Because HR also rises with exercise, caffeine, etc., we interpret it alongside
HRV and step count for a fuller stress picture \cite{Iqbal2022}. \\ \midrule
Respiration Rate & breaths per minute &
Breathing rate increases under acute stress (tachypnea) and slows during
recovery \cite{tiwari2019breathing}. It adds insight into autonomic function
and the effect of relaxation techniques, though speech and activity can also
alter it \cite{Fincham2023}. \\ \midrule
Sleep Summary & 0–100 scale &
Garmin derives a nightly sleep score from duration, stress during sleep, and
stage composition \cite{Firstbeat2019,FirstbeatSleepDetection}. We include it
because next-day stress responses could be influenced by prior-night sleep
quality\cite{wen2024nightly}. Scores are labeled Excellent (90–100),
Good (80–89), Fair (60–79), and Poor (0–59). \\
\bottomrule
\end{tabularx}
\caption{Physiological features used in this work, along with their units and
the rationale for each feature’s role in stress monitoring and recovery.}
\label{tab:design_choices}
\end{table*}

% 2. CALCULATIONS AND FORMULAS
\subsubsection{Calculations and Formulas}
\label{sec:calculations_formulas}
We derive HRV as the root-mean-square of successive differences (RMSSD) from the raw BBI series. RMSSD is a frequently used metric in prior work involving wearable devices \cite{finseth2023real, sheridan2021heart, jaafar2021analysis, sheridan2020heart}.

%The primary calculation for heart rate variability (HRV) in our pipeline is based on RMSSD (Root Mean Square of Successive Differences), 
%which is derived from raw beat-to-beat interval (BBI) data. %After exploring different methods for computing HRV, we found that 
RMSSD is computed using the following equations:

\begin{align}
\text{RMSSD} &= \sqrt{\frac{1}{N} \sum_{i=1}^{N} \left(\text{BBI}_i - \text{BBI}_{i-1}\right)^2}, \label{eq:rmssd}
\end{align}

In Equation~\eqref{eq:rmssd}, $\text{BBI}_i -
\text{BBI}_{i-1}$ is the difference
between consecutive beat-to-beat intervals, representing the instantaneous variability between
heartbeats. Here, $\text{BBI}_i$ is the $i$th interval (measured in milliseconds). The variable $N$ is the total number of BBI intervals within the analysis window, with a minimum requirement of $N \ge 20$. To ensure data continuity, intervals are only included if there are no significant recording
gaps (> 30 minutes) in the data, which could indicate device removal or sensor issues. This minimum count threshold is necessary to ensure that the RMSSD
calculation is statistically robust—fewer than 20 intervals could make the variance overly sensitive to noise or transient artifacts, leading to unreliable HRV estimates. The RMSSD provides a robust measure of HRV; higher RMSSD values suggest better recovery and a more balanced autonomic nervous system \cite{ebersole2020contribution}. \\

RMSSD is computed from a 15 minute sliding window that advances every 30 seconds. A
30 second step size captures sufficient heartbeats for a reliable
ultra-short HRV estimate—while still tracking rapid autonomic changes,
mirroring other wearable-stress studies
\cite{Dalmeida2021HRVStress,Tervonen2021UltraShort,%
Velmovitsky2023AppleWatch,Liu2023DriverStress,Darwish2025ThreeStage}.

\noindent\textbf{HRV RMSSD Smoothing Window}

The RMSSD is computed over a rolling window of 15 minutes from the raw BBI
data, with the result re-sampled every 30 seconds. This window captures
short-term variations in HRV while ensuring that enough data points are
available for a reliable estimate. Our analysis (see
Figure~\ref{fig:rolling_hrv_search} in Appendix~A) shows that a 15-minute
window with a 30-second step provides a good balance between temporal
resolution and statistical robustness. \\

\noindent\textbf{Visualization Smoothing}

For visualization purposes, both BBI and RMSSD data are further smoothed to
enhance readability and highlight underlying trends. When plotting
trajectories, we apply a centered 5-point rolling average to the data. The
smoothed value at index $i$ for a time series
$\{\text{RMSSD}_1, \text{RMSSD}_2, \ldots, \text{RMSSD}_n\}$ (sampled at
30-second intervals) is given by:
\begin{equation}
\text{RMSSD}_{\text{vis}}(i) =
\frac{1}{\min(5,\, 2i-1,\, 2(n-i)+1)}
      \sum_{j=\max(1,\, i-2)}^{\min(n,\, i+2)} \text{RMSSD}_j.
\end{equation}

The same formula is applied to the BBI data (with $\text{BBI}$ substituted for
$\text{RMSSD}$), resulting in a centered moving average that spans
approximately 2 minutes (5 data points × 30 seconds). This procedure reduces
high-frequency fluctuations while preserving meaningful patterns.

% 6. USABILITY AND INTERACTION FEEDBACK
\subsection{Qualitative Measures}
\label{sec:usability_interaction_feedback}

%In addition to building a pipeline that allows users to visualize their own physiological data, 
This pilot study is designed to observe user behavior and reflections through the following measures:
\begin{enumerate}
    \item \textbf{Ease of Use:} How intuitively users can navigate and interact with the visualization tool.
    \item \textbf{Frequency of Use:} How often the tool is used, indicating user engagement.
    \item \textbf{Tagging Frequency:} The average number of tags entered per day, indicating user engagement.
    \item \textbf{User Satisfaction:} Feedback on the effectiveness of the tool to help users understand their physiological responses to interventions.
    \item \textbf{Recommendation Likelihood:} Whether users would recommend the tool to friends and family, indicating overall satisfaction and perceived value.
\end{enumerate}

This feedback will guide future improvements and iterations of the system, helping to refine both the visualization and the ML components towards a useful system that can give personalized and actionable insights.

% ==========================================================
\section{Findings}
% ==========================================================
\label{sec:results_overview}

\subsection{Participant Retention and Data-Collection Compliance Was High}
%A total of seven student participants were recruited from two academic institutions.  
All seven participants completed Phase One of the study, which lasted two weeks, and six elected to continue into Phase Two, which lasted another two weeks.  
Most participants preferred using the annotation form to log events and interventions, rather than syncing their calendars or logging directly through the visualization tool.

\textbf{High data-collection compliance.}  
Participants wore the device for most of each day, averaging 20.1 h in Phase 1 and 19.1 h in Phase 2 (see Appendix A, Figures \ref{fig:wear_time_p1}–\ref{fig:wear_time_p2}). Participants were encouraged to log one to two events or interventions per day and generally complied, although some opted to log at times that were more convenient rather than immediately after an event, as recommended. In total, 269 interventions and events were recorded.

% --------------------------------------------------------------
\subsection{RQ1 — Self-tagged Interventions Enabled Pattern Discovery in Physiological Data}
\label{sec:results_rq1}

\textbf{Hypothesis H1 (Pattern Discovery).}  Integrating self‑tagged context with wearable sensor outputs will uncover patterns that correlate specific activities or interventions with improvements in physiological markers related to stress.

%comment on different people are using different approaoaches 

\textbf{Intervention diversity--What did participants tag?} The 269 tags of the participants were split into nine categories. We considered different ways of categorizing these tags, and found that the distributions were generally similar. Table~\ref{tab:intervention_categories} describes each of the intervention categories used in this study. Figure~\ref{fig:intervention_categories} in the appendix visualizes the data in Table~\ref{tab:intervention_categories}.

\begin{table*}[htbp]
    \small
    \setlength{\tabcolsep}{6pt}
    \renewcommand{\arraystretch}{1.15}
    \centering
    \begin{tabularx}{\textwidth}{@{} l X p{0.5in} p{0.5in} @{}}
    \toprule
    \textbf{Category} & \textbf{Description} & \textbf{Number Logged} & \textbf{Median Duration} (minutes) \\
    \midrule
    (1) Socializing / Social Interaction & Any in-person or virtual activity that fosters connection with others (e.g. chatting with friends, group meals, or social events). & 52 & 60 \\
    (2) Academic \& Educational & Coursework, study sessions, lectures, or professional skill-building that primarily engages cognitive effort. & 44 & 75\\
    (3) Spirituality / Mindful Activities & Prayer, meditation, journaling, or routines that cultivate presence or reflection. & 37 & 5 \\
    (4) Food / Drink / Nutrition & Preparing, consuming, or enjoying food and beverages, from snacks and coffee breaks to full meals or cooking sessions. & 32 & 75 \\
    (5) Physical Activity\,--\,Cardio & Aerobic exercises such as running, cycling, brisk walking, or HIIT that raise heart rate. & 28 & 62.5 \\
    (6) Physical Activity\,--\,Non-cardio & Movement with lower cardiovascular demand, including stretching, yoga, light strength training, or mobility work. & 16 & 137.5  \\
    (7) Rest \& Recovery & Deliberate downtime for recuperation, such as napping, quiet rest, passive relaxation, or foam rolling. & 27 & 30 \\
    (8) Healthcare / Therapy & Medical or therapeutic appointments and self-care actions (e.g., counseling sessions, physiotherapy, doctor visits). & 10 & 60 \\
    (9) Other & Miscellaneous activities not captured by the preceding categories. & 7 & 23 \\
    \bottomrule
    \end{tabularx}
    \caption{Description of the nine intervention categories used to group self-tagged interventions.}
    \label{tab:intervention_categories}
\end{table*}

\textbf{Intervention duration.} Table~\ref{tab:intervention_categories} shows the median duration of interventions and events in each category. \emph{Physical Activity - Non-Cardio} interventions took the longest, with the median intervention taking more than 2 hours, and exhibited the highest variability in duration. In contrast, \emph{Mindful Activities} were the quickest interventions, with most being a little over 5 minutes per intervention. Despite these differences in duration, longer interventions did not always correspond with stronger physiological effects; even brief activities, when matched to a user's unique needs, can yield positive effects. \\

\textbf{When were interventions logged?} 
Logging of intervention start times peaked on Sunday around 09:00 and Wednesday around 15:00, while overnight hours (00:00 – 06:00) saw almost no entries (see Appendix A, Figure \ref{fig:logged_intervention_by_day}).

\textbf{Agreement across physiological measures.}

The three measures for stress: HRV, HR, and Garmin stress scores do not always move in perfect alignment. Depending on the event or intervention, each metric corresponds to different aspects and time scales of the stress-response trajectory \cite{Shaffer2017}. \emph{Mindful Activities} were associated with consistent reductions in heart rate (-0.68 to -2.97 bpm) and a short-term drop in stress score (-2 points at 15 minutes post-intervention), suggesting some positive effects (see Table~\ref{tab:combined_hrv_hr_stress} in the appendix for full results). However, HRV decreased simultaneously (-6.61 to -13.22\,ms), which may seem contradictory to these improvements. A drop in HRV can be expected during a supposedly `calming' mindful activity. This is because many mindfulness practices require sustained concentration and that cognitive effort can briefly activate the sympathetic branch of the autonomic nervous system (the fight-or-flight pathway), reducing beat-to-beat variability even as paced breathing slows the average heart rate \cite{iskaf2024breath, nijjar2014modulation, peressutti2012does}. In our study, interventions categorized as \emph{Mindful Activities} primarily involved journaling and prayer, both of which demand substantial cognitive engagement.
\emph{Rest and Recovery} is the only intervention category that shows consistent positive effects across all metrics: increasing HRV, decreasing HR, and decreasing stress score. Because the three physiological metrics rarely align perfectly across all points in the stress trajectory, they serve as complementary indicators. Together, they offer a more comprehensive view of physiological responses to stress, highlighting the value of using multiple measures rather than relying on a single indicator for analyzing stress and recovery. \\

\textbf{Intervention preferences and physiological impact.}
Interventions recorded most frequently by participants, such as \emph{Socializing} (20.6\%) and \emph{Mindful Activities} (14.6\%) were not always the most physiologically beneficial. At the population level, \emph{Socializing} produced mixed results. Interventions in this category were associated with a decrease in heart rate (-4.35 to -4.95 bpm) and stress score (-6.36 to -7.25 points), but also with slight reductions in HRV (-0.81 to -2.18 ms). Similarly, \emph{Mindful Activities} resulted in a sharp decrease in HRV (-6.61 to -13.22 ms) and slight changes in the stress score (-2.00 to 1.07 points), while also reducing heart rate (-0.68 to -2.97 bpm). Interestingly, \emph{Rest and Recovery}, which accounted for only 10.7\% of all events and interventions tagged, resulted in the most consistent improvements in all variables.

These patterns highlight a potential disconnect between participant preferences and interventions that produce the strongest physiological recovery indicators. Notably, some of the most effective strategies, such as \emph{Rest and Recovery}, were the least frequently tagged. This may reflect differences in how consciously certain interventions were tagged. For example, participants may have been more likely to tag intentional activities, such as socializing or journaling, while overlooking more passive behaviors such as resting. These findings open opportunities for future iterations of the system to gently nudge users toward underutilized but effective strategies, such as deliberate rest, while tailoring recommendations to account for individual variability in response and preference. \\

\textbf{Temporal variation in physiological impact.}
The data reveal meaningful differences in how various interventions influence physiological stress markers over time and across measures.  \emph{Socializing} is associated with sustained reductions in stress score, from –6.81 points at 15 minutes to –7.25 points at 120 minutes post-intervention. In contrast, \emph{Mindful Activities} produced the most immediate reduction in stress score (–2 points at 15 minutes), which then reversed, reaching a slight increase (+1.07 points) by 120 minutes. Further illustrating these temporal dynamics in the physiological data, \emph{Non-cardio Exercise} showed an initial sharp decline in HRV (-10.16 ms), likely due to sympathetic nervous system activation during exertion, while improvements in stress score peaked (at -8.31 points) later, 120 minutes post-intervention. These patterns highlight the importance of considering not just the type of intervention, but also the timing of its physiological effects.

% To examine whether these temporal patterns around the interventions are statistically meaningful, we compared each intervention category against null distributions (Figure~\ref{fig:distributions}). These null distributions were derived by initially targeting a uniform sample of random timestamps across all user data, which, after filtering for valid data and successful calculation of physiological changes, provided the basis for comparison.

\begin{figure}[htbp]
    \centering
    \begin{subfigure}[b]{0.48\textwidth}
        \includegraphics[width=\textwidth]{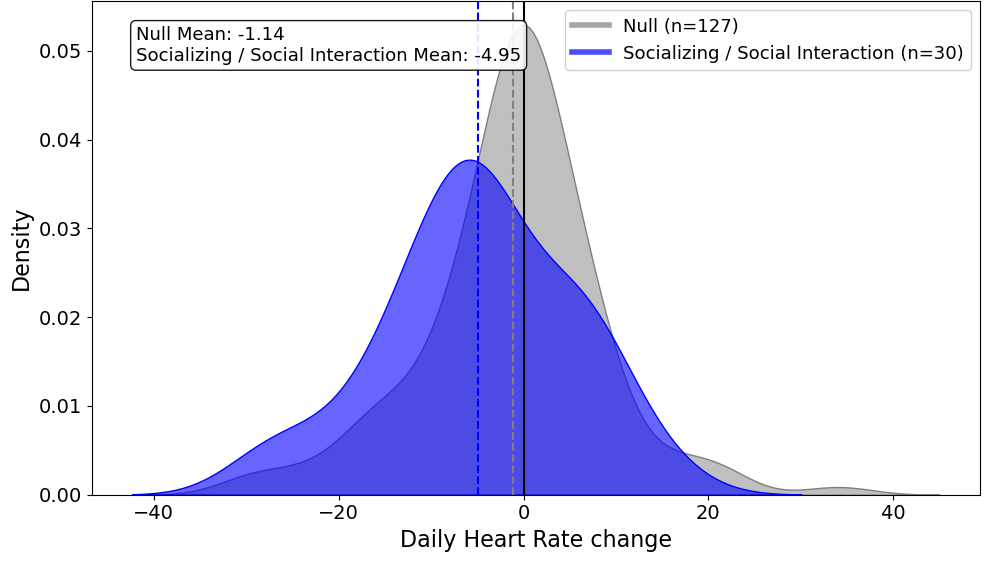}
        \caption{Socializing vs Null: Heart Rate (30min window)}
        \label{fig:dist_social_hr}
    \end{subfigure}
    \hfill
    \begin{subfigure}[b]{0.48\textwidth}
        \includegraphics[width=\textwidth]{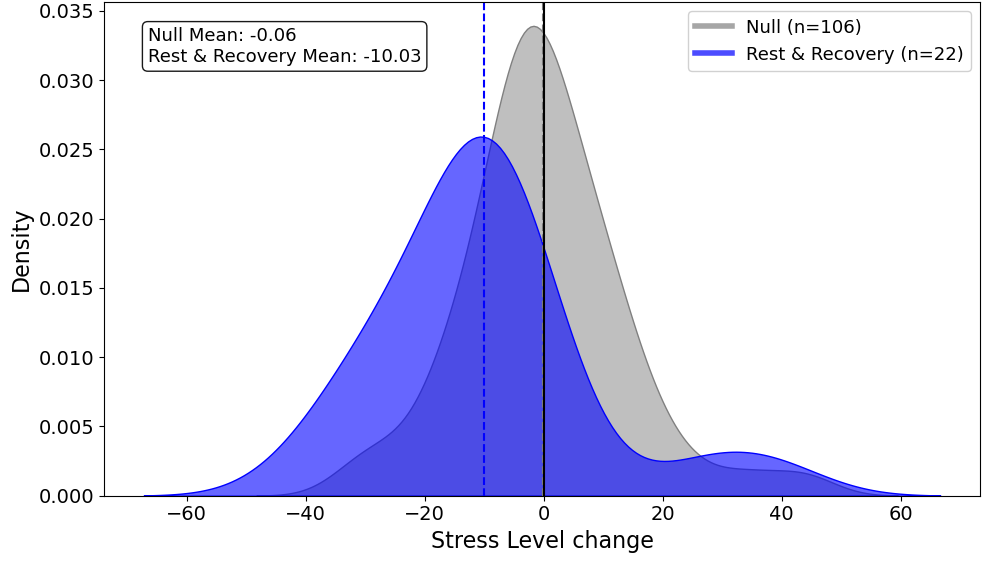}
        \caption{Rest \& Recovery vs Null: Stress Level (30min window)}
        \label{fig:dist_rest_stress}
    \end{subfigure}
    
    \begin{subfigure}[b]{0.48\textwidth}
        \includegraphics[width=\textwidth]{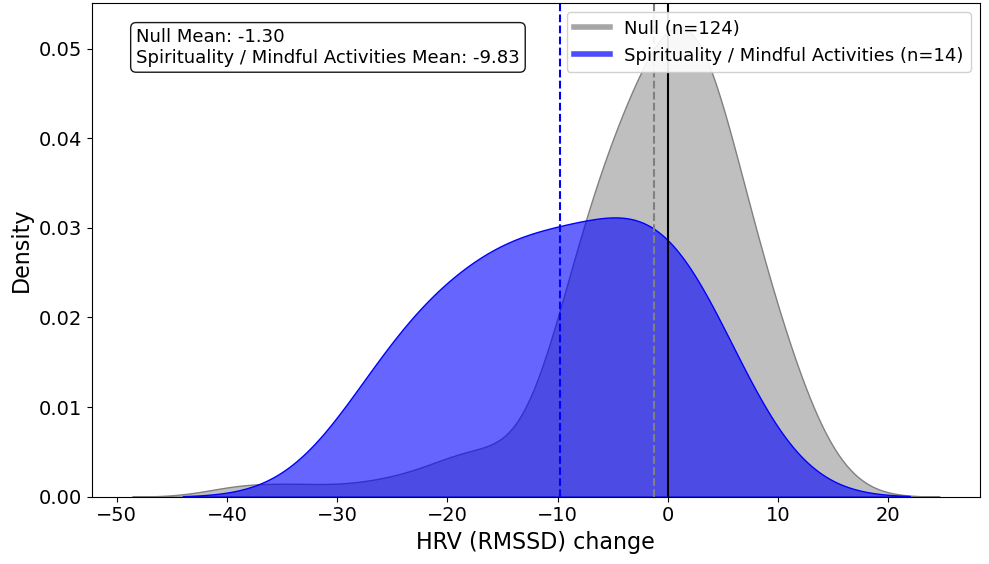}
        \caption{Spirituality / Mindful Activities vs Null: HRV (30min window)}
        \label{fig:dist_mind_hrv}
    \end{subfigure}
    \hfill
    \begin{subfigure}[b]{0.48\textwidth}
        \includegraphics[width=\textwidth]{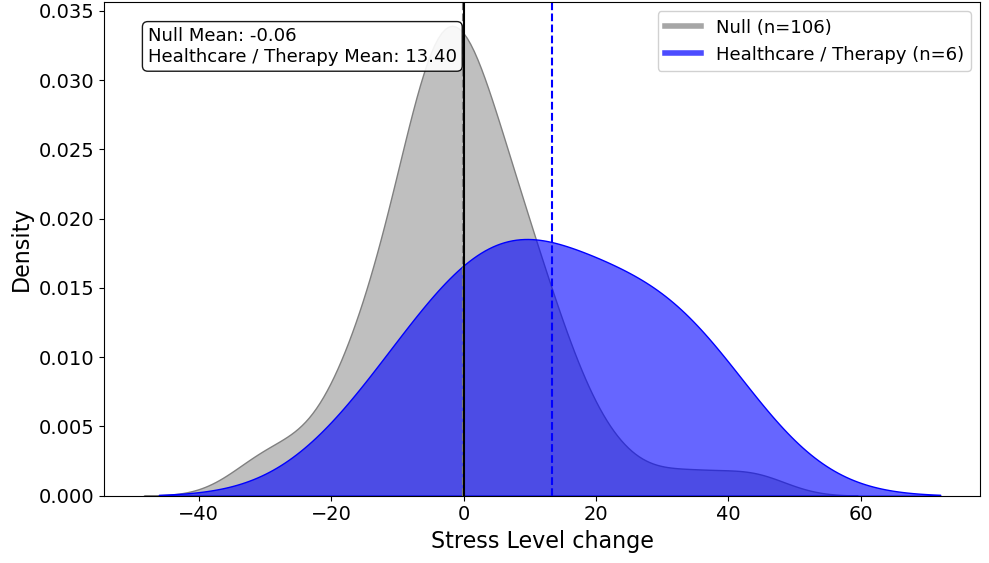}
        \caption{Healthcare/Therapy vs Null: Stress Level (30min window)}
        \label{fig:dist_health_stress}
    \end{subfigure}
    
    \caption{Distribution comparisons between intervention categories and null distributions. 
    (a) Socializing shows a significant shift toward reduced heart rate ($p=0.004$, $d=-0.60$). 
    (b) Rest \& Recovery demonstrates significant stress reduction ($p=0.0227$, $d=-0.66$).
    (c) Spirituality / Mindful Activities shows decreased HRV initially ($p=0.005$, $d=-1.00$).
    (d) Healthcare/Therapy shows a marginal increase in stress ($p=0.0735$, $d=1.28$). 
    Gray distributions represent random timestamps (null distribution), while blue distributions show the intervention category.}
    \label{fig:distributions}
\end{figure}

To examine whether these temporal patterns around the interventions are statistically meaningful, we compared each intervention category against null distributions derived from  random timestamps (constrained to 8 am - 11 pm local time) uniformly sampled across all user data (Figure~\ref{fig:distributions}). The goal with this approach is to distinguish changes that follow a specific intervention from those that occur naturally over the day. The \emph{Socializing} category shows a statistically significant reduction in heart rate at the 30-minute window ($p=0.004$, Cohen's $d=-0.60$, Figure~\ref{fig:dist_social_hr}), with a distribution clearly shifted toward negative values compared to the null distribution. Similarly, \emph{Rest and Recovery} demonstrates a significant reduction in stress levels over the same timeframe ($p=0.0227$, Cohen's $d=-0.66$, Figure~\ref{fig:dist_rest_stress}). In contrast, \emph{Spirituality / Mindful Activities} shows a significant decrease in HRV at the 30-minute window ($p=0.005$, Cohen's $d=-1.00$, Figure~\ref{fig:dist_mind_hrv}), suggesting initial nervous system activation during focused practice. Furthermore, the \emph{Healthcare/Therapy} category showed increases in stress levels at the 30-minute window that, while not statistically significant, still suggest a meaningful pattern in the data ($p=0.0735$, Cohen's $d=1.28$, Figure~\ref{fig:dist_health_stress}), with a distribution notably shifted toward positive values. Additional distribution plots for 120-minute effects of both \emph{Rest and Recovery} and \emph{Healthcare/Therapy}, where the effects are more pronounced and statistically significant, are provided in the appendix (Figure~\ref{fig:dist_rest_health_120}). These distribution comparisons further support our observation that different intervention types are associated with different temporal patterns in physiological responses.

Our analysis involved multiple statistical comparisons across physiological variables, intervention categories, and time windows. To account for potential false positives, we interpret results in the context of both p-values and effect sizes. All comparisons shown in Figure~\ref{fig:distributions} achieved p-values below 0.1 and medium to large effect sizes (|$d$| > 0.60), suggesting these effects are substantial. Future iterations of the study with a larger participant cohort could incorporate formal corrections for multiple comparisons, such as Bonferroni or false discovery rate (FDR) adjustments, to further validate these patterns.

Figure~\ref{fig:hrv_time_series} in the appendix provides additional temporal insight by showing HRV trajectories over time for each intervention category. Although some categories such as \emph{Rest and Recovery} show relatively smooth and sustained increases in HRV, others such as \emph{Physical Activity} or \emph{Mindful Activities} exhibit more volatile patterns, further illustrating the variability in physiological responses between different types of interventions.

% ------------------------------------------------------------
% \subsection{RQ2 — Usability and Insight Discovery}
\subsection{RQ2 — Self-tagged Interventions Support Reflection, Insight, and Personalized Stress Management}
\label{sec:results_rq2}

\textbf{Hypothesis H2 (Enhanced Self-Management).}  
Interactive plots that overlay self-tags on physiology will (i) be easy to use and (ii) help participants surface insights they can act on.

\begin{figure}[ht]
  \centering
  \begin{subfigure}[t]{0.495\textwidth}
    \includegraphics[width=\textwidth]{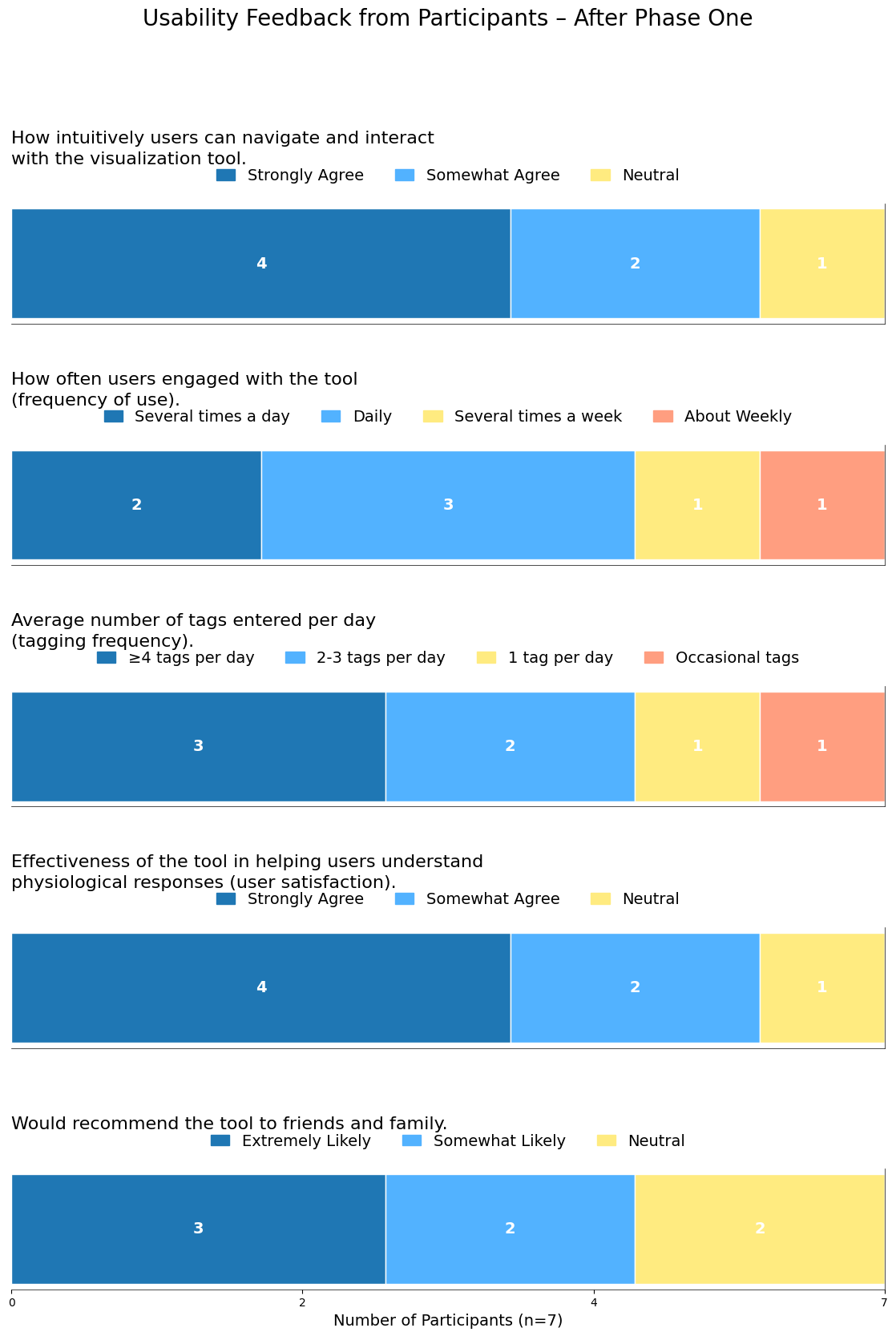}
    \caption{Usability metrics after Phase One (\(n=7\)).}
    \label{fig:usability_p1}
  \end{subfigure}
  \hfill
  \begin{subfigure}[t]{0.495\textwidth}
    \includegraphics[width=\textwidth]{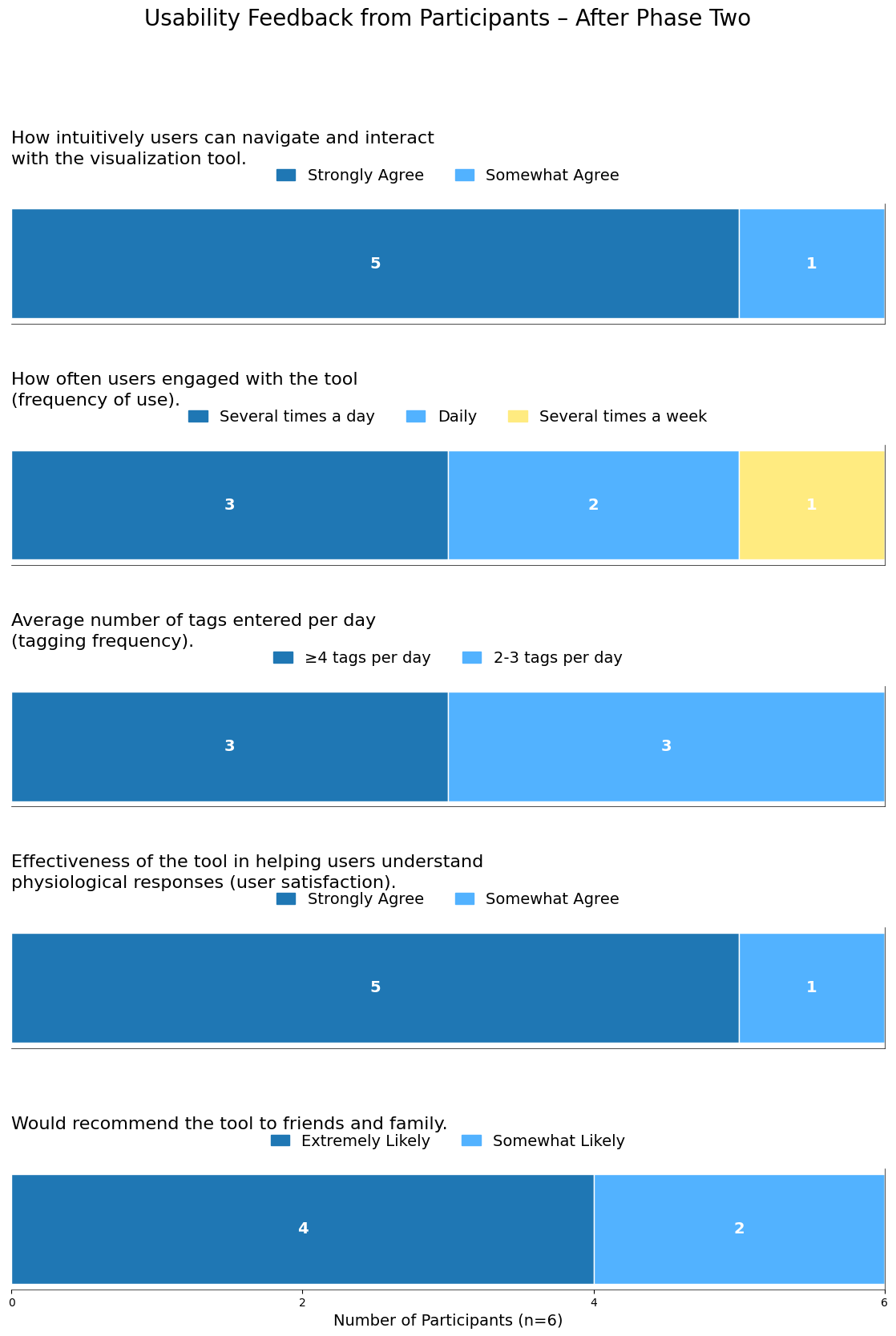}
    \caption{Usability metrics after Phase Two (\(n=6\)).}
    \label{fig:usability_p2}
  \end{subfigure}
  \caption{Responses to the post-study survey after Phase One and Phase Two.}
  \label{fig:usability_combined}
\end{figure}

\textbf{High usability.}  
Figure~\ref{fig:usability_p1} shows that Phase One participants rated tagging as \emph{manageable}, the interface \emph{intuitive}, and the plots \emph{insightful}; most used the tool daily. After phase two ended, we conducted 30-minute exit interviews with four participants in the study who responded to having interest. The participants were asked the same questions around three open-ended themes: 1) their experience using the visualization tool and 2) insights they gained about stress from using the tool.   One interviewee noted:

\begin{quote}
"The website’s flow was very user-friendly.  
The left-hand navigation bar made the different parts of the tool easy to find."
\end{quote}

\textbf{Built-in reflection.}  
Two participants said that logging events immediately, rather than batching tags later, created a brief pause for self-reflection. One participant mentioned:

\begin{quote}
"I treated the logging process as a moment of reflection.  I would pause after an intervention to decide whether it really helped or whether I should do less of it next time."
\end{quote}

\textbf{Insight discovery for better stress management.}   
Several participants reported concrete takeaways that emerged from using the tool.

One participant focused on their physiological response to one-on-one meetings:

\begin{quote}
“The visualization showed me I get more stressed about meetings than I need to be. During the meeting, my body gradually realizes that everything is fine and calms down again. Next time I can try some self-talk beforehand so I don’t get as stressed at the beginning.”
\end{quote}

Although the participant logged only one instance of a 1-to-1 meeting, Appendix A (Figure \ref{fig:interview_meeting}) supports the pattern: HR and stress metrics spike at the start and then drift back toward baseline by the end of the meeting.

One participant also focused on the aggregated effects of all the interventions they tried and their overall impact over the course of the study. The participant noted: 

\begin{quote}
"An insight I gained was that all the interventions I tried helped reduce my stress over the course of the study. I feel much less stressed overall at the end of the study than I did at the beginning. I can’t say that any one intervention worked better for me; rather, they all worked together to help me lower my stress over time."
\end{quote}

These comments and feedback suggest that the tool not only revealed patterns; it led to concrete self-management strategies.  
For one participant, recognizing the pre-meeting stress spike prompted a plan to use calming self-talk before future meetings.  
For another, seeing the cumulative downward trend in stress reinforced the value of maintaining a range of intervention routines over time rather than looking for a single 'best' technique.

\textbf{Iterative improvement.} Following additional visualization updates, feedback shows improved satisfaction compared with Phase One (Figure \ref{fig:usability_p2}), especially for daily engagement, tagging frequency, and willingness to recommend—supporting \textbf{H2}.

% interventions generally happen for some time of time ----generally last for 

%intervention effects have 

%sample plot for how figures might fit in the story

\subsection{Physiological Response to Stress Varies by Individual, Personality, and Perception}
\label{sec:results_exploratory}

\begin{figure}[htbp]
  \centering
  \includegraphics[width=0.96\textwidth]{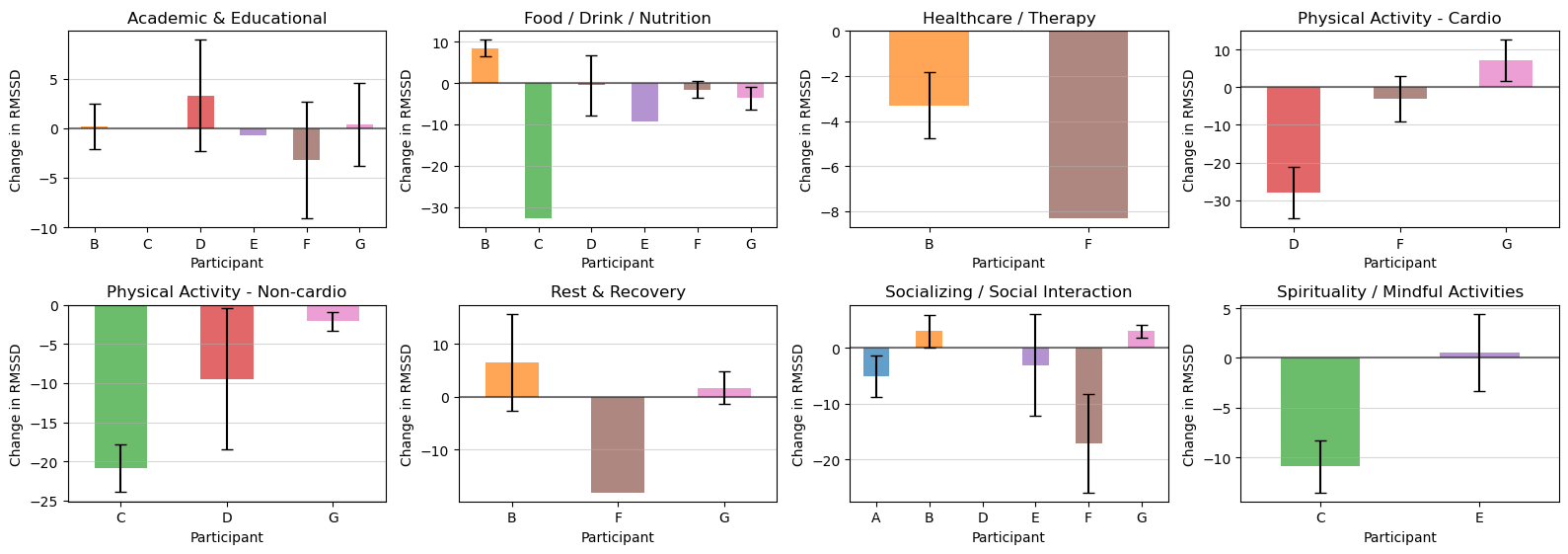}
  \caption{Average changes in HRV by participant and category. Change was calculated as the difference between the average HRV in the 15-minute window before the intervention and the 30-minute window after the intervention}
    \figsubcaption{These plots show the average change in HRV for each participant and each intervention category.} % Comment on participant preferences? Tendency for HRV to increase/decrease? If certain interventions are more effective?
  \label{fig:participant_rmssd_30}
\end{figure}

\textbf{Individual variability.} Across all intervention categories, the magnitude of physiological response varied substantially between the participants (Figure~\ref{fig:participant_rmssd_30}).  While some individuals (e.g. participant B) exhibited pronounced increases in HRV immediately after most interventions, others (e.g. participants C and F) showed muted or even negative responses. Additionally, individual-level trends reinforce our earlier finding that different intervention categories vary in their magnitude of their physiological impact. For example, although both \emph{Healthcare / Therapy} and \emph{Non-cardio Physical Activity} resulted in a decrease in HRV 30 minutes post-intervention, the decrease was more pronounced following non-cardio physical activity. Figure~\ref{fig:participant_rmssd_30_yscale} in the appendix presents the same data as Figure~\ref{fig:participant_rmssd_30} but uses a consistent y-axis scale to better highlight these differences across intervention types. These patterns highlight the need to learn person-specific and category-specific baselines and response profiles in future work. \\

\textbf{Personality signals.}
\label{sec:personality_signals}
Qualitative interviews suggest that self-identified extroverts experienced greater HRV gains from social interventions, while introverts tended toward neutral or negative shifts (anecdotal, $n{=}3$). These self‑reports align with physiological variations—particularly HRV changes during socializing events—seen between participants. Figure~\ref{fig:participant_rmssd_30} illustrates this variability: some participants experienced HRV improvements or close to neutral effects after social interactions, while others showed declines. This variation hints at possible moderation by factors such as personality traits, social context, or baseline stress state. A validated personality scale (for example, the Big Five Inventory\cite{john1999big}) could be explored in follow-up work to go beyond anecdotal evidence and formally test these trait–state interactions.

\begin{figure}[htbp]
  \centering
  \includegraphics[width=0.48\textwidth]{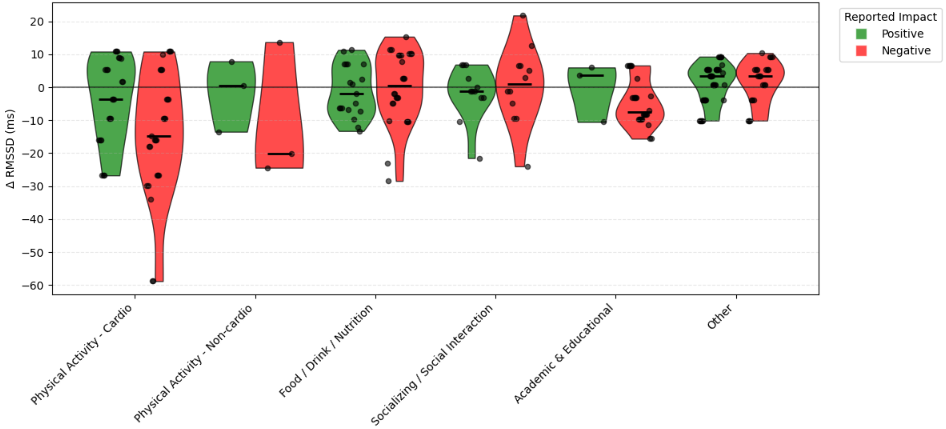}
  \caption{Agreement between perceived and measured HRV change (60 minute window).}
    \figsubcaption{Violin plot showing the change in average HRV in the 60-minute window before the intervention and the 60-minute window after the intervention.}
  \label{fig:sentiment_analysis}
\end{figure}

\textbf{User reporting versus physiological data.}  
Figure~\ref{fig:sentiment_analysis} compares the self-reported stress impact ratings of the participants (positive, neutral, or negative) with the measured HRV change (\(\Delta\) RMSSD) over a 60-minute window. In general, the accuracy of the participants' perception varied between the types of intervention, highlighting a meaningful but imperfect alignment between subjective experience and physiological stress response.

In particular, physical activity (cardio) interventions often had short-term negative HRV changes despite the participants labeling them as positive. This suggests a potential mismatch between perceived benefits and acute physiological response, possibly reflecting short-term sympathetic activation followed by long-term recovery benefits. Future work might explore longitudinal HRV recovery windows or delayed annotations to capture this nuance more accurately.

In contrast, academic or educational activities often showed small HRV improvements in positively tagged instances, although the number of those logged are limited. These early signals suggest that intellectually engaging contexts may sometimes support stress regulation, but a larger sample size is needed to draw firmer conclusions.

For socializing interventions, variation in HRV response was especially wide, with some positive and negative tags producing similar physiological profiles. This reinforces earlier anecdotal findings (Section~\ref{sec:personality_signals}) around individual personality differences, where extroverts reported more benefit from social interaction than introverts.

These findings point to a broader challenge: self-perception does not always map cleanly to physiological state. Future iterations could prompt participants to log the perceived impact of stress immediately after the intervention, reducing retrospective bias or emotional drift. Additionally, these variations underscore the need for personalized models and real-time annotation tools that can better capture how individuals interpret and respond to different stress-related events.

\section{Discussion and Conclusion}

\textbf{Interpretation of findings.} 
Our pilot supports our hypotheses 1) that pairing continuous wearable signals with a participant-tagged context can expose stress patterns and 2) we can build towards a personalized visualization tool that participants can use to identify when and how interventions affect their physiological response. \\

\textbf{RQ1/H1 supported.} 
Overlaying tags with sensor data revealed interpretable patterns—especially for social, rest, and cardio activities—that would be invisible in raw streams.

\textbf{Divergence across metrics.} 
Our analysis highlights some discrepancies between HRV, HR, and Stress Score as physiological measures of stress. While one metric might suggest that an intervention had a positive effect, the other measures might show the opposite trend. This emphasizes the importance of multi-metric tracking to ensure a more comprehensive understanding of stress responses, as relying on a single metric could lead to misleading conclusions about participant well-being. 

Future research should strongly consider incorporating multiple physiological measures, which will allow for more nuanced interpretation of intervention effectiveness. For example, in our study, each metric captures different aspects of the stress response. HRV captures autonomic adaptability, heart rate reflects autonomic arousal, and stress score combines multiple physiological measures with proprietary algorithms. By examining multiple signals, researchers can detect and analyze more complex and layered patterns in stress responses.

\textbf{Presence of temporal trends.}
We also found temporal dynamics in how different interventions affect stress-related physiological measures. Not all interventions had immediate effects. \emph{Non-cardio Exercise} resulted in a sharp decline of HRV at 15 minutes post-intervention, while the greatest reduction to stress score for the same intervention category occurred 120 minutes post-intervention. Furthermore, the net effects of interventions were not consistent over time. For example, \emph{Mindful Activities} led to an initial decrease in stress score 15 minutes post-intervention, but this effect gradually reversed, resulting in an increased stress score by 120 minutes post-intervention. These lagged effects suggest that evaluating interventions based solely on immediate or fixed post-activity windows may miss important delayed responses. 

This finding supports our design decision to allow users to customize pre- and post-intervention time windows in the visualization tool, enabling them to understand how interventions unfold over different durations. Since interventions operate on varying timescales, fixed time windows risk obscuring meaningful trends. Future systems should offer flexible or adaptive time windows, potentially recommending time ranges based on intervention type or user-specific patterns. Additional research could explore methods to identify optimal time windows across different physiological metrics or intervention types, such as detecting inflection points in the stress response trajectory.

\textbf{Disconnect between physiological impact and participant preferences and perception.} 
Our findings reveal a mismatch between the interventions participants chose most frequently and those that resulted in the strongest physiological benefits. For example, \emph{Socializing} and \emph{Academic Activities} were the two most common interventions, yet they did not consistently lead to improvements across HRV, HR, or Stress Score. On the other hand, less common interventions such as \emph{Rest and Recovery} showed the most sustained and consistent positive physiological effects. This suggests that users may gravitate toward familiar or habitual behaviors rather than those that are most restorative. Additionally, we observe very little correlation between perceived stress impact and physiological response. Participants sometimes rated interventions as positive or negative despite worsening physiological measures.  

From a design perspective, these findings suggests that stress-tracking systems could support more informed decision-making by highlighting discrepancies between perceived and physiological effects. Such systems could also encourage users to try interventions that may be more effective, even if they are less familiar or frequently used. Future research could investigate how users form and update beliefs about what interventions work for them, and how systems can support that process through personalized recommendations or reflection prompts. 
\\

% This highlights the need for tools that help users better understand their subjective experiences with objective data. \\

\textbf{RQ2/H2 supported.} The tagging workflow was adopted and users reported high engagement. The system was perceived as intuitive and became more efficient to use over time, especially as participants became familiar with the interface and workflow.

\textbf{Usability and interaction feedback.}  
Survey responses indicate that the web tool was easy to navigate, tagging became faster over time, and visualizations were perceived as helpful. Gains in satisfaction and willingness to recommend between phase one and two reflect a general increase in engagement, even without the presence of push notifications or gamification, suggesting that the tool has intrinsic value.

\textbf{Role of visualizations in insight generation.}
The exit interviews provide anecdotal evidence that the visualizations played a key role in helping participants reflect on their physiological responses. Several users noted that the tool helped them recognize which interventions were most effective and why certain strategies were more effective than others. One participant was able to connect changes in their Stress Score to a group work session, while another identified that their interventions were more effective on days following high-quality sleep. These examples suggest that interactive visualizations not only helped participants interpret noisy physiological data but also supported the generation of meaningful, personalized insights. While the tool is individually used, it enables richer self-understanding that can positively influence social interaction, helping individuals recognize when they are stressed, how they recover, and how their stress responses differ from others. This creates new opportunities for shared awareness, coordination, and mutual support around stress in work and other collaborative settings. \\

\textbf{Limitations.}  
The study’s scope was intentionally meant to be a pilot study, serving primarily as a proof of concept. First, the sample was small and drawn from university settings, limiting generalizability. Second, self‑tagging introduces timing errors and subjective bias that may blur true sensor–context relationships. Third, the four‑week duration is too short to evaluate long‑term adherence or durable stress‑management change. Finally, although anecdotal interview comments hinted that introversion–extroversion may shape responses to social interventions, we did not administer a validated personality scale, so these observations remain suggestive. \\

\textbf{Future work.}  
Next steps will focus on scaling the participant pool, collecting validated personality measures, and extending the deployment period. A larger and more diverse cohort will allow us to test whether the introversion–extroversion pattern persists and to explore other personality traits. Adding a more in-depth personality assessment during onboarding of the study could let us correlate personality types with physiological responses to specific intervention types, especially social versus solitary activities. Longer‑term deployments will reveal whether personalized insights translate into sustained behavior change, while lightweight prompting and semi‑automated tag suggestions can reduce logging burden without sacrificing user engagement. We could also allow users to log emotions alongside interventions, enabling more nuanced comparisons between perceived emotional states and physiological data. \\

\textbf{Conclusion.}  
Although limited in scale, this study lays the groundwork for truly personalized stress‑management tools. By blending objective physiology, subjective context, and individual personality traits, future systems could recommend the right intervention at the right moment for each user—moving the field beyond one‑size‑fits‑all advice toward precision self‑care. More broadly, our work shows how personalized insights from wearable data can support reflection and improve stress self-management, ultimately fostering awareness in social and collaborative contexts.

%\clearpage

%\section*{References}
\bibliographystyle{ACM-Reference-Format}
\bibliography{references}

\appendix
\renewcommand{\thesection}{\Alph{section}}
% \section*{Appendix A. Screenshots from Visualization Tool}
% \addcontentsline{toc}{section}{Appendix A. Screenshots from Visualization Tool}
\input{appendix}
% help prevent extra space/pages caused by delayed float placement.
\FloatBarrier
\end{document}

%% file: appendix.tex
\appendix
\renewcommand{\thesubsection}{\Alph{section}\arabic{subsection}}

\clearpage
\section{Appendix A. Visualization Tool Screenshots}

\subsection{Data Collection}
\FloatBarrier
% Google Form for Data Collection (Full-Page Float)
\begin{figure}[htbp]
    \centering
    \includegraphics[width=0.6\linewidth]{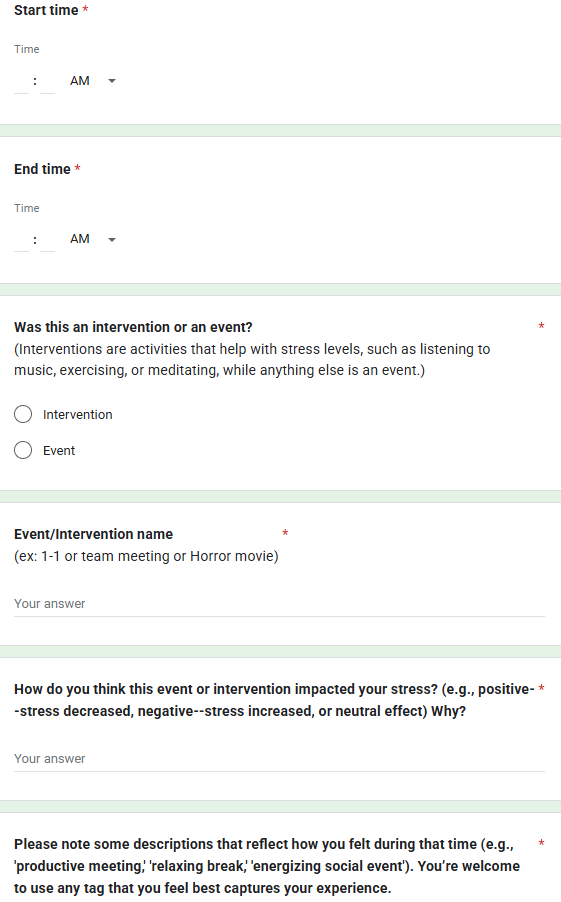}
    \caption{Google Form used for collecting data.}
    \label{fig:google_form}
\end{figure}

% User Interface Screenshots: Activity Labeling & Intervention Selection
\begin{figure}[htbp]
    \centering
    \begin{subfigure}[t]{0.9\linewidth}
        \centering
        \includegraphics[width=\linewidth]{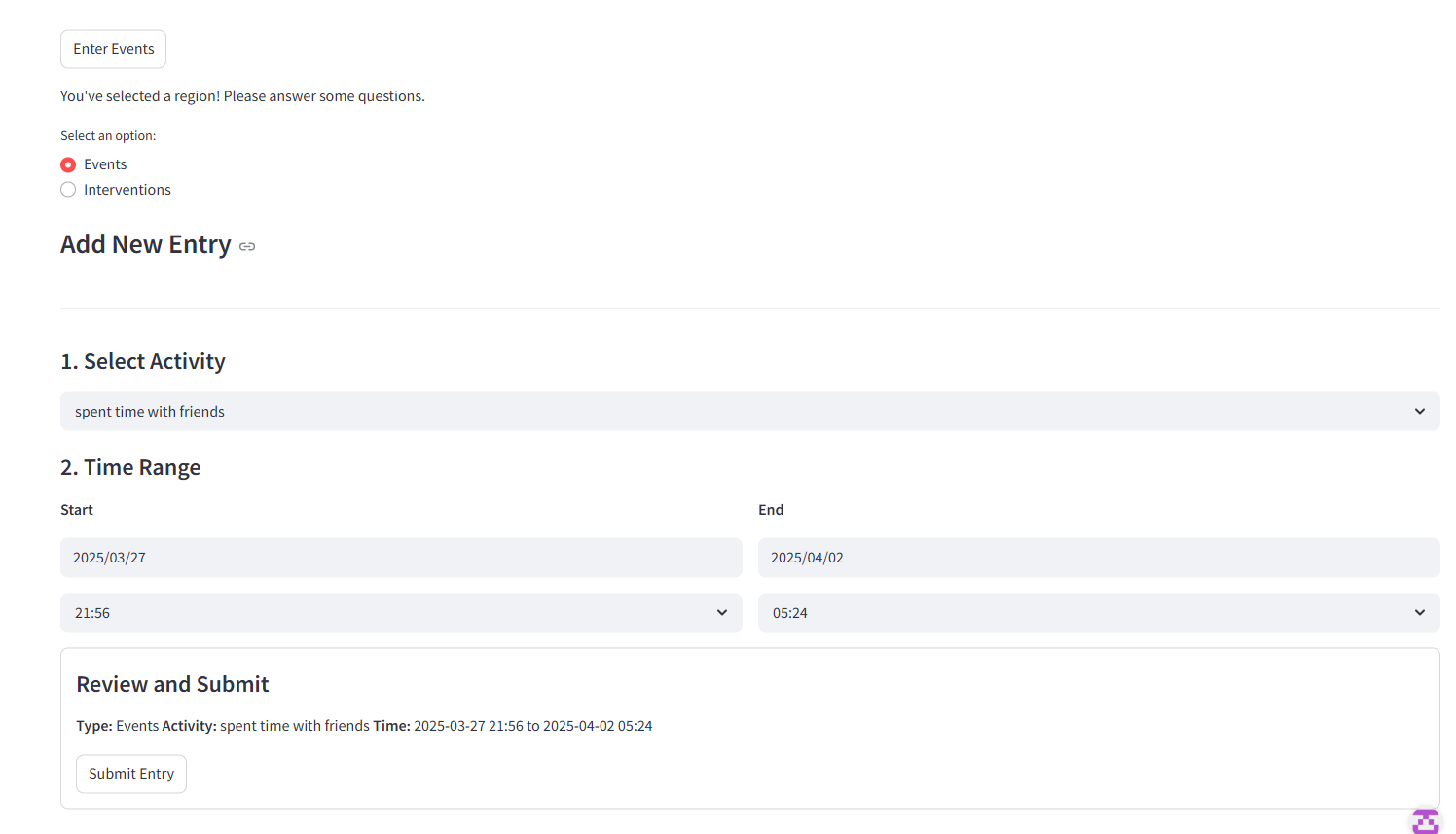}
        \caption{Activity Labeling: Users can fine-tune the time range and select or add an activity.}
        \label{fig:annotations_labeling}
    \end{subfigure}
    \vspace{1em}  % add a small vertical space between subfigures
    \begin{subfigure}[t]{0.9\linewidth}
        \centering
        \includegraphics[width=\linewidth]{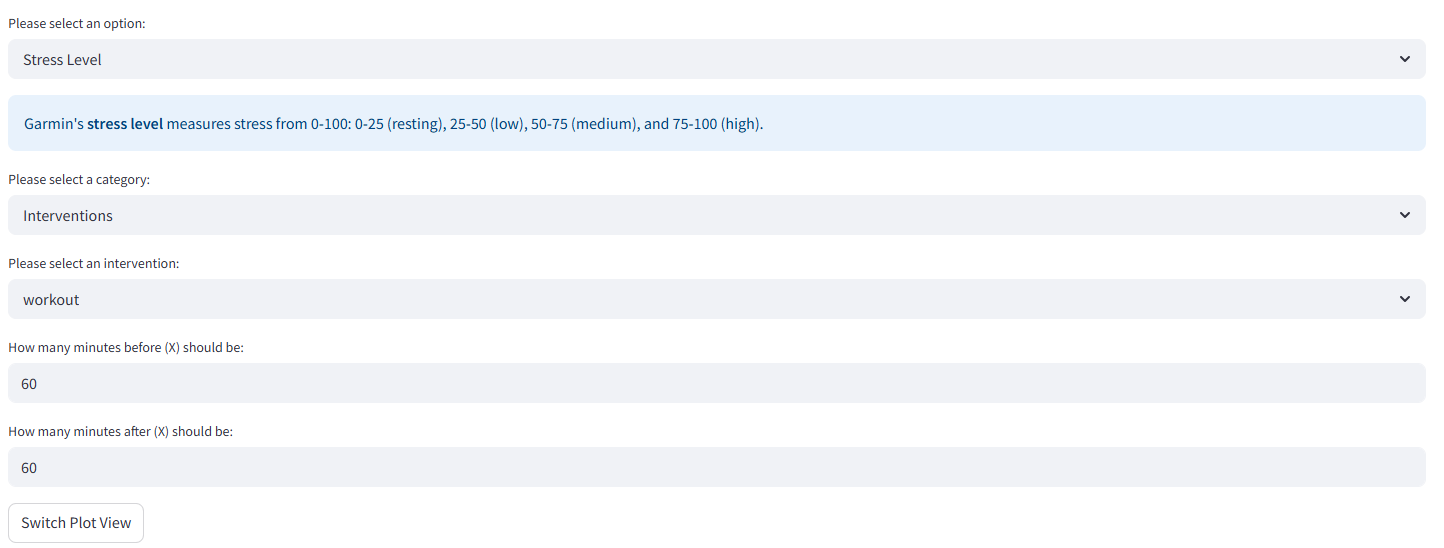}
        \caption{Intervention Selection: Users choose the specific intervention they wish to analyze.}
        \label{fig:intervention_selection}
    \end{subfigure}
    \caption{User Interface for entering annotations and selecting interventions.}
\end{figure}

% Annotations upload interface
\begin{figure}[htbp]
    \centering
    \includegraphics[width=\linewidth]{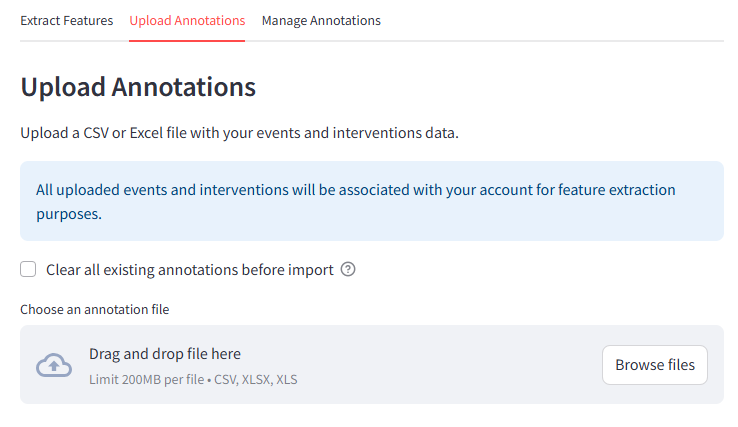}
    \caption{User interface for uploading data.}
    \label{fig:data_upload}
\end{figure}

\FloatBarrier
\subsection{User Interface}
% Home screen
\begin{figure}[htbp]
    \centering
    \includegraphics[width=0.48\textwidth]{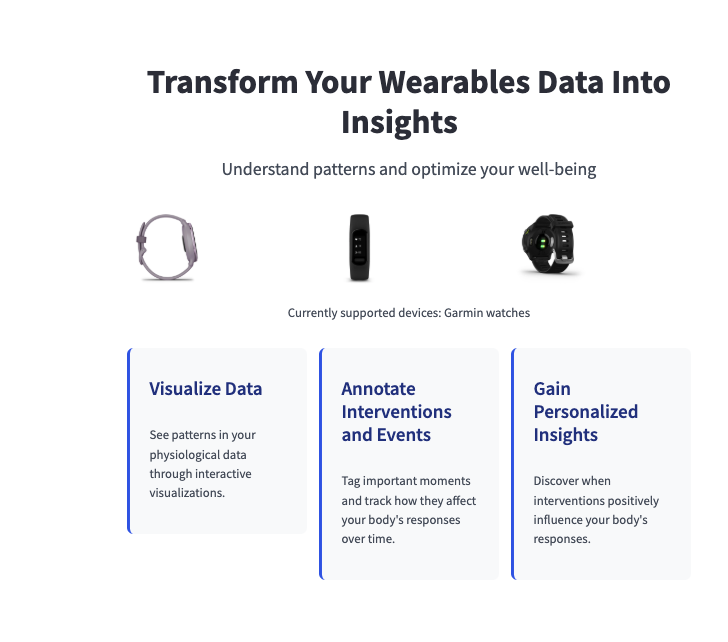}
    \caption{Home screen of the website}
    \label{fig:home_screen}
\end{figure}

% User Authentication Screens
\begin{figure}[htbp]
    \centering
    \begin{subfigure}[t]{\linewidth}
        \centering
        \includegraphics[width=0.7\linewidth]{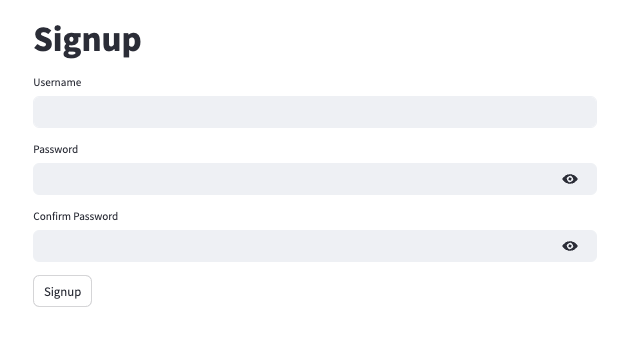}
        \caption{Signup Page}
        \label{fig:signup}
    \end{subfigure}
    \begin{subfigure}[t]{\linewidth}
        \centering
        \includegraphics[width=0.7\linewidth]{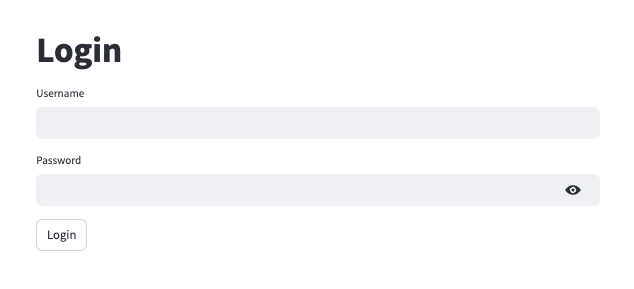}
        \caption{Login Page}
        \label{fig:login}
    \end{subfigure}
    \caption{User authentication screens.}
\end{figure}

% Annotated Visualizations for Selected Physiological Variables
\begin{figure}[htbp]
    \centering
    \begin{subfigure}[t]{0.48\linewidth}
        \centering
        \includegraphics[width=\linewidth]{img/show_annotations/hr.png}
        \caption{Heart Rate}
    \end{subfigure}
    \hfill
    \begin{subfigure}[t]{0.48\linewidth}
        \centering
        \includegraphics[width=\linewidth]{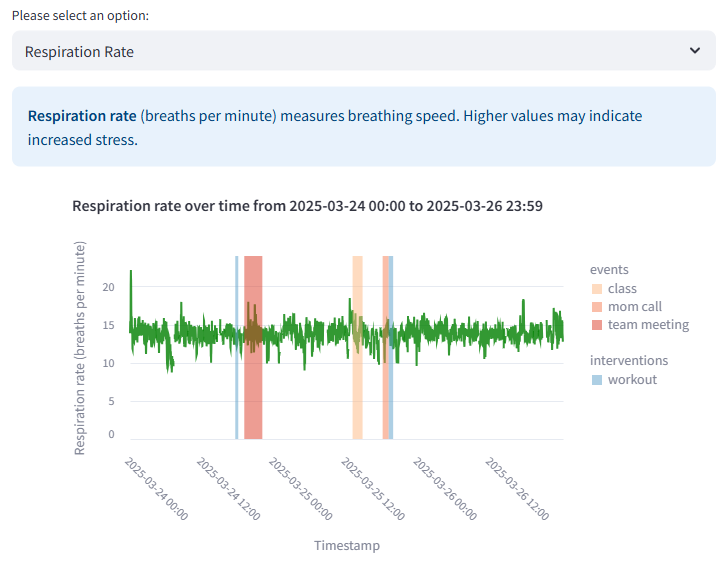}
        \caption{Respiration Rate}
    \end{subfigure}

    \vspace{0.5em}

    \begin{subfigure}[t]{0.6\linewidth}
        \centering
        \includegraphics[width=\linewidth]{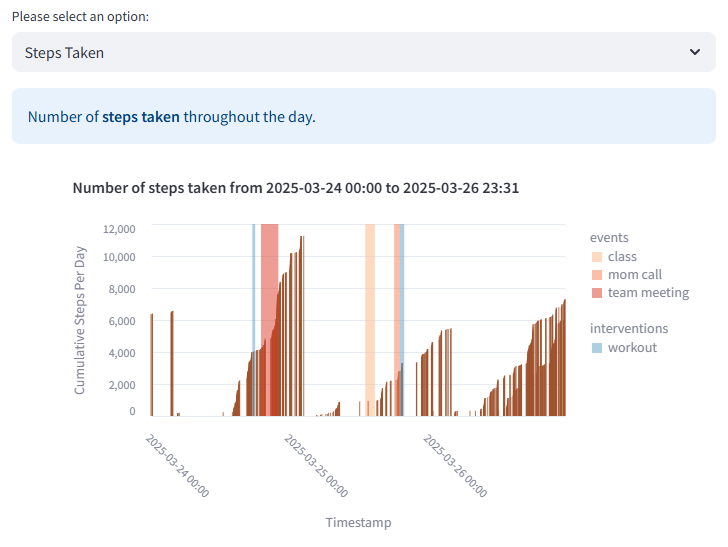}
        \caption{Steps Taken}
    \end{subfigure}

    \caption{Annotated visualizations for a user's physiological variables.}
    \label{fig:more_annotated_visuals}
\end{figure}

\begin{figure}[htbp]
    \centering
    \includegraphics[width=0.48\textwidth]{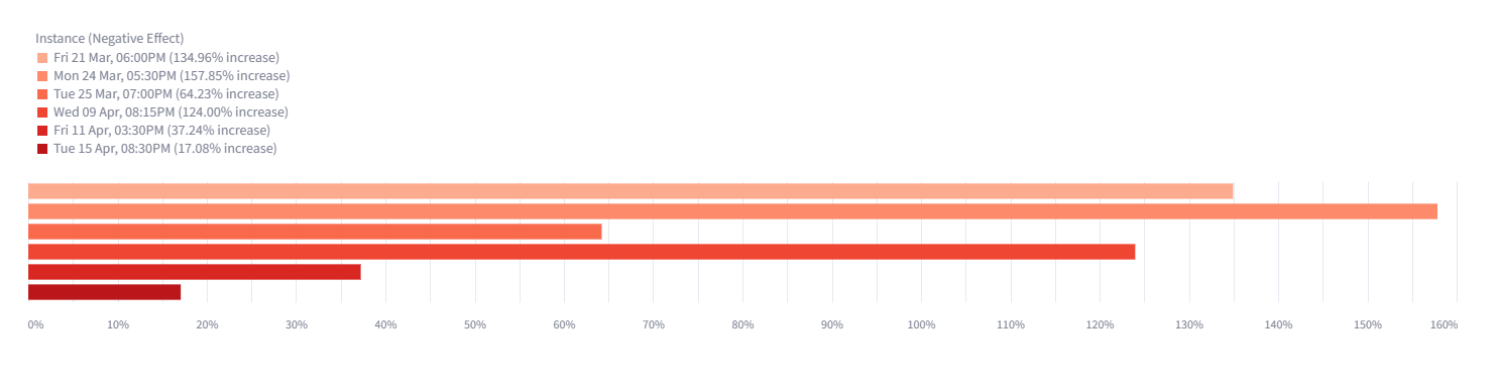}
    \caption{Aggregate Comparison}
    \figsubcaption{This visualization displays the percent change in Stress Score for each instance of a user's intervention, allowing for easy comparison across repeated occurrences.}
    \label{fig:percent_comparison}
\end{figure}

\begin{figure}[htbp]
    \centering
    \includegraphics[width=0.48\textwidth]{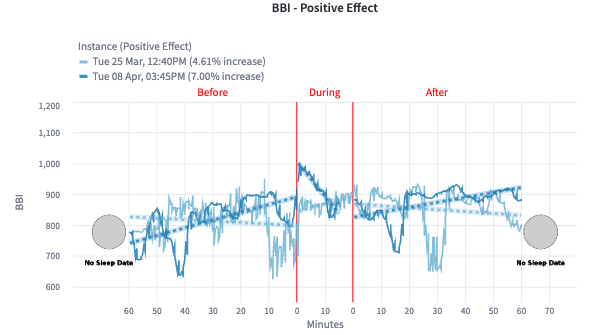}
    \caption{Compare Interventions}
    \figsubcaption{This visualization displays a user's data  (overlaid on each other) for the chosen variable before, during, and after the intervention, as well as the sleep quality of the previous and next night.}
    \label{fig:intervention_aggregate}
\end{figure}

% Visualization property selection
\begin{figure}[htbp]
    \centering
    \includegraphics[width=0.3\textwidth]{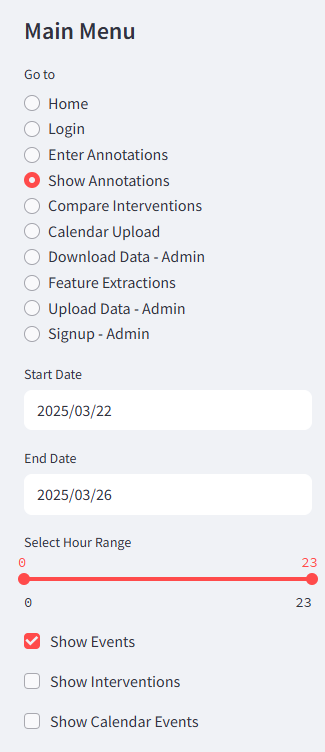}
    \caption{Visualization Options}
    \label{fig:time_event_selection}
\end{figure}

% Event/Intervention Management
\begin{figure}[htbp]
    \centering
    \begin{subfigure}[t]{\linewidth}
        \centering
        \includegraphics[width=0.6\linewidth]{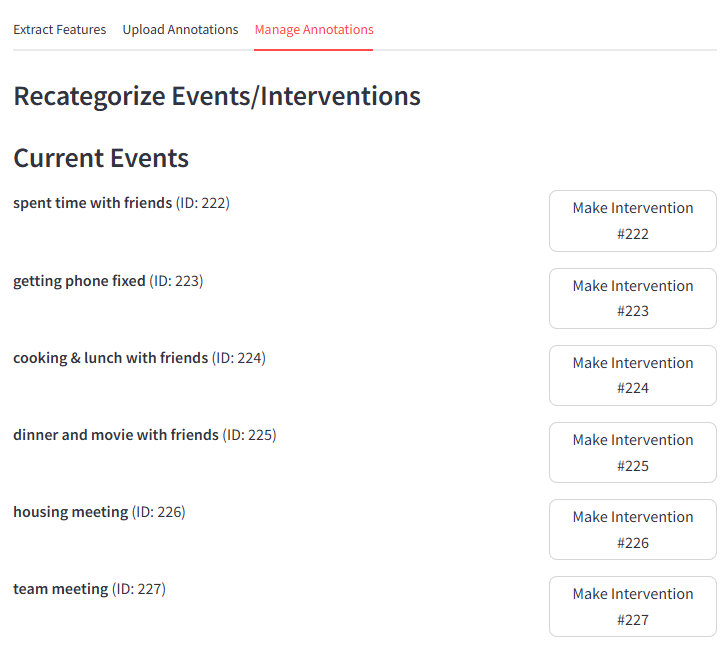}
        \caption{Managing Events}
    \end{subfigure}
    \begin{subfigure}[t]{\linewidth}
        \centering
        \includegraphics[width=0.6\linewidth]{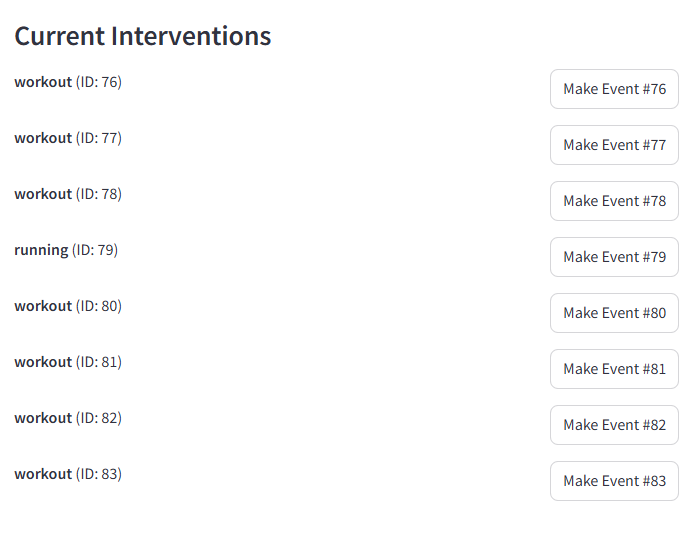}
        \caption{Managing Interventions}
    \end{subfigure}
    \caption{Users can manage their events and interventions easily.}
    \label{fig:manage_events_interventions}
\end{figure}

\FloatBarrier
\subsection{Feature Extraction}

\begin{figure}[htbp]
    \centering
    \includegraphics[width=0.48\textwidth]{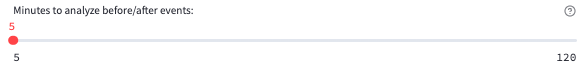}
    \caption{Analysis Options}
    \label{fig:analysis_options}
\end{figure}

\begin{figure}[htbp]
    \centering
    \includegraphics[width=\linewidth]{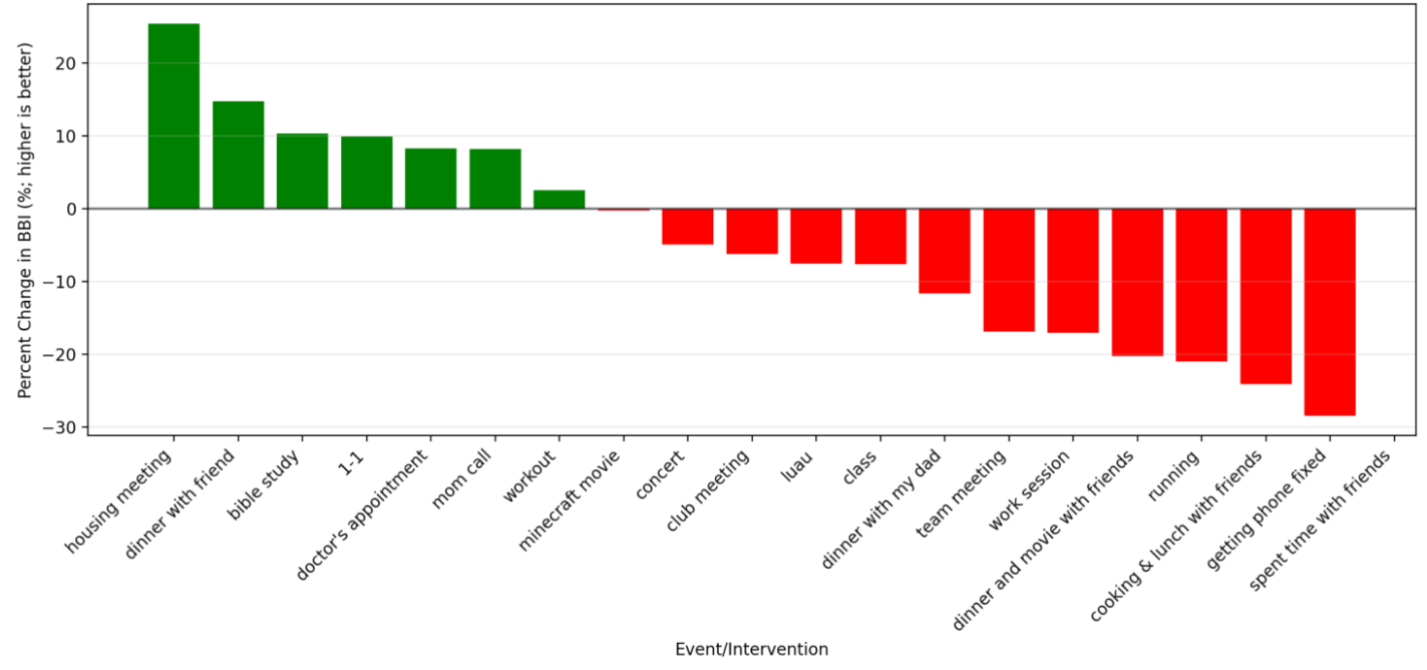}
    \caption{A user's average BBI percentage change for each activity}
    \label{fig:bbi_change_analysis}
\end{figure}

\begin{figure}[htbp]
    \centering
    \includegraphics[width=\linewidth]{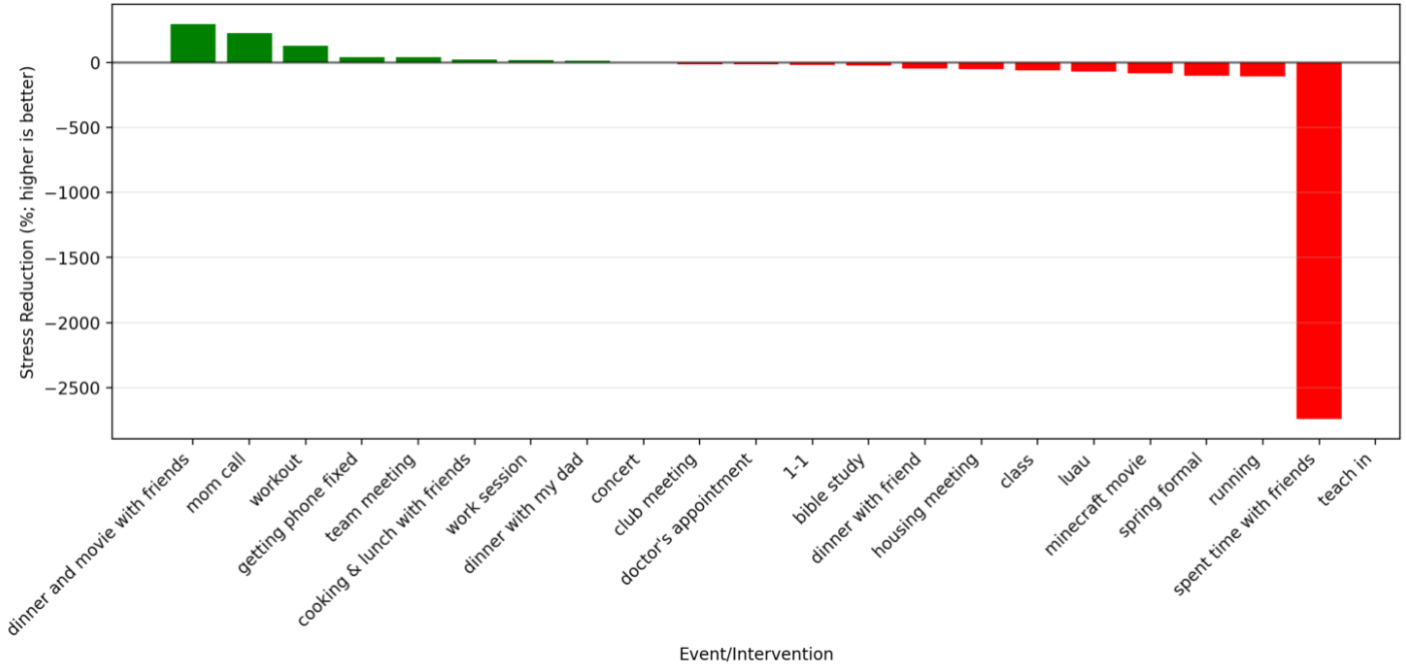}
    \caption{A user's average stress percentage change for each activity}
    \label{fig:stress_metric_analysis}
\end{figure}

\begin{figure}[htbp]
    \centering
    \includegraphics[width=\linewidth]{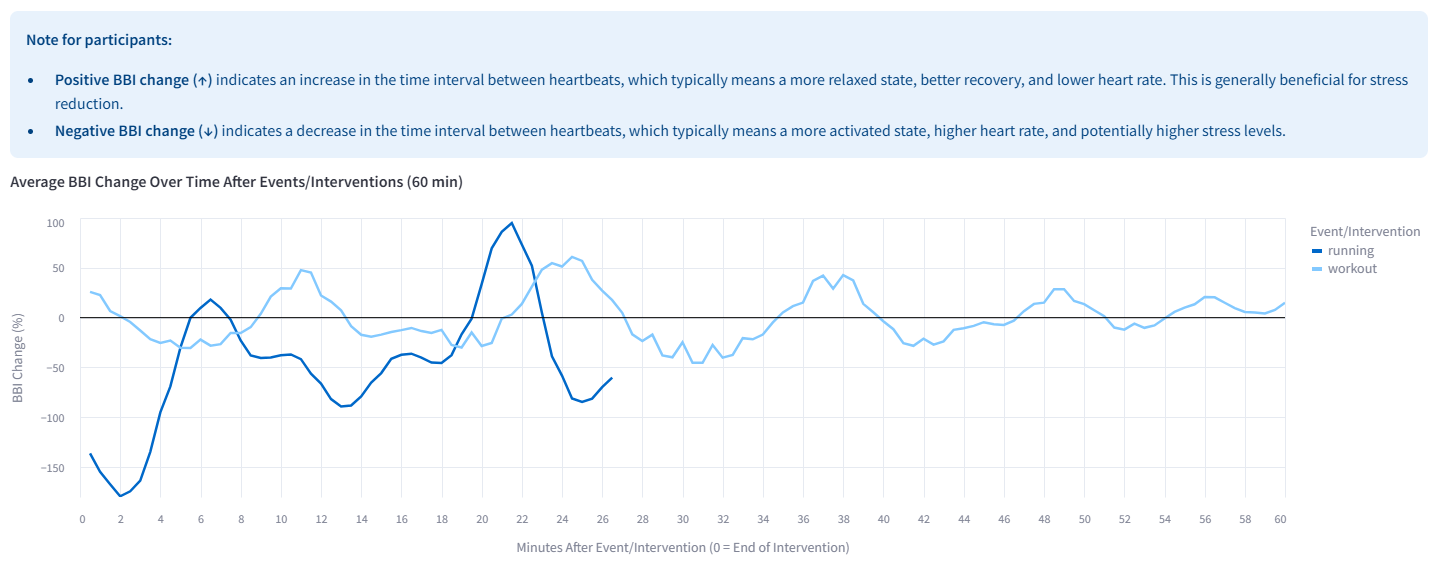}
    \caption{A user's average BBI percentage change for each activity}
    \label{fig:continuous_bbi_change_analysis}
\end{figure}

\begin{figure}[htbp]
    \centering
    \includegraphics[width=0.48\textwidth]{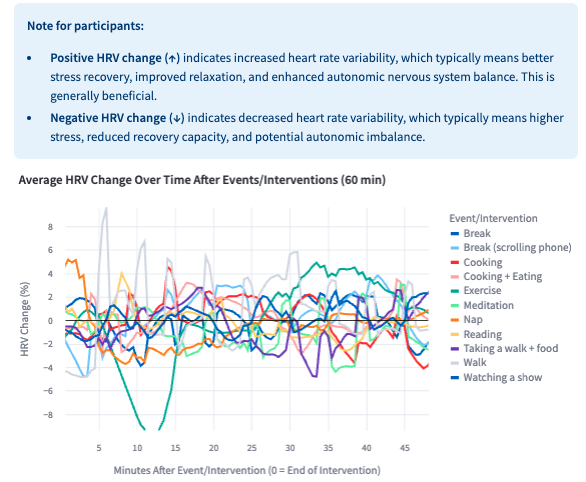}
    \caption{Continuous HRV Change}
    \figsubcaption{One user's average HRV changes after an intervention are displayed. Different instances of the same intervention are averaged. This user tagged many different types of interventions.}
    \label{fig:continuous_hrv_change_analysis}
\end{figure}

\begin{figure}[htbp]
    \centering
    \includegraphics[width=0.48\textwidth]{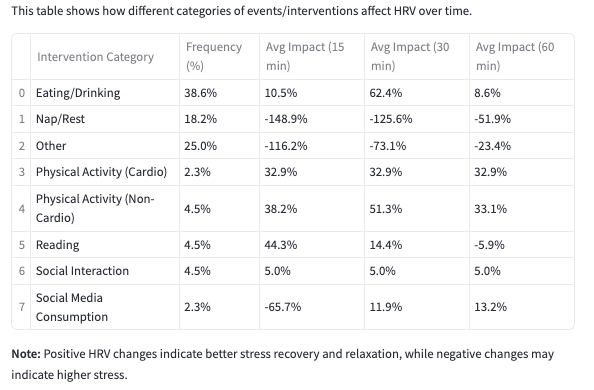}
    \caption{Intervention Category Impact Table}
    \figsubcaption{This table presents data from the user shown in Figure~\ref{fig:continuous_hrv_change_analysis}. Each intervention was categorized by our ML model and the table reports the average impact of each intervention over varying time windows after it occurred, along with how frequently each intervention was used.}
    \label{fig:hrv_intervention_category_impact_table}
\end{figure}

\begin{figure}[htbp]
    \centering
    \includegraphics[width=\linewidth]{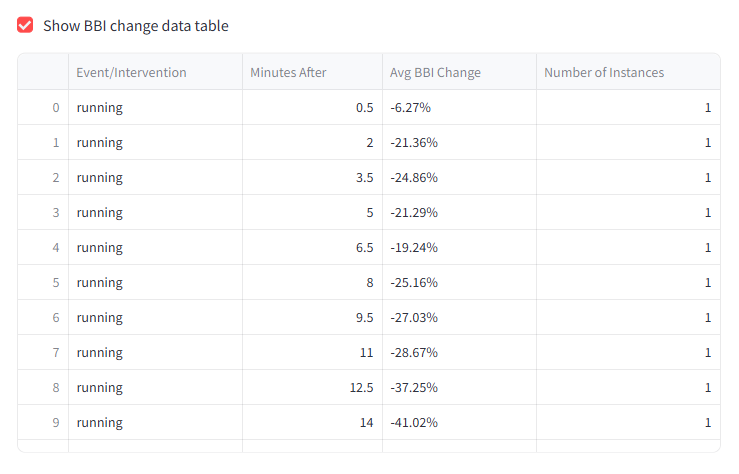}
    \caption{A table displaying the average BBI change for each instance of an activity} % TODO: what does minutes after category mean?
    \label{fig:bbi_intervention_category_impact_table}
\end{figure}

\begin{figure}[htbp]
    \centering
    \includegraphics[width=\linewidth]{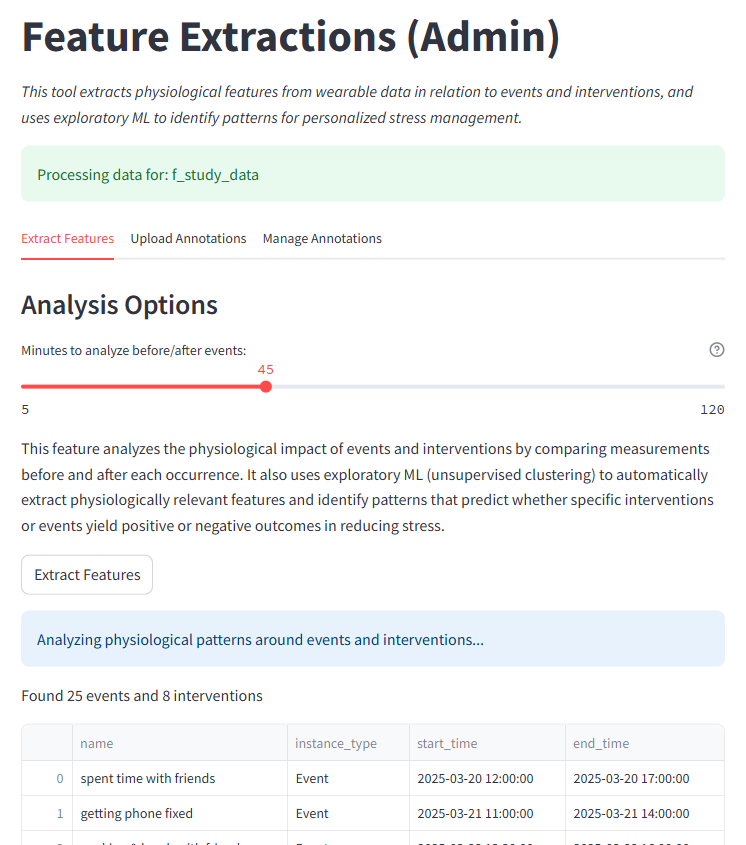}
    \caption{Users can determine the number of minutes to analyze before and after events.}
    \label{fig:select_minutes_slider}
\end{figure}

\begin{figure}[htbp]
    \centering
    \includegraphics[width=\linewidth]{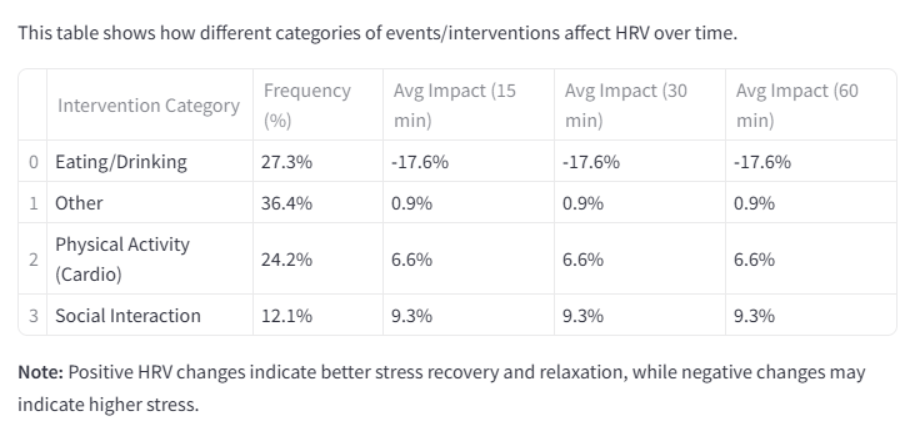}
    \caption{A user's summary of how different categories of events/interventions affect HRV over time.}
    \label{fig:intervention_category_impact_table}
\end{figure}

% \FloatBarrier
%  \subsection{Signal Search and Analysis}

\clearpage
\FloatBarrier
\section{Appendix B. Supplementary Results}

\FloatBarrier
\subsection{Data Smoothing Visualizations}

\begin{figure*}[htbp]
    \centering
    \includegraphics[width=\textwidth,height=0.85\textheight,keepaspectratio]{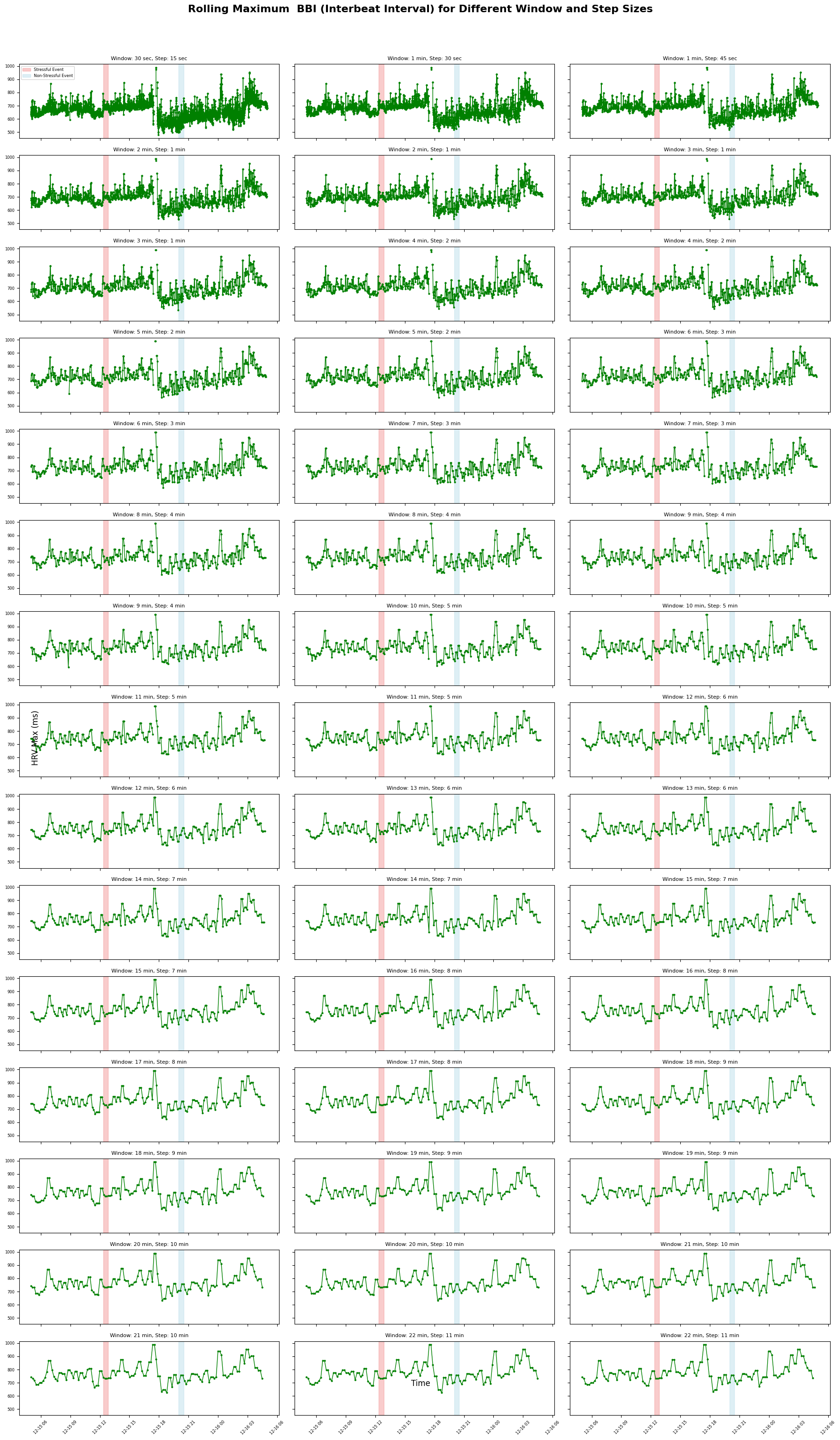}
    \caption{BBI Robustness Analysis and Search.}
    \label{fig:rolling_bbi_max_search}
\end{figure*}

\begin{figure*}[htbp]
    \centering
    \includegraphics[width=\textwidth,height=0.85\textheight,keepaspectratio]{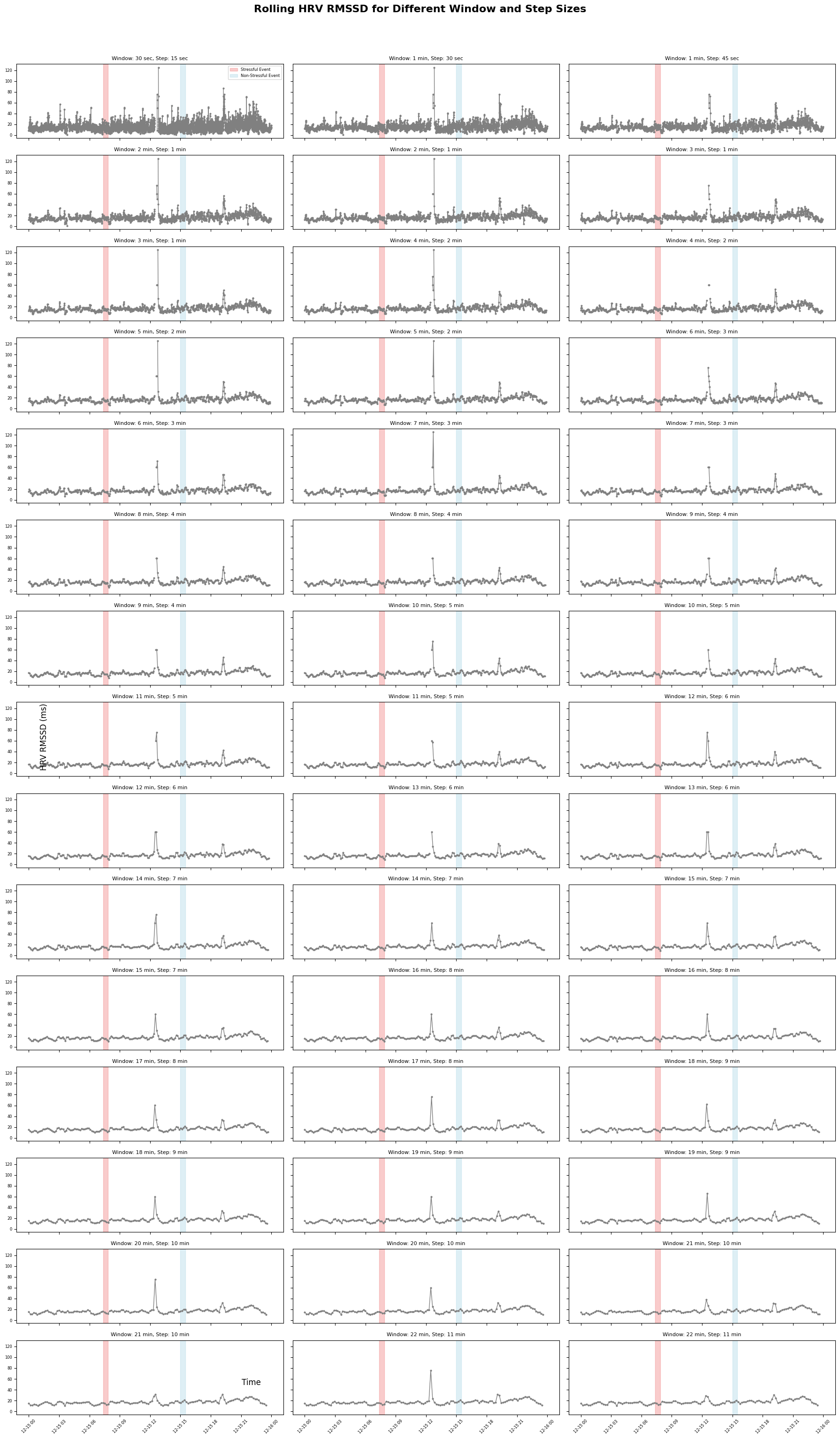}
    \caption{HRV Robustness Analysis and Search.}
    \label{fig:rolling_hrv_search}
\end{figure*}

\FloatBarrier
\subsection{Utilization Metrics}

\begin{figure}[htbp]
  \centering
  \includegraphics[width=0.45\textwidth]{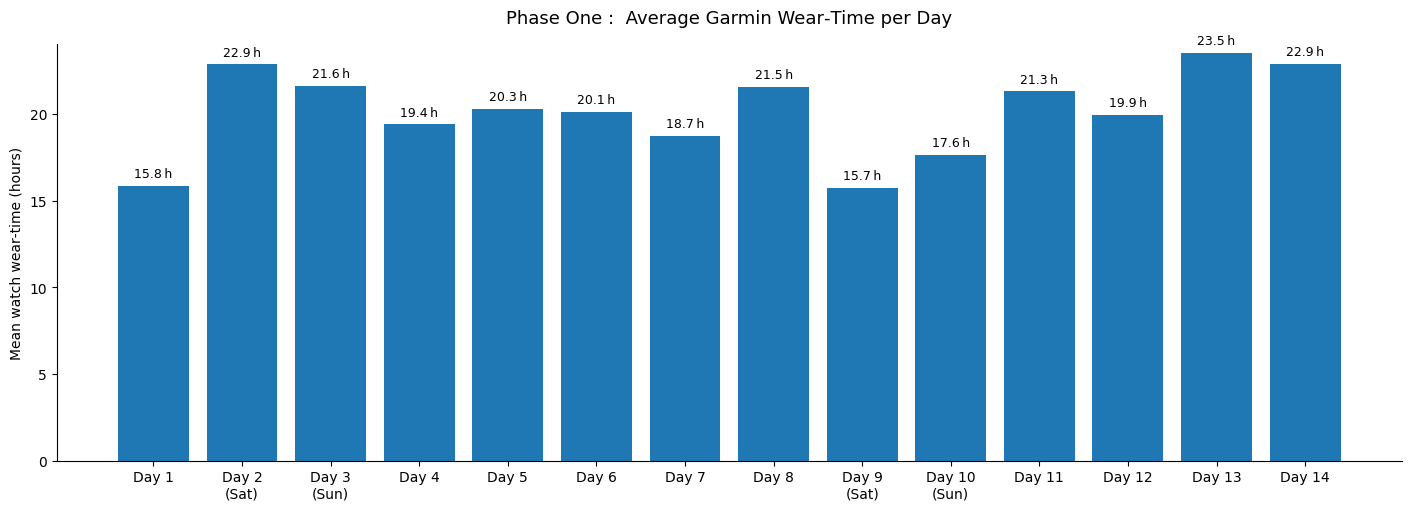}
  \caption{Mean watch wear‑time (hours per day) during Phase One.}
  \figsubcaption{Participants wore the device for the vast majority of each day.}
  \label{fig:wear_time_p1}
\end{figure}

\begin{figure}[htbp]
    \centering
    \includegraphics[width=0.48\textwidth]{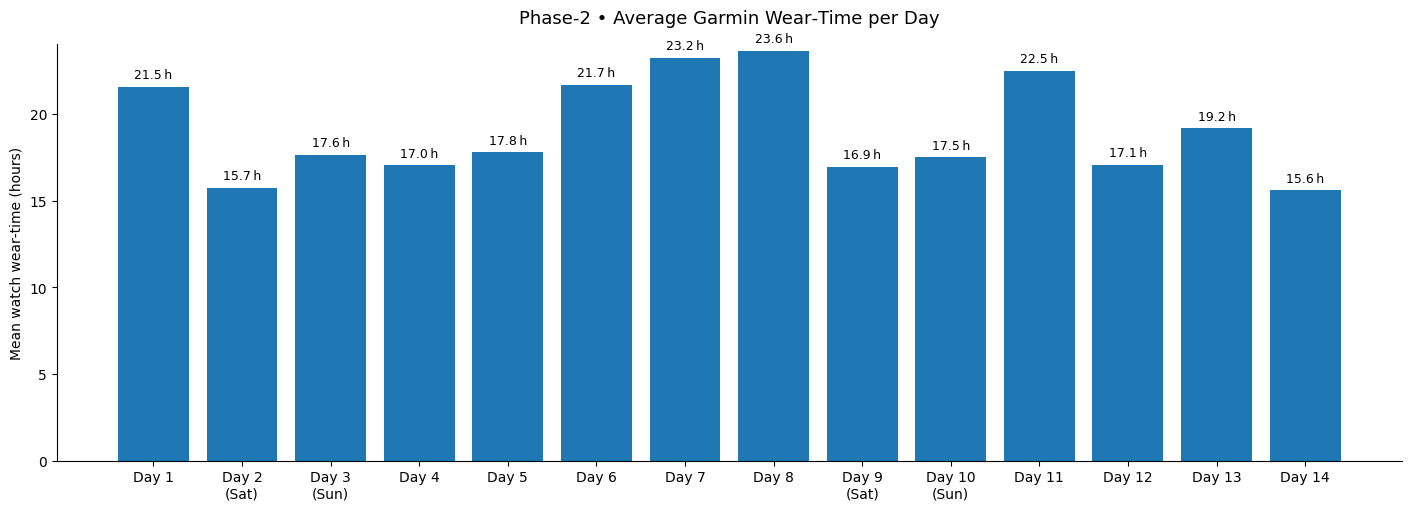}
    \caption{Mean watch wear‑time in hours per day in the phase one of the study}
    \smallskip
    \small \textit{Participants adhered to the recommendation to wear their watches for most hours each day during phase two of the study.  }
    \label{fig:wear_time_p2}
\end{figure}

\begin{figure}[htbp]
  \centering
  \includegraphics[width=0.48\textwidth]{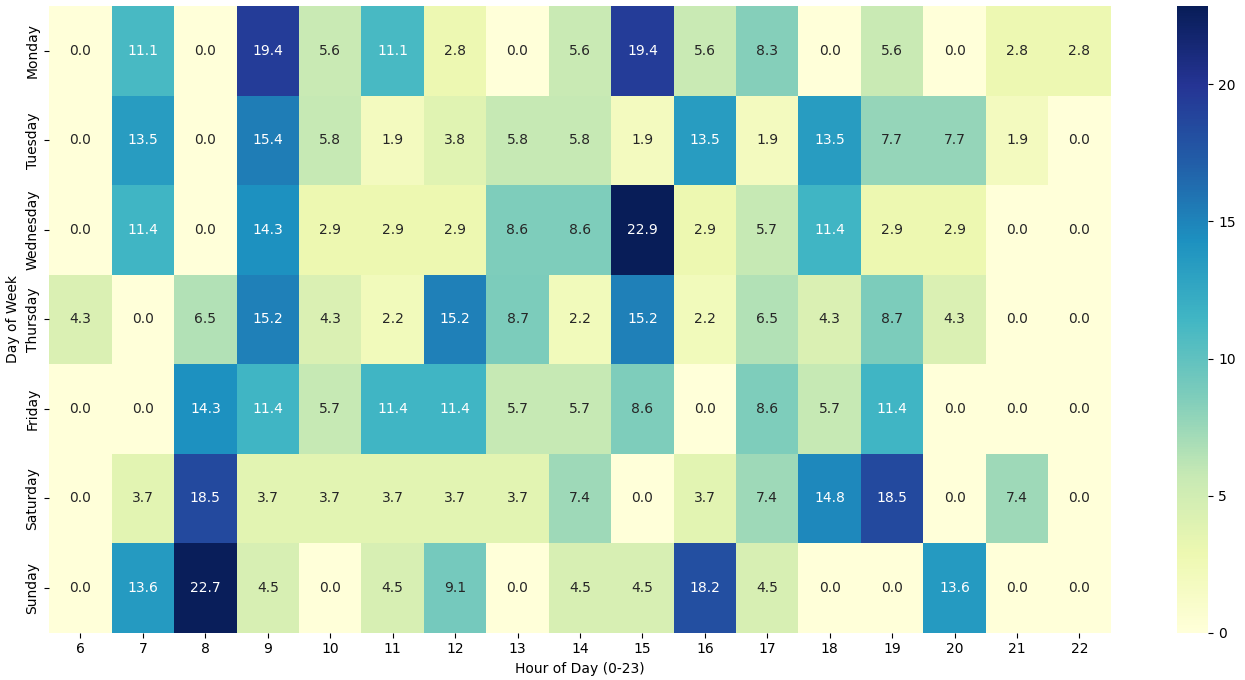}
  \caption{Logged interventions by day and hour.}
    \figsubcaption{The heatmap shows the average number of interventions logged  at specific times across all participants. Darker cells indicate a higher number of interventions logged.}
  \label{fig:logged_intervention_by_day}
\end{figure}

\FloatBarrier
\subsection{Aggregated Results}

\begin{table*}[htbp]
    \centering
    \resizebox{\textwidth}{!}{
    \begin{tabular}{@{}lcccccccccccccc@{}}
        \toprule
        \textbf{Intervention Category} & \textbf{Freq.\ (\%)} 
        & \multicolumn{3}{c}{\textbf{15 min}} 
        & \multicolumn{3}{c}{\textbf{30 min}} 
        & \multicolumn{3}{c}{\textbf{60 min}} 
        & \multicolumn{3}{c}{\textbf{120 min}} \\
        \cmidrule(lr){3-5}\cmidrule(lr){6-8}\cmidrule(lr){9-11}\cmidrule(lr){12-14}
        & & \textbf{HRV} & \textbf{HR} & \textbf{Stress} 
          & \textbf{HRV} & \textbf{HR} & \textbf{Stress} 
          & \textbf{HRV} & \textbf{HR} & \textbf{Stress} 
          & \textbf{HRV} & \textbf{HR} & \textbf{Stress}  \\
        \midrule
        Socializing / Social Interaction    & 20.6\% & -0.814 & -4.483 & -6.814 & -1.725 & -4.950 & -6.810 & -2.184 & -4.346 & -6.361 & -1.044 & -4.816 & -7.251 \\
        Academic \& Educational              & 17.4\% & -0.574 &  4.349 &  4.522 & -0.598 &  4.009 &  2.387 & -1.065 &  4.921 &  2.123 & -1.123 &  6.128 &  1.056 \\
        Spirituality / Mindful Activities   & 14.6\% & -6.605 & -2.971 & -2.002 & -9.825 & -0.683 &  0.022 & -11.700 & -1.093 &  0.919 & -13.215 & -1.439 &  1.074 \\
        Food / Drink / Nutrition            & 12.6\% & -2.748 &  1.010 &  0.816 & -2.511 &  1.143 &  2.342 & -2.349 &  2.441 &  4.394 & -1.207 &  0.406 &  1.466 \\
        Physical Activity – Cardio          & 11.1\% & -7.819 & -4.822 & -4.435 & -5.761 & -5.097 & -3.831 & -3.546 & -3.006 & -0.634 & -2.547 & -4.808 & -6.361 \\
        Rest \& Recovery                     & 10.7\% &  0.360 & -1.848 & -11.926 &  0.255 & -2.516 & -10.033 &  1.294 & -5.264 & -10.489 &  2.756 & -7.351 & -13.826 \\
        Physical Activity – Non-cardio      & 6.3\% & -10.155 & -2.257 & -6.046 & -7.011 & -0.955 & -5.137 & -3.207 & -1.591 & -6.953 & -2.075 &  1.918 & -8.314 \\
        Healthcare / Therapy                & 4.0\% & -4.542 & 10.040 & 13.522 & -4.230 & 11.191 & 13.401 & -3.164 &  7.674 & 11.383 & -1.731 &  7.733 & 10.716 \\
        Other                               & 2.8\% & -2.554 & 12.376 & -0.783 & -2.431 & 12.963 & -5.007 & -4.445 &  4.158 & -13.054 & -5.281 &  3.598 &  2.226 \\
        \bottomrule
\end{tabular}
    }
    \caption{Average changes in HRV, heart rate (bpm), and Stress Score (Garmin 0–100 scale) measured at 15, 30, 60, and 120 minutes after the intervention. Each change was calculated by comparing the mean value within each post-intervention time window to the mean value in the 15-minute window before the intervention. Finally, changes in the physiological variables were averaged across interventions in the same category.}
    \figsubcaption{This table displays the average change in HRV, heart rate, and Stress Score for each intervention category, along with how frequently each intervention occurred. Values are shown across multiple post-intervention time windows, enabling comparison of short and longer-term physiological effects.}
    \label{tab:combined_hrv_hr_stress}
\end{table*}

\begin{figure}[ht]
  \centering

  \begin{minipage}[t]{0.48\textwidth}
    \centering
    \includegraphics[width=\linewidth]{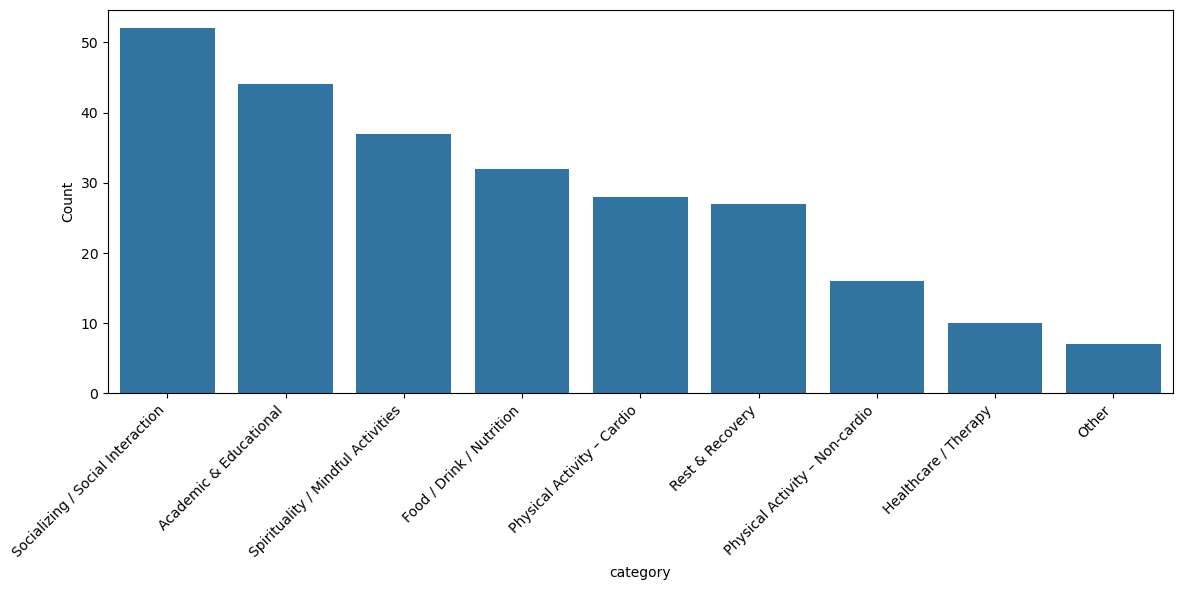}
    \caption{Distribution of intervention categories.}
    \figsubcaption{
      The number of tagged activities belonging to each intervention category.
    }
    \label{fig:intervention_categories}
  \end{minipage}%
  \hfill
  \begin{minipage}[t]{0.48\textwidth}
    \centering
    \includegraphics[width=\linewidth]{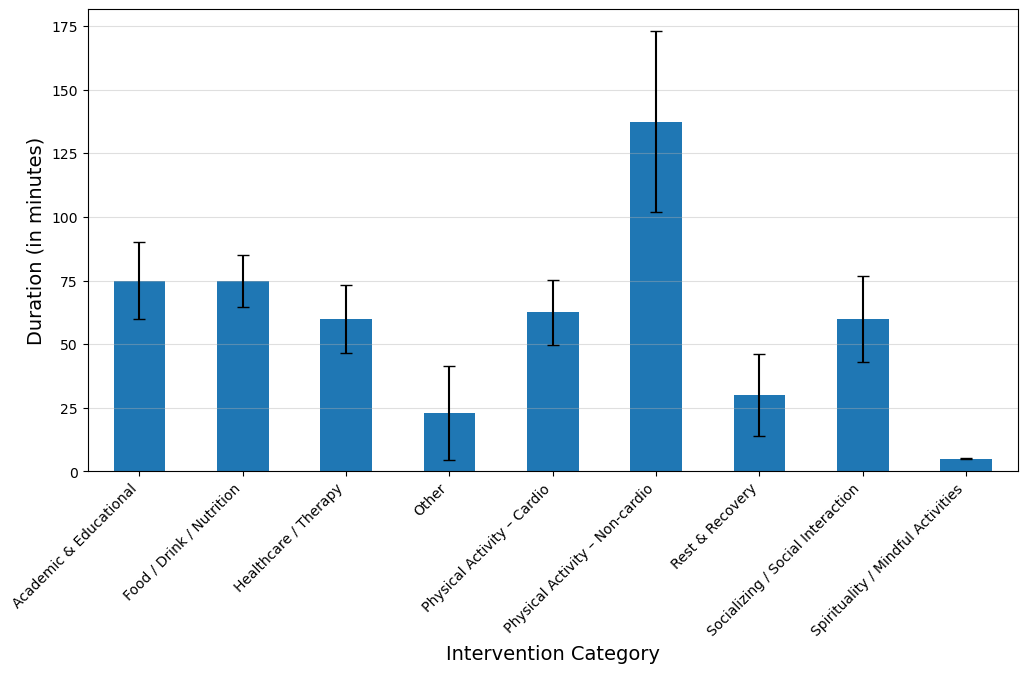}
    \caption{Duration of intervention/event (in minutes), by category.}
    \figsubcaption{
      The median and standard error of duration for activities belonging to each intervention category.
    }
    \label{fig:duration_categories}
  \end{minipage}

\end{figure}

\begin{figure}[htbp]
  \centering
  \includegraphics[width=\linewidth]{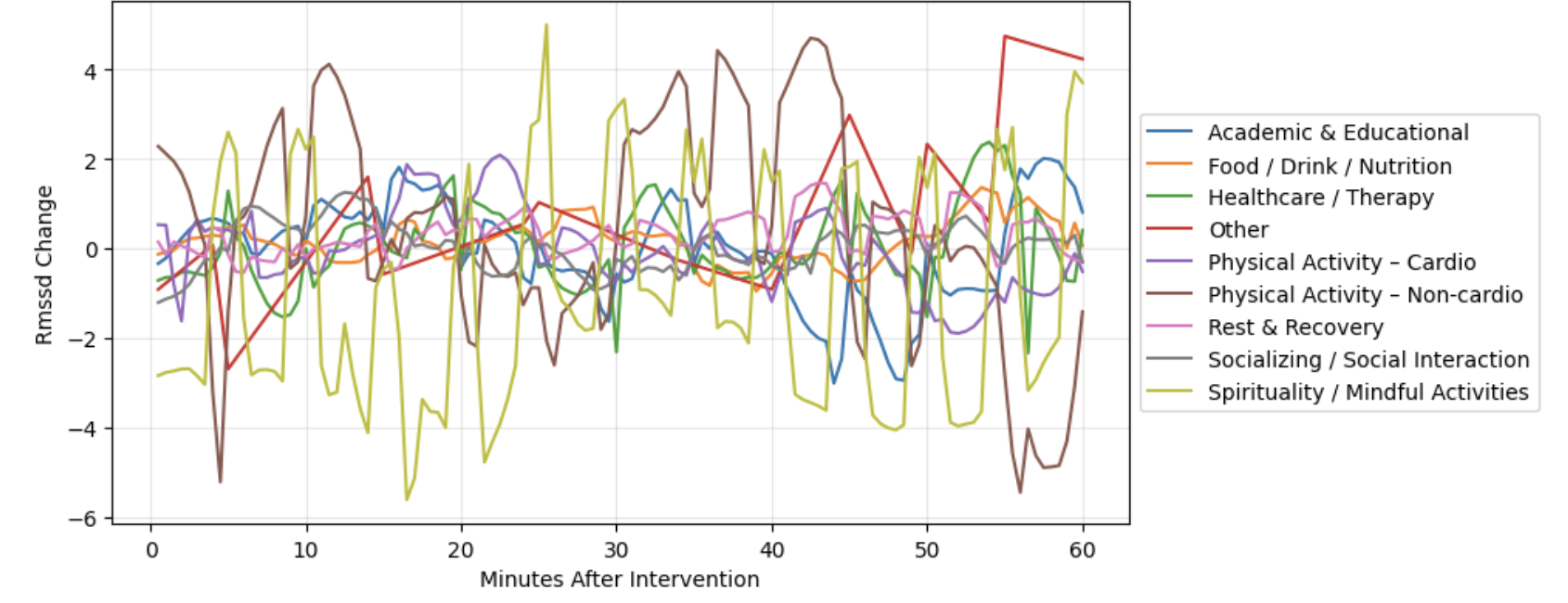}
  \caption{Population-level average change in HRV (RMSSD) 60 minutes post-intervention.}
    \figsubcaption{The plot above shows the mean trajectory for change in HRV (RMSSD) for all activities in each intervention category.}
  \label{fig:hrv_time_series}
\end{figure}

\begin{figure}[htbp]
    \centering
    \includegraphics[width=0.48\textwidth]{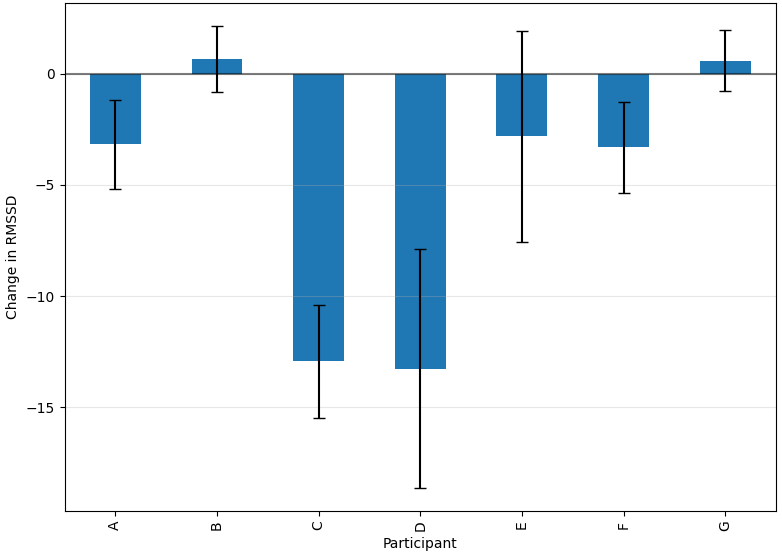}
    \includegraphics[width=0.48\textwidth]{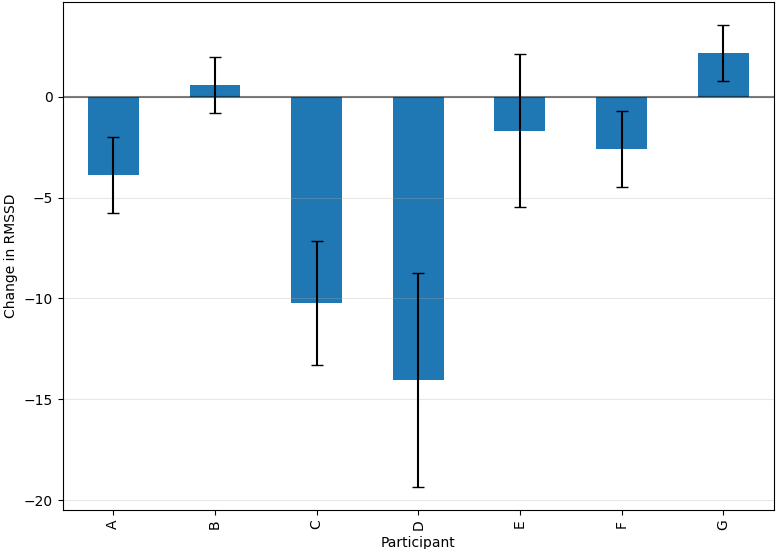}
    \caption{View Annotations}
    \smallskip
    \small \textit{Average change in HRV 30 and 60 minutes before and after intervention by participant.}
    \label{fig:avg_hrv_changes_participants}
\end{figure}

\begin{figure}[htbp]
    \centering
    \includegraphics[width=0.48\textwidth]{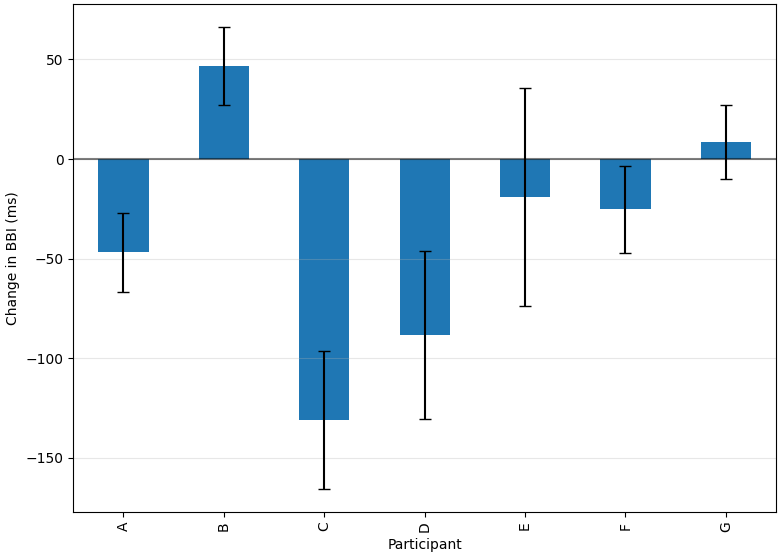}
    \includegraphics[width=0.48\textwidth]{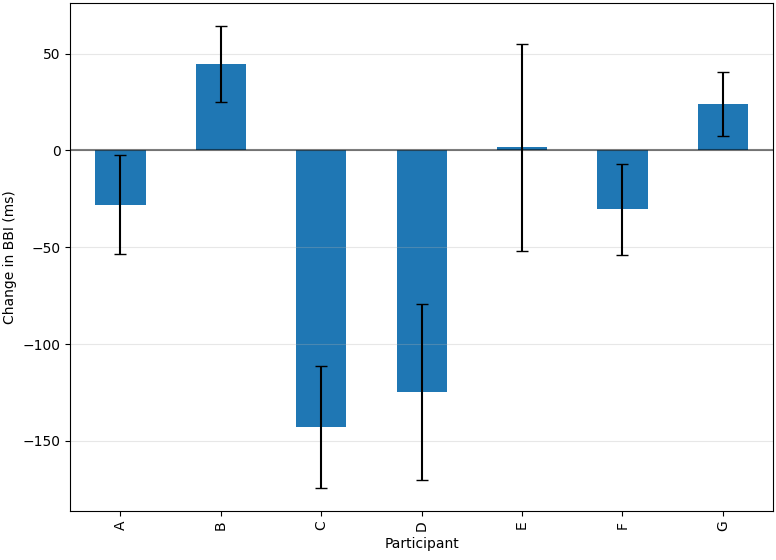}
    \includegraphics[width=0.48\textwidth]{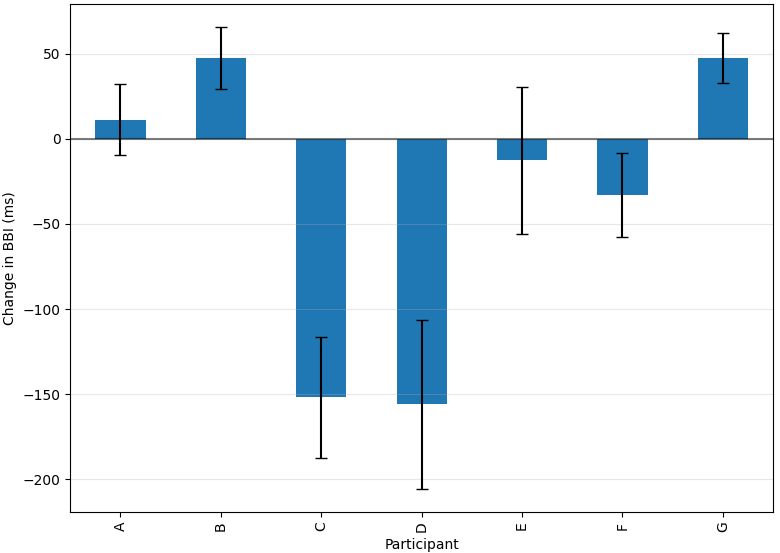}
    \caption{View Annotations}
    \smallskip
    \small \textit{Average change in BBI 15, 30, 60 minutes before and after intervention by participant.}
    \label{fig:avg_bbi_changes_participants}
\end{figure}

\begin{figure}[htbp]
    \centering
    \includegraphics[width=0.48\textwidth]{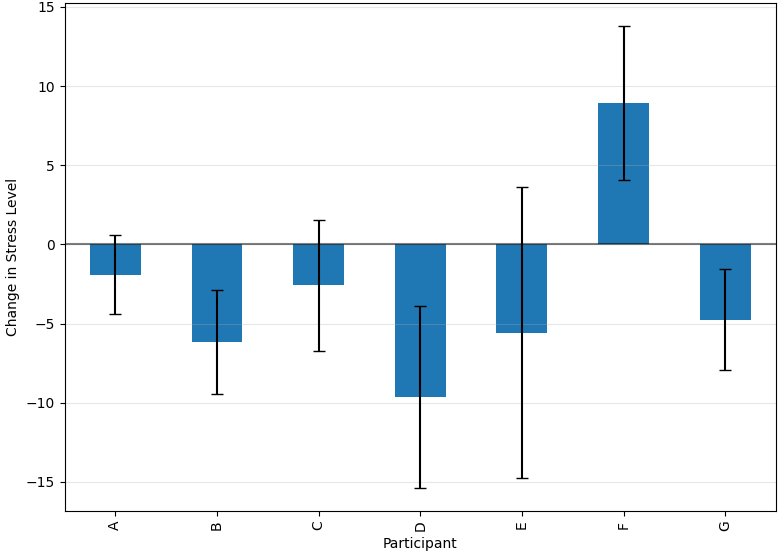}
    \includegraphics[width=0.48\textwidth]{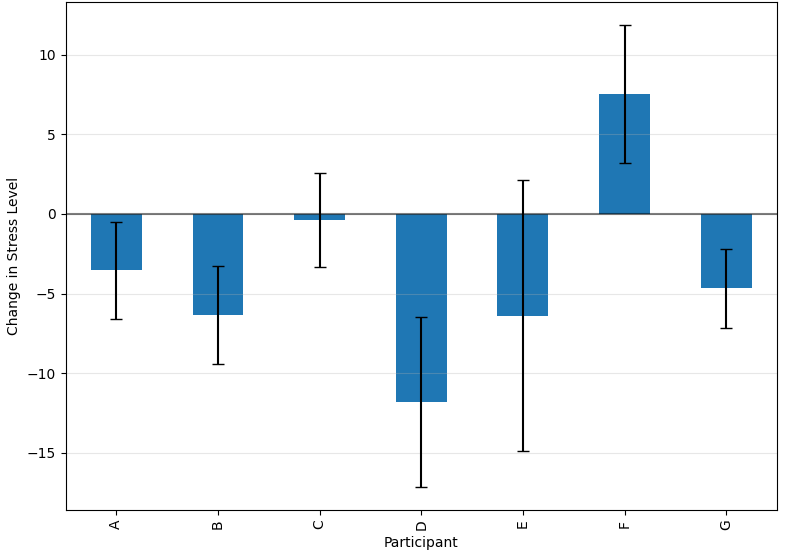}
    \includegraphics[width=0.48\textwidth]{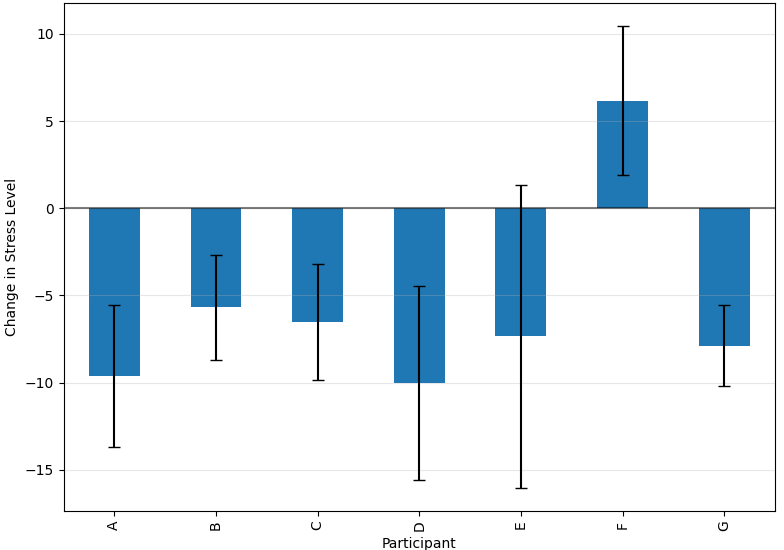}
    \caption{View Annotations}
    \smallskip
    \small \textit{Average change in Garmin Stress 15, 30, 60 minutes before and after intervention by participant.}
    \label{fig:avg_stress_changes_participants}
\end{figure}

\begin{figure}[htbp]
    \centering
    \includegraphics[width=0.48\textwidth]{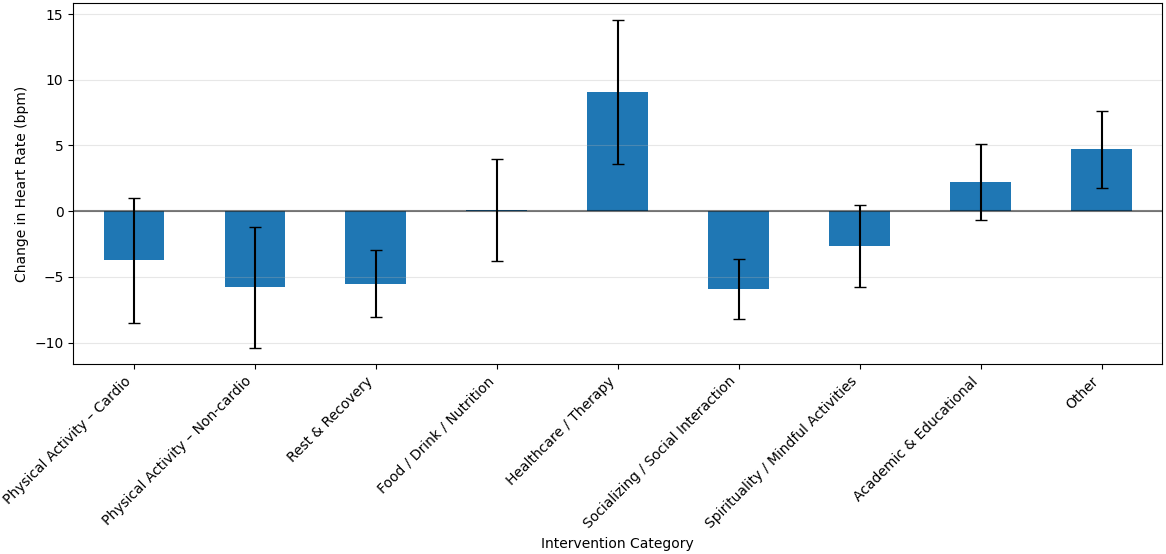}
    \includegraphics[width=0.48\textwidth]{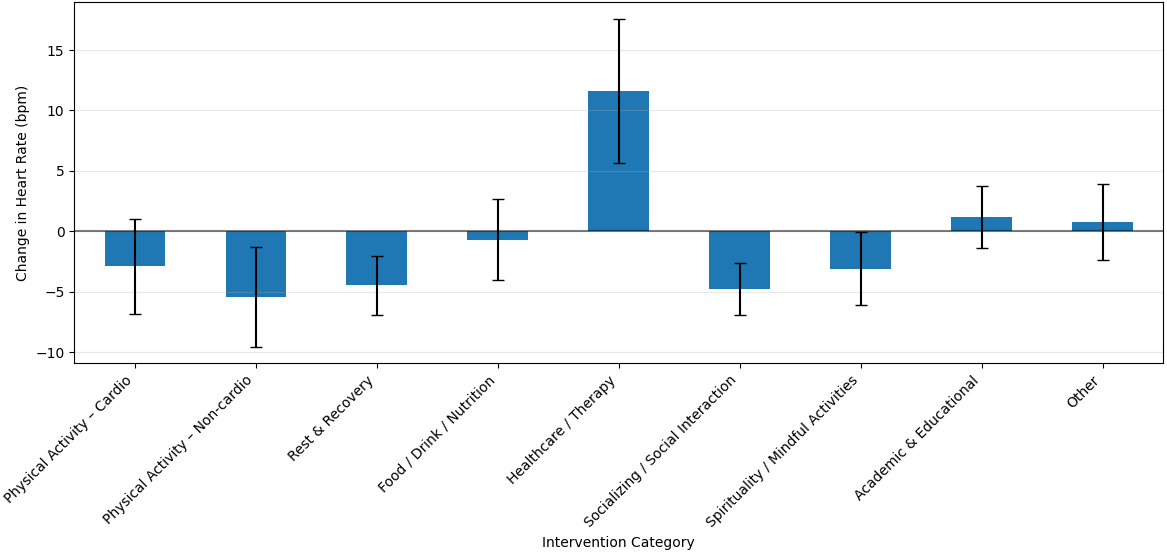}
    \includegraphics[width=0.48\textwidth]{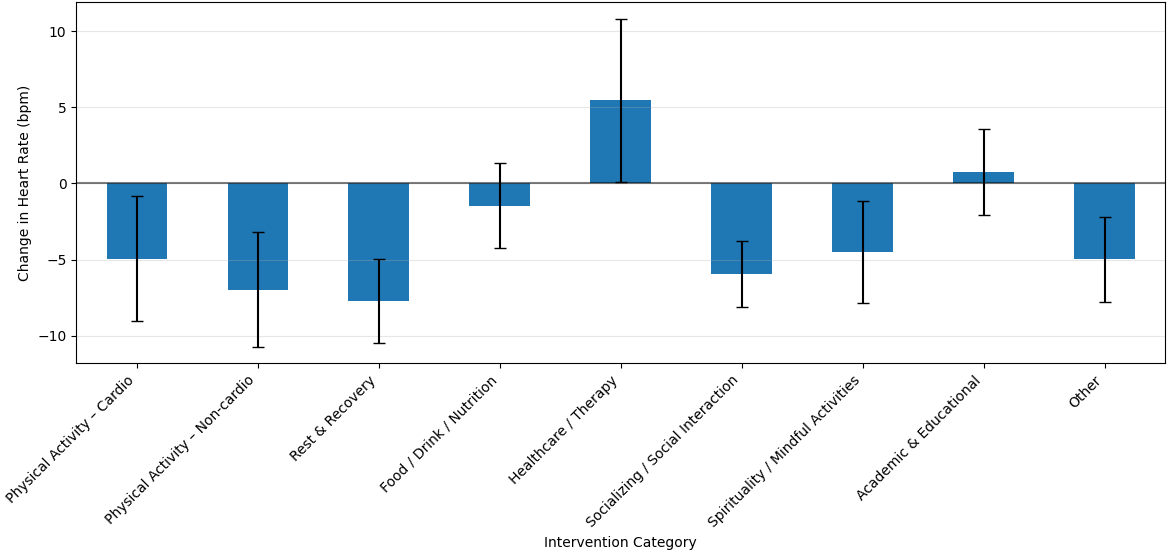}
    \includegraphics[width=0.48\textwidth]{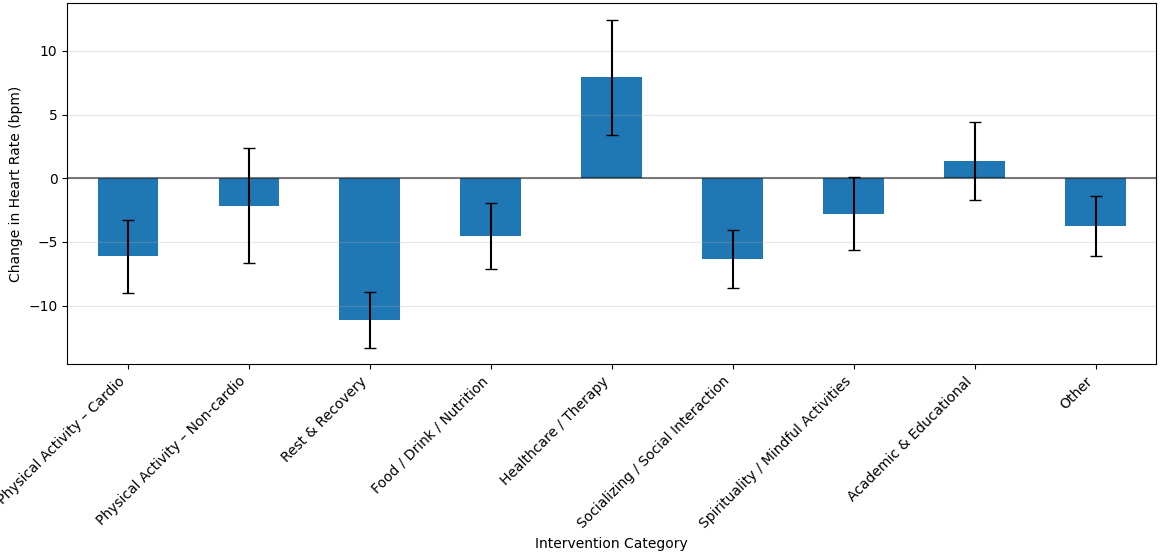}
    \caption{View Annotations}
    \smallskip
    \small \textit{Average change in HR by category for 15, 30, 60, 120 minutes before and after intervention by participant.}
    \label{fig:change_hr_category}
\end{figure}

\begin{figure}[htbp]
    \centering
    \includegraphics[width=0.48\textwidth]{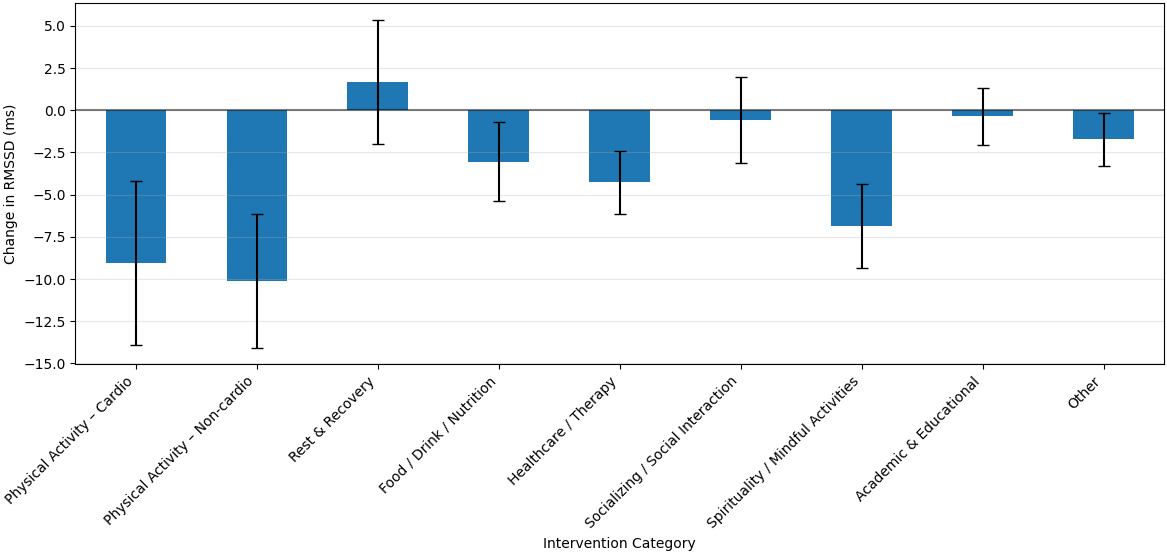}
    \includegraphics[width=0.48\textwidth]{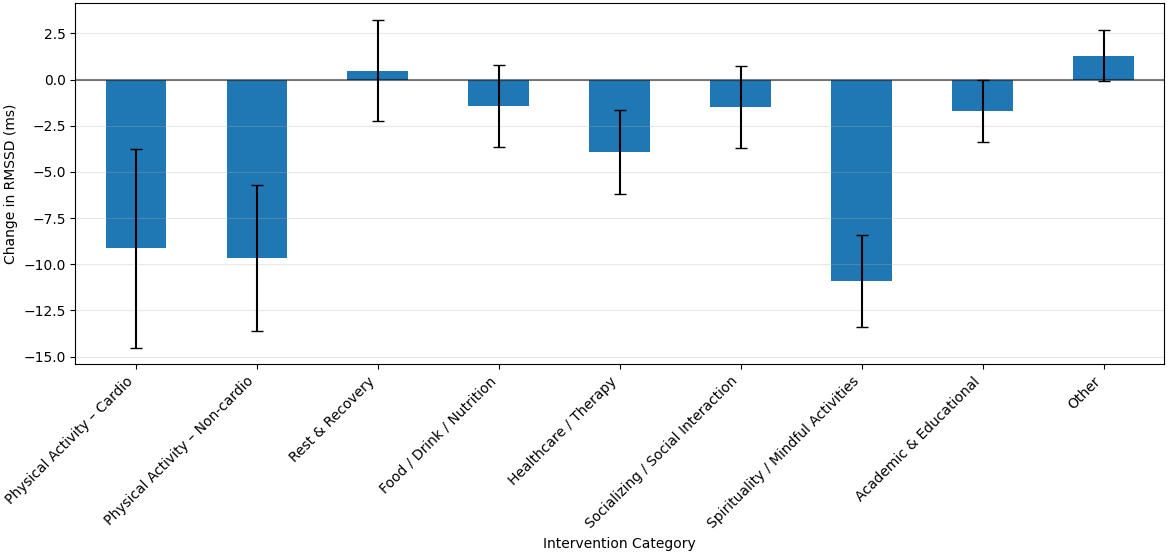}
    \includegraphics[width=0.48\textwidth]{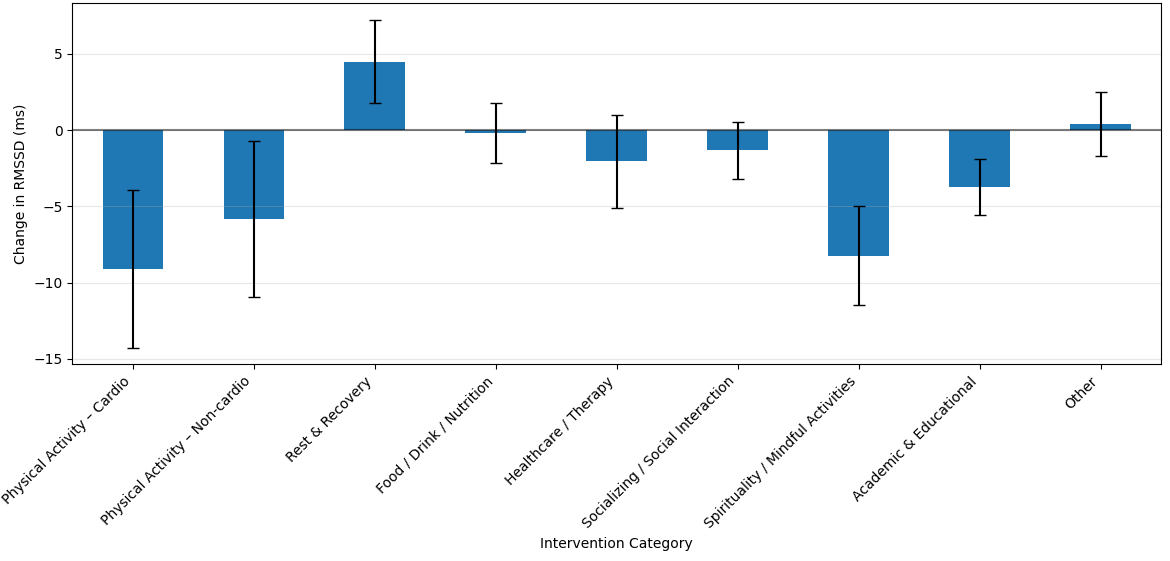}
    \includegraphics[width=0.48\textwidth]{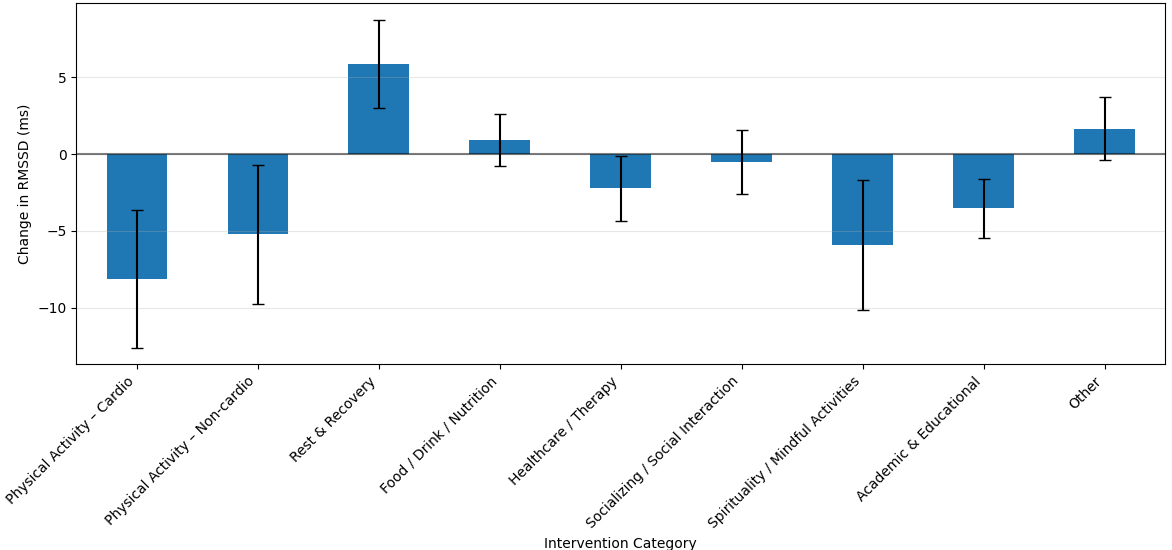}
    \caption{View Annotations}
    \smallskip
    \small \textit{Average change in HRV by category for 15, 30, 60, 120 minutes before and after intervention by participant.}
    \label{fig:change_hrv_category}
\end{figure}

\begin{figure}[htbp]
    \centering
    \includegraphics[width=0.48\textwidth]{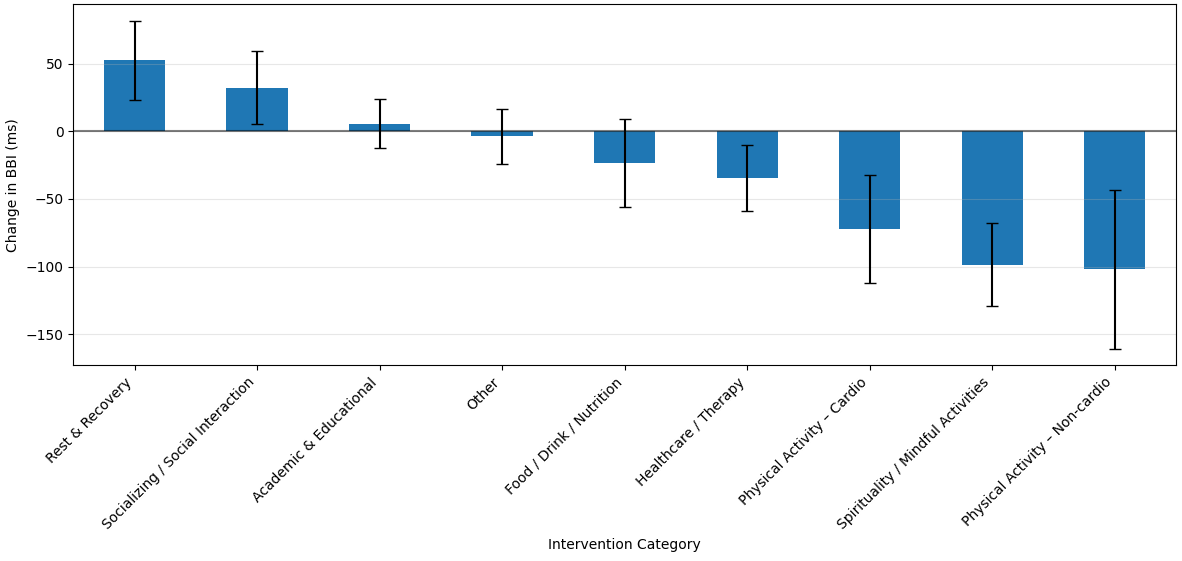}
    \includegraphics[width=0.48\textwidth]{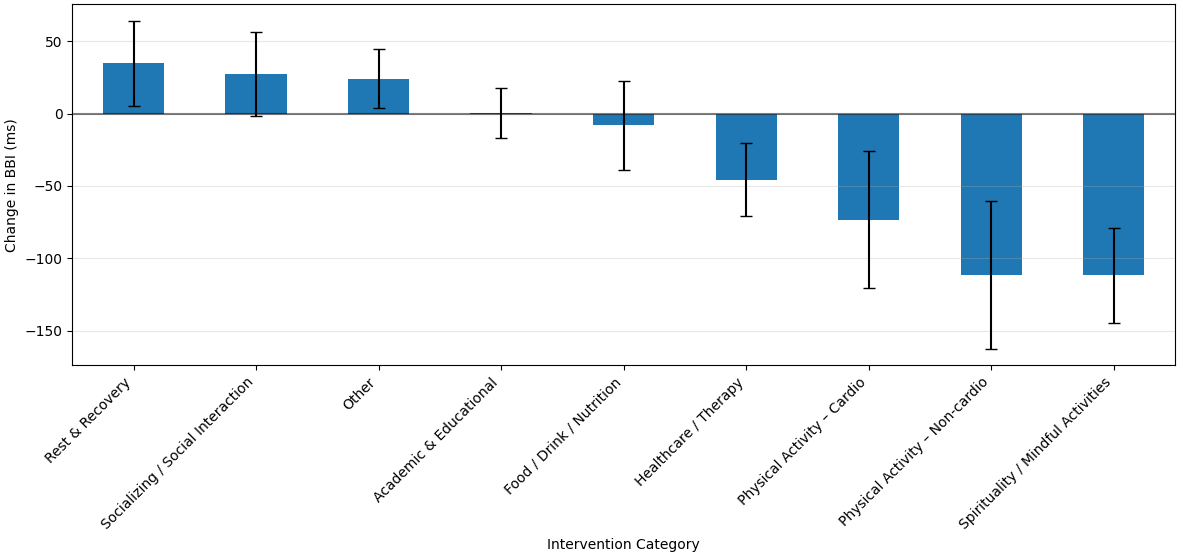}
    \includegraphics[width=0.48\textwidth]{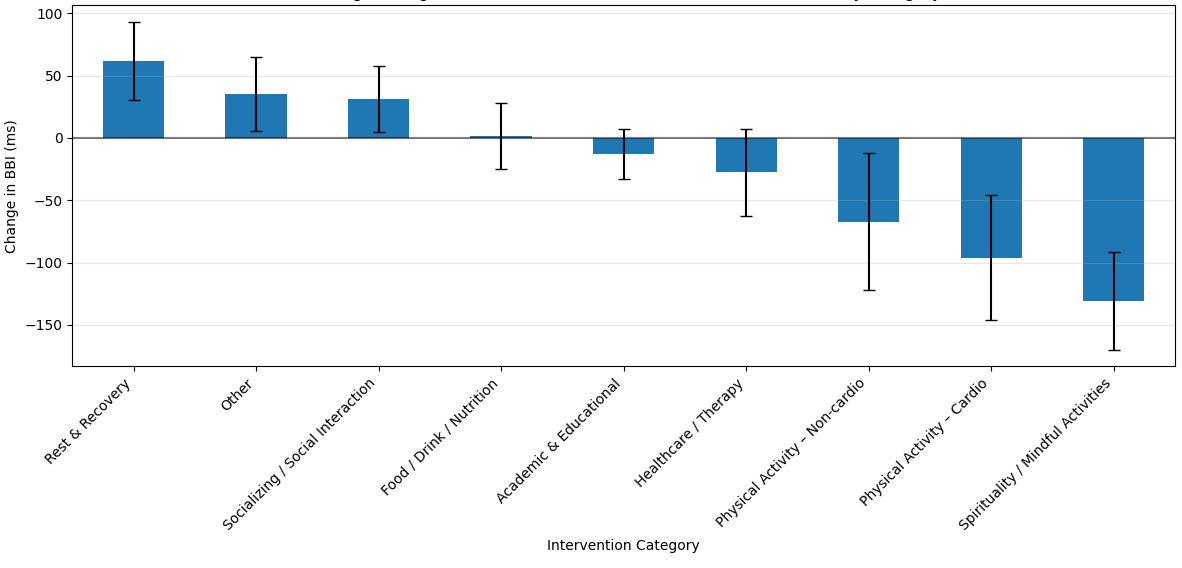}
    \caption{View Annotations}
    \smallskip
    \small \textit{Average change in BBI by category for 15, 30, 60 minutes before and after intervention by participant.}
    \label{fig:change_bbi_category}
\end{figure}

\begin{figure}[htbp]
    \centering
    \includegraphics[width=0.48\textwidth]{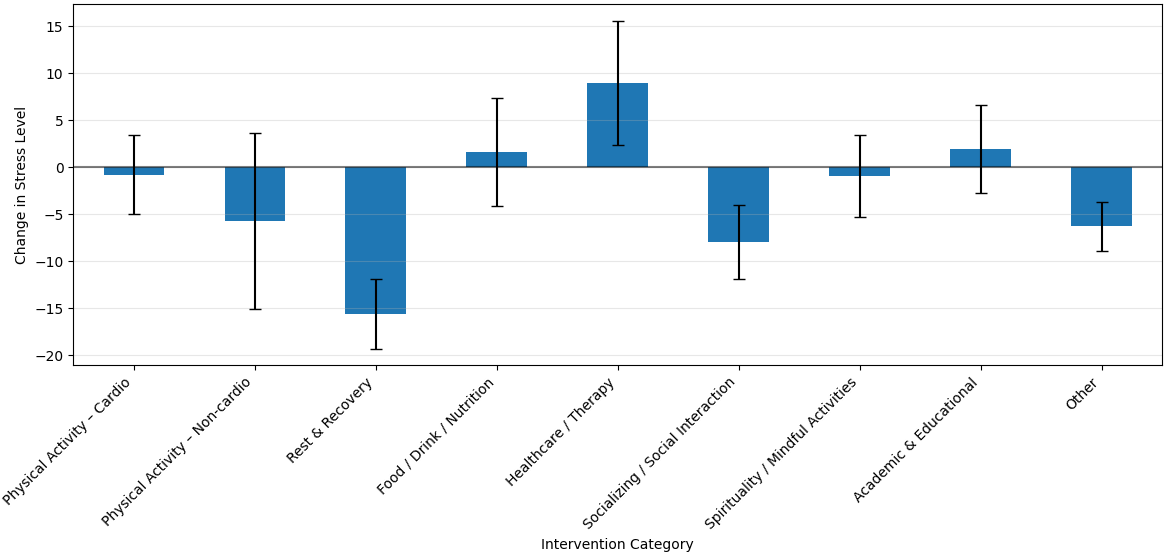}
    \includegraphics[width=0.48\textwidth]{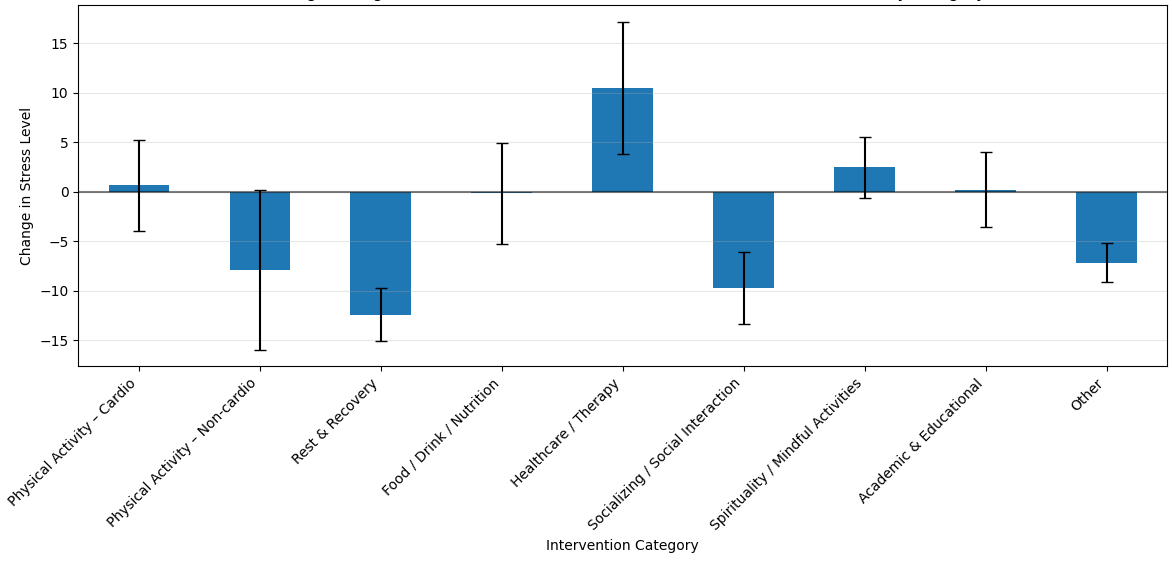}
    \includegraphics[width=0.48\textwidth]{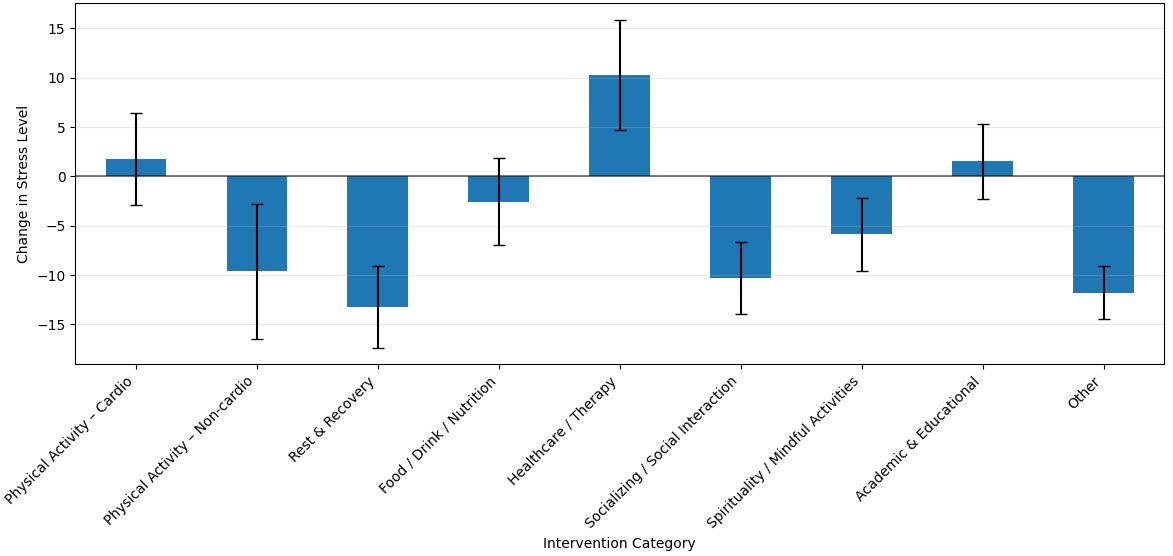}
    \includegraphics[width=0.48\textwidth]{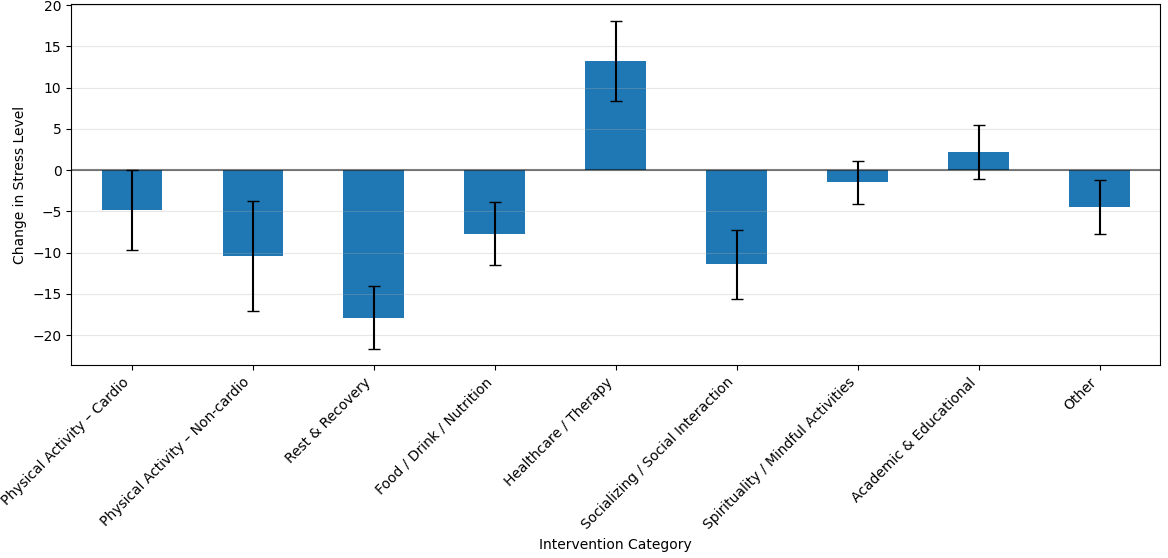}
    \caption{View Annotations}
    \smallskip
    \small \textit{Average change in Garmin Stress by category for 15, 30, 60 minutes before and after intervention by participant.}
    \label{fig:change_stress_category}
\end{figure}

\begin{figure}[htbp]
    \centering
    \includegraphics[width=0.48\textwidth]{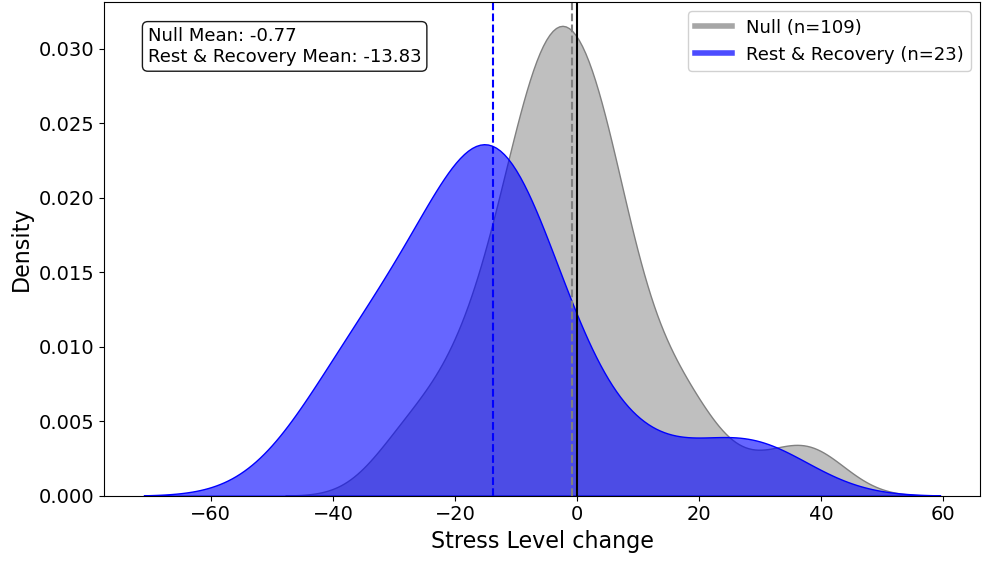}
    \includegraphics[width=0.48\textwidth]{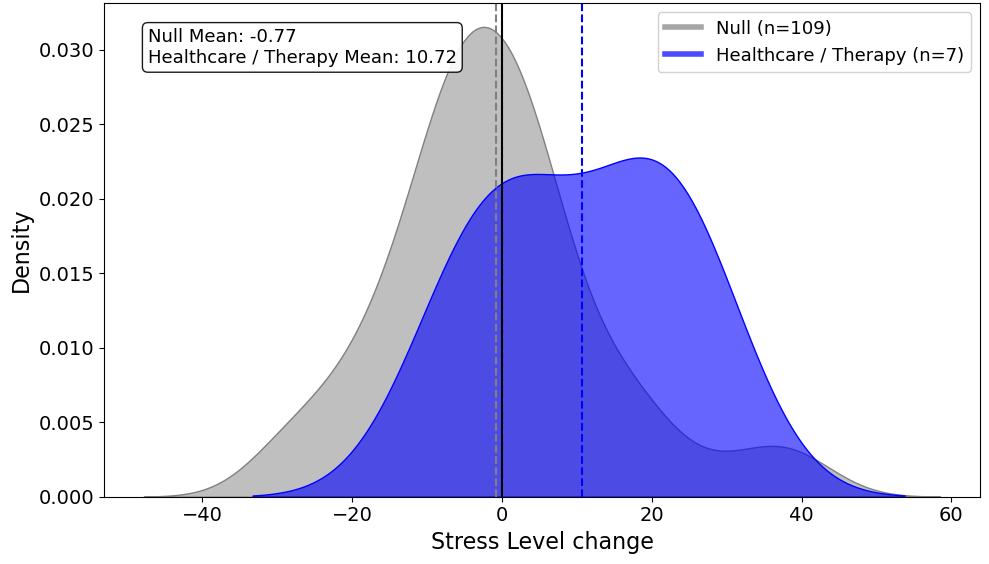}
    \caption{View Annotations}
    \smallskip
    \small \textit{Distribution of change in stress level 120 minutes post-intervent for activities in the Rest \& Recoveries and Healthcare / Therapy categories.}
    \label{fig:dist_rest_health_120}
\end{figure}

% \begin{figure}[htbp]
%   \centering
%   \includegraphics[width=0.96\textwidth]{img/results/participant_category_hrv_30.png}
%   \caption{Average changes in HRV (RMSSD) by participant and category. Change was calculated as the difference between the average HRV (RMSSD) in the 15-minute window before the intervention and the 30-minute window after the intervention}
%     \figsubcaption{These plots show the average change in HRV (RMSSD) for each participant and each intervention category.} % Comment on participant preferences? Tendency for HRV to increase/decrease? If certain interventions are more effective?
%   \label{fig:participant_rmssd_30}
% \end{figure}

\begin{figure}[htbp]
  \centering
  \includegraphics[width=0.96\textwidth]{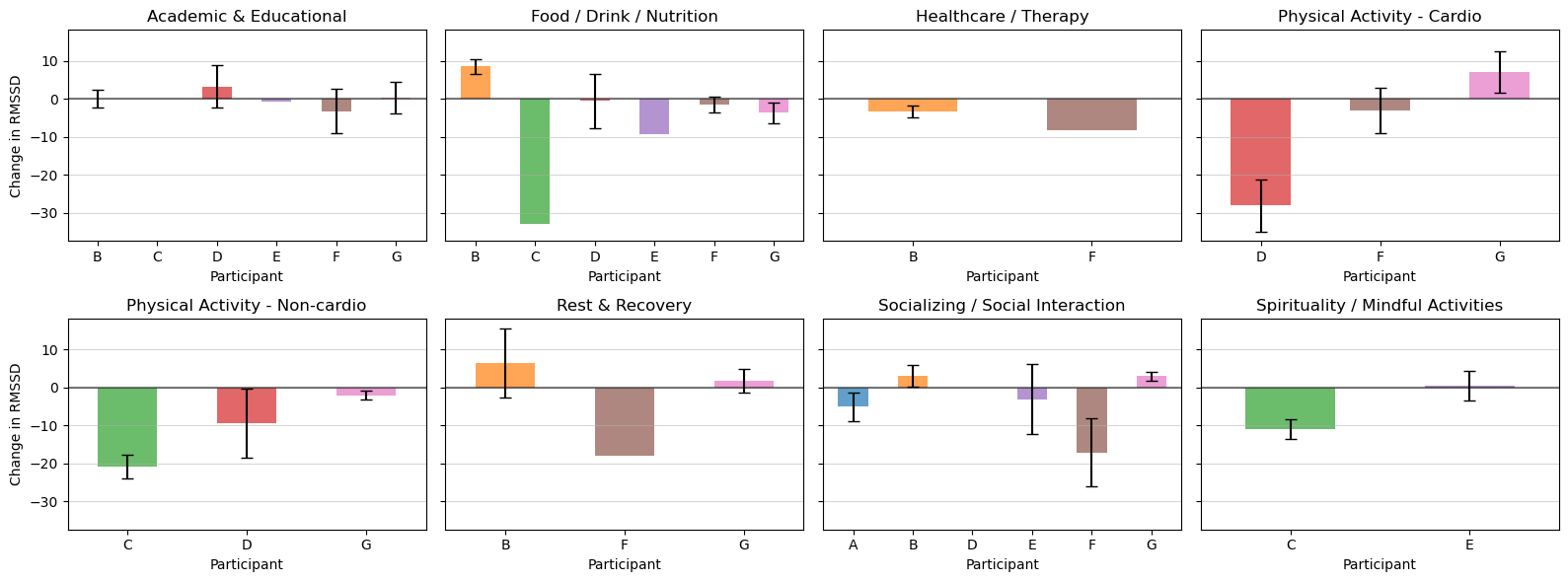}
  \caption{Average changes in HRV (RMSSD) by participant and category. Change was calculated as the difference between the average HRV (RMSSD) in the 15-minute window before the intervention and the 30-minute window after the intervention}
    \smallskip
    \small \textit{This plot presents the same data in Figure~\ref{fig:participant_rmssd_30} but uses a consistent y-axis scale to highlight differences in the magnitude of the physiological impact.}
  \label{fig:participant_rmssd_30_yscale}
\end{figure}

\begin{figure}[htbp]
    \centering
    \includegraphics[width=0.48\textwidth]{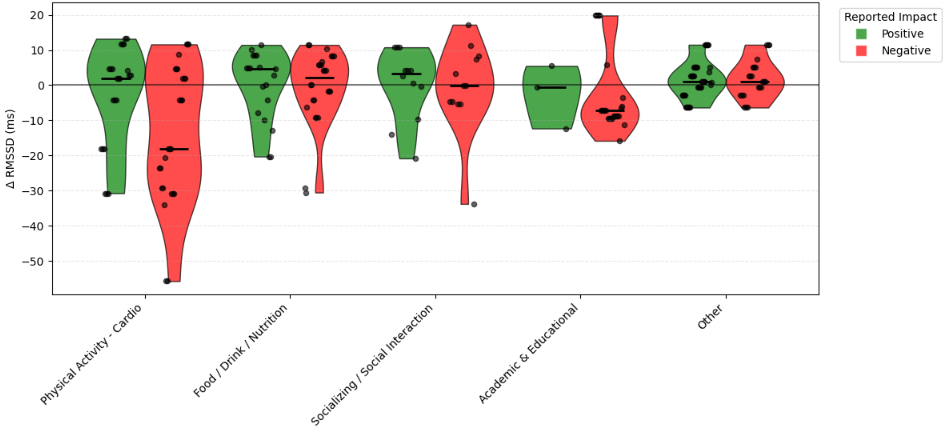}
    \includegraphics[width=0.48\textwidth]{img/results/perceived_hrv_60.png}
    \includegraphics[width=0.48\textwidth]{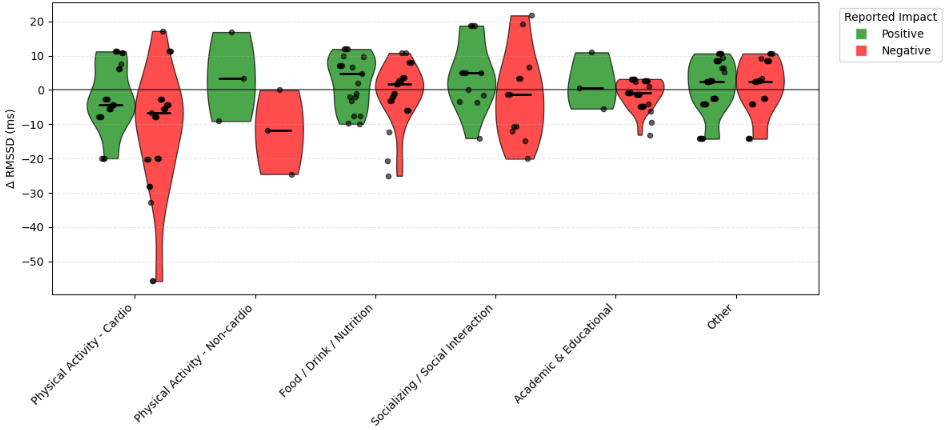}
    \caption{View Annotations}
    \smallskip
    \small \textit{Difference in perceived HRV impact of an intervention vs. actual impact for 15, 30, 60, and 120 minutes.}
    \label{fig:perceived_hrv}
\end{figure}

\begin{figure}[htbp]
    \centering
    \includegraphics[width=0.48\textwidth]{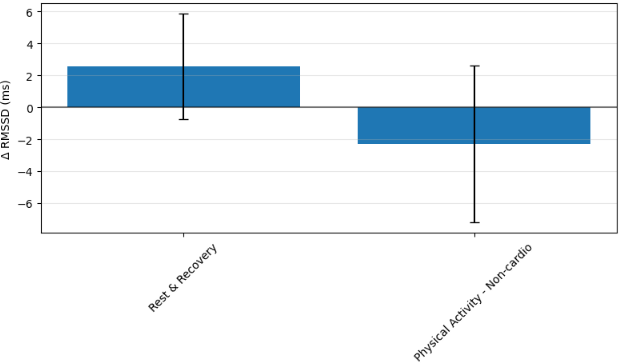}
    \includegraphics[width=0.48\textwidth]{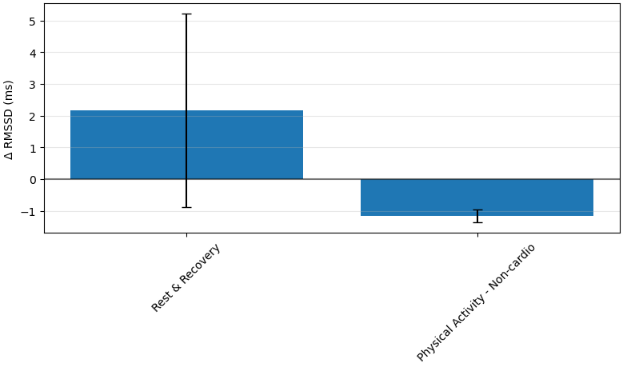}
    \includegraphics[width=0.48\textwidth]{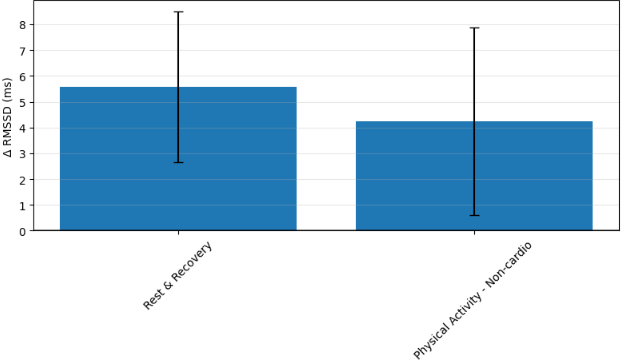}
    \includegraphics[width=0.48\textwidth]{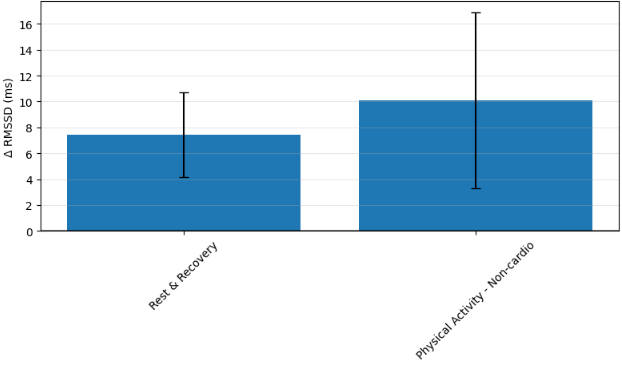}
    \caption{View Annotations}
    \smallskip
    \small \textit{Change in HRV for one user at 15, 30, 60 and 120 minutes after an intervention in a certain category.}
    \label{fig:g_hrv}
\end{figure}

\begin{figure}[htbp]
    \centering
    \includegraphics[width=\linewidth]{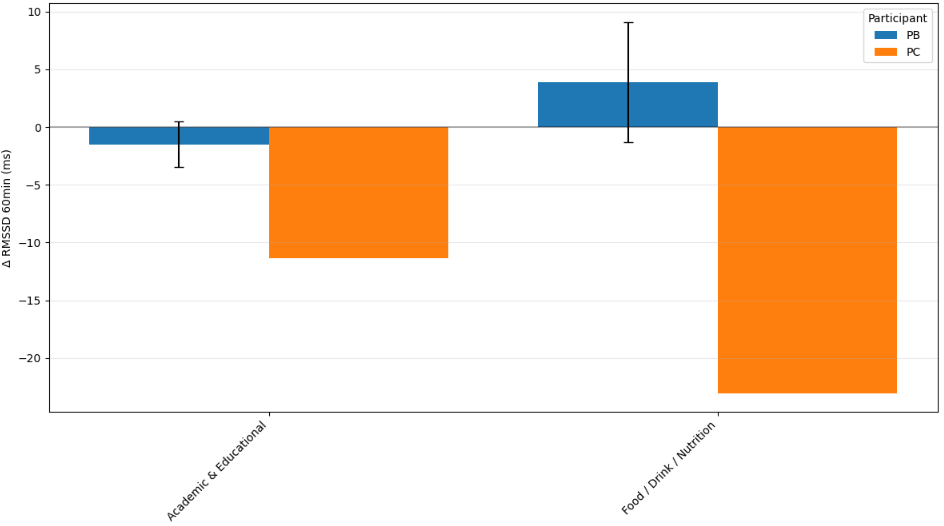}
    \caption{Comparing two users HRV change 60 minutes after an intervention by intervention category.}
    \label{fig:b_vs_c}
\end{figure}

\begin{figure}[htbp]
    \centering
    \includegraphics[width=0.48\textwidth]{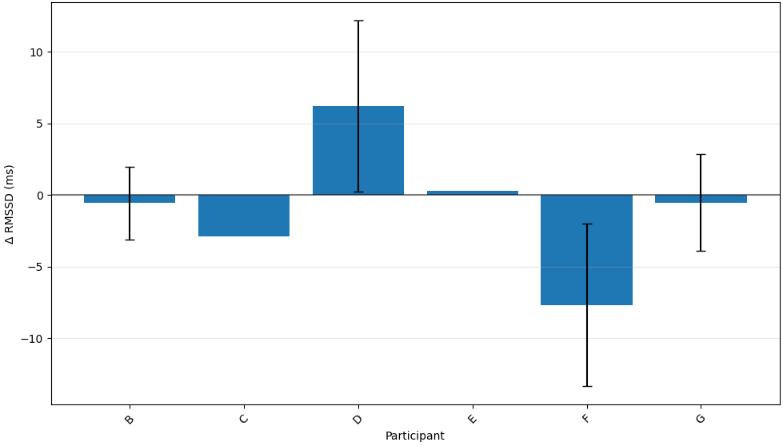}
    \includegraphics[width=0.48\textwidth]{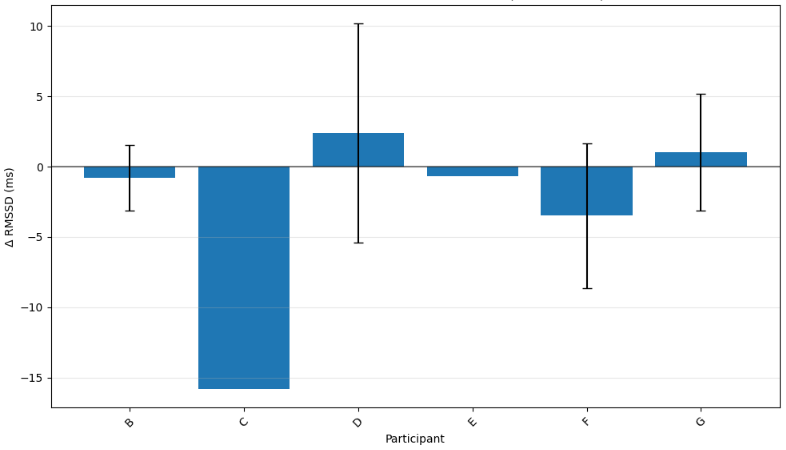}
    \includegraphics[width=0.48\textwidth]{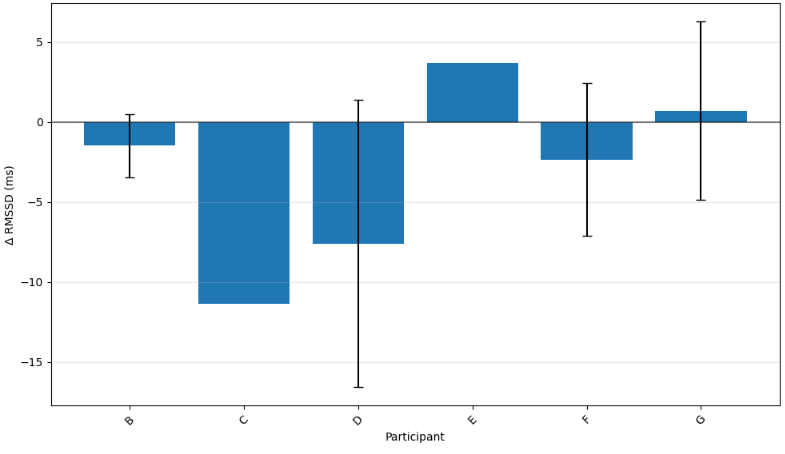}
    \includegraphics[width=0.48\textwidth]{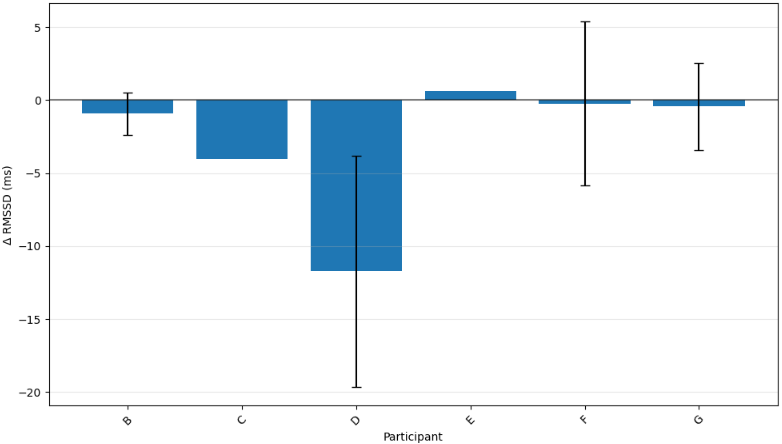}
    \caption{View Annotations}
    \smallskip
    \small \textit{Change in HRV at 15, 30, 60 and 120 minutes after an academic activity.}
    \label{fig:academic_hrv}
\end{figure}

\begin{figure}[htbp]
    \centering
    \includegraphics[width=0.48\textwidth]{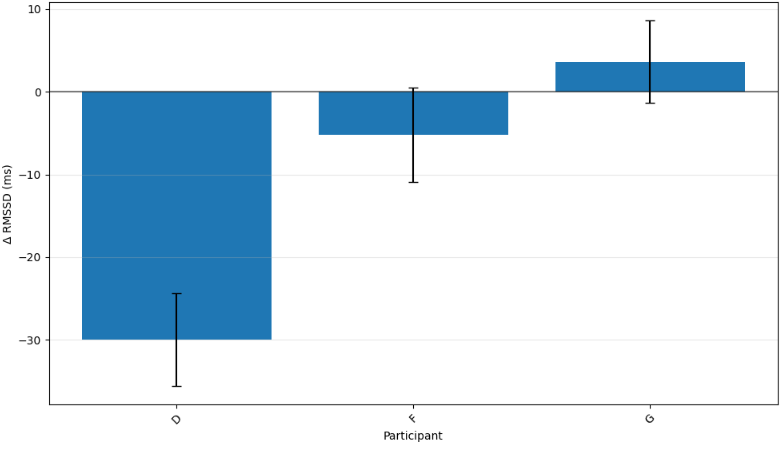}
    \includegraphics[width=0.48\textwidth]{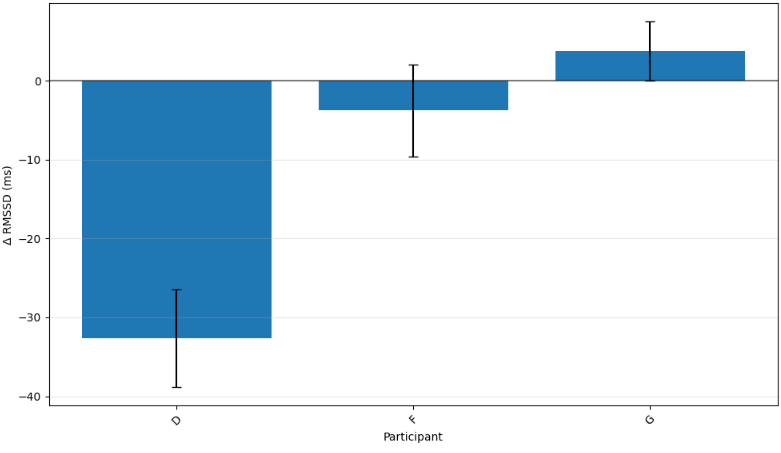}
    \includegraphics[width=0.48\textwidth]{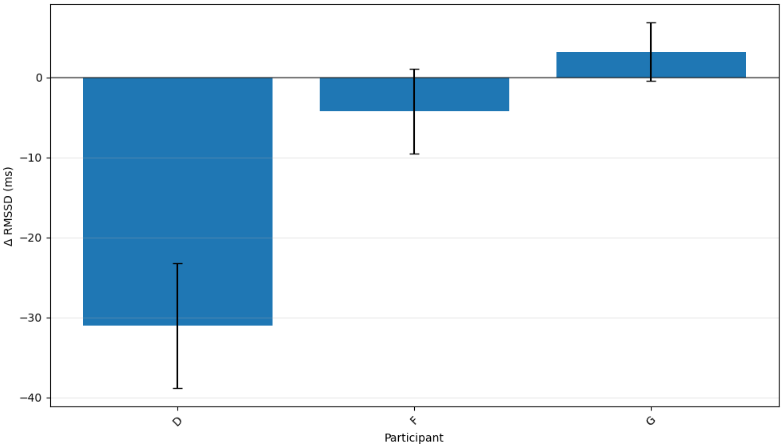}
    \includegraphics[width=0.48\textwidth]{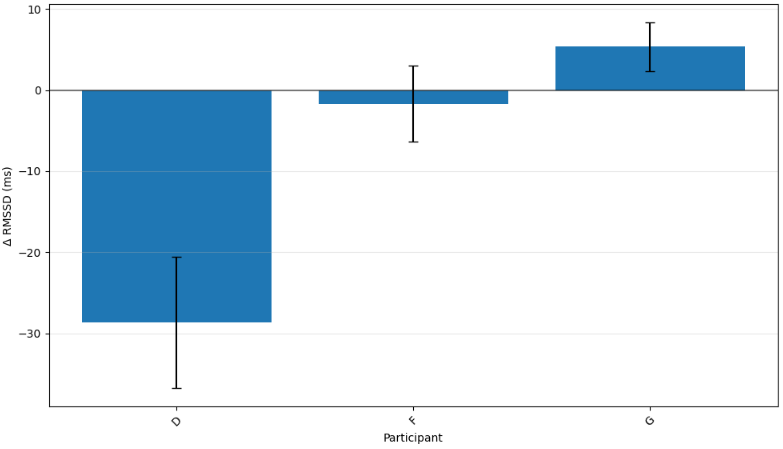}
    \caption{View Annotations}
    \smallskip
    \small \textit{Change in HRV at 15, 30, 60 and 120 minutes after cardio exercise.}
    \label{fig:cardio_hrv}
\end{figure}

\begin{figure}[htbp]
    \centering
    \includegraphics[width=0.48\textwidth]{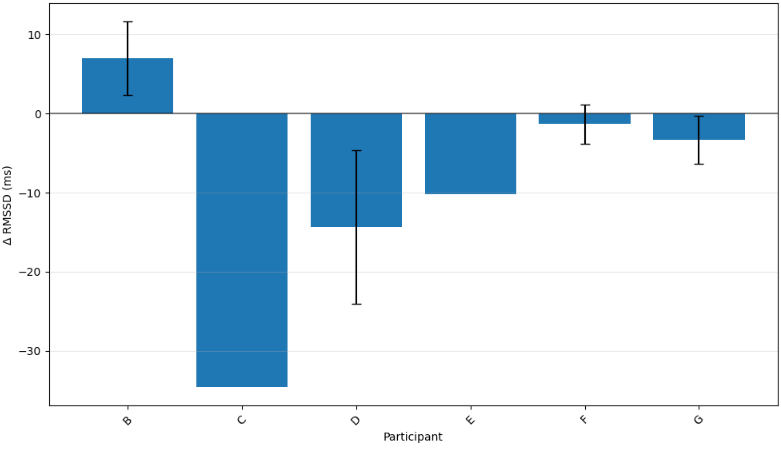}
    \includegraphics[width=0.48\textwidth]{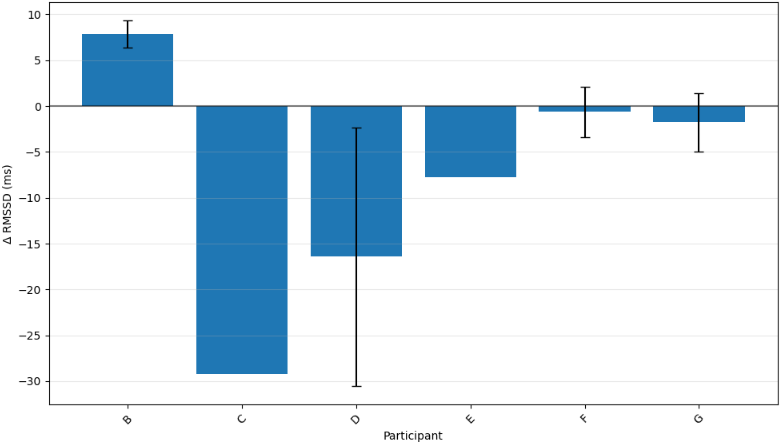}
    \includegraphics[width=0.48\textwidth]{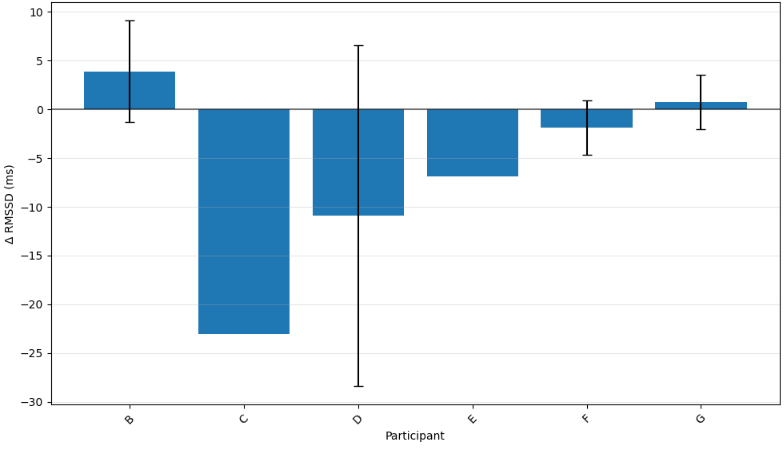}
    \includegraphics[width=0.48\textwidth]{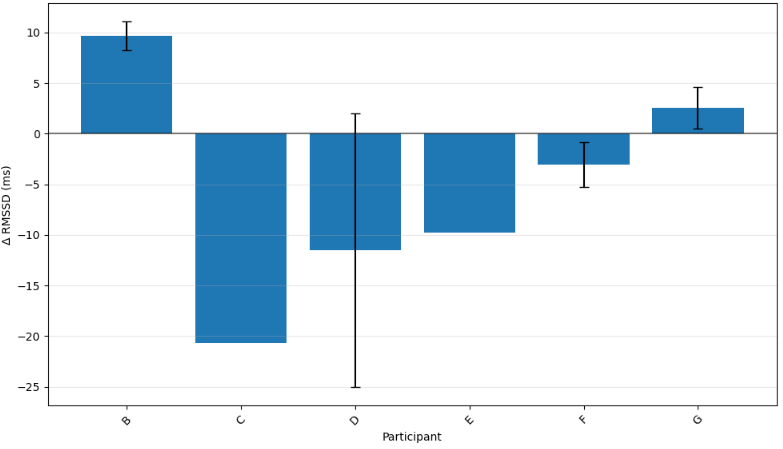}
    \caption{View Annotations}
    \smallskip
    \small \textit{Change in HRV at 15, 30, 60 and 120 minutes after food-related activity.}
    \label{fig:food_hrv}
\end{figure}

\begin{figure}[htbp]
    \centering
    \includegraphics[width=0.48\textwidth]{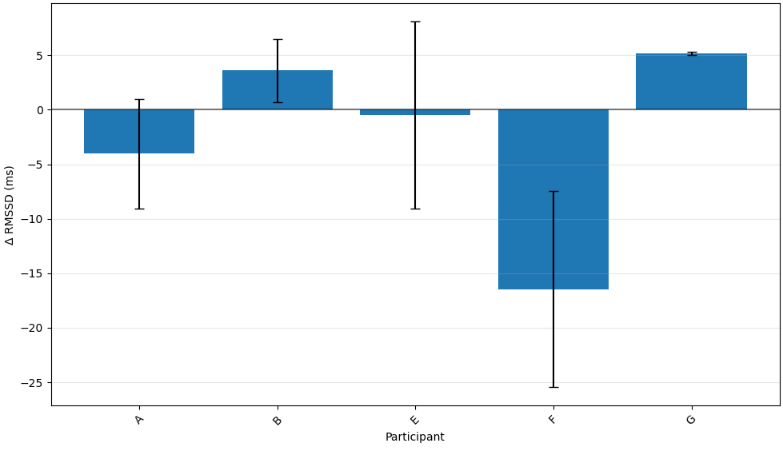}
    \includegraphics[width=0.48\textwidth]{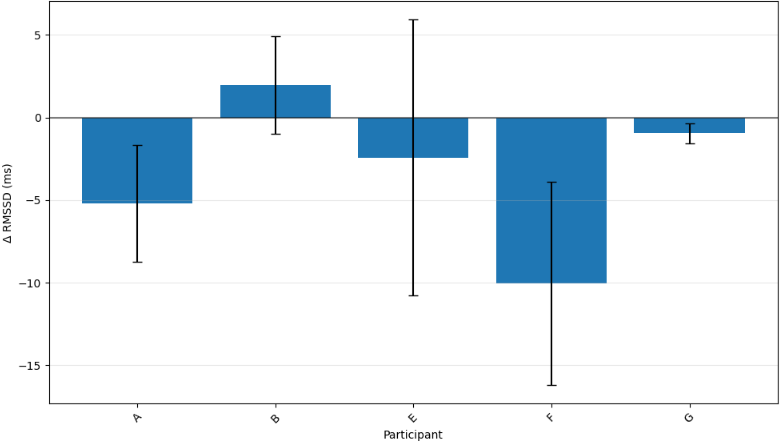}
    \includegraphics[width=0.48\textwidth]{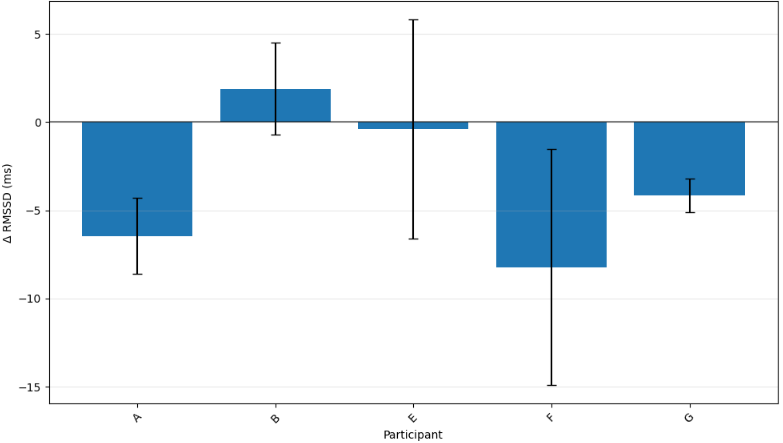}
    \includegraphics[width=0.48\textwidth]{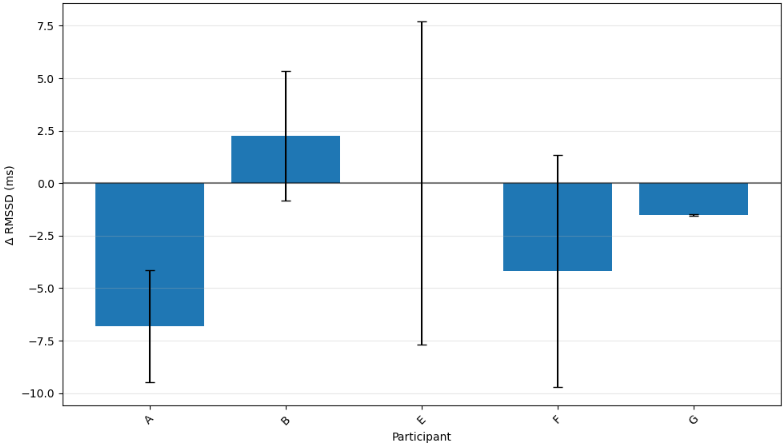}
    \caption{View Annotations}
    \smallskip
    \small \textit{Change in HRV at 15, 30, 60 and 120 minutes after social activity.}
    \label{fig:social_hrv}
\end{figure}

\begin{figure}[htbp]
    \centering
    \includegraphics[width=0.48\textwidth]{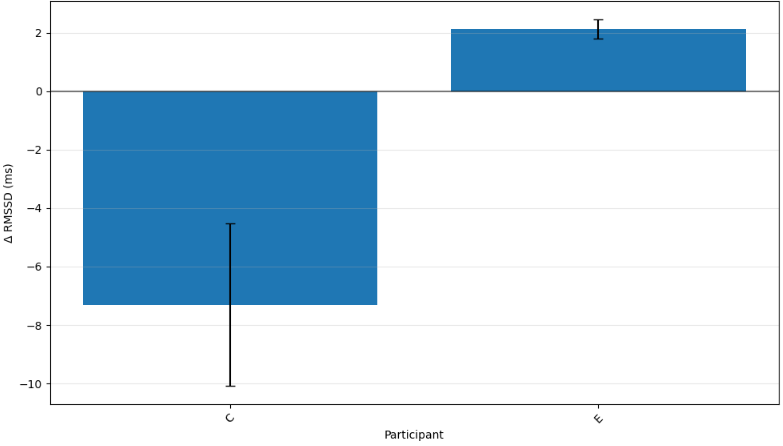}
    \includegraphics[width=0.48\textwidth]{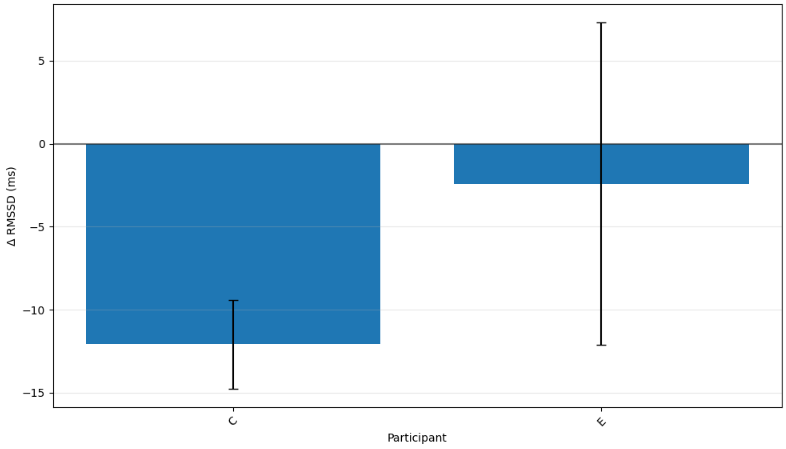}
    \includegraphics[width=0.48\textwidth]{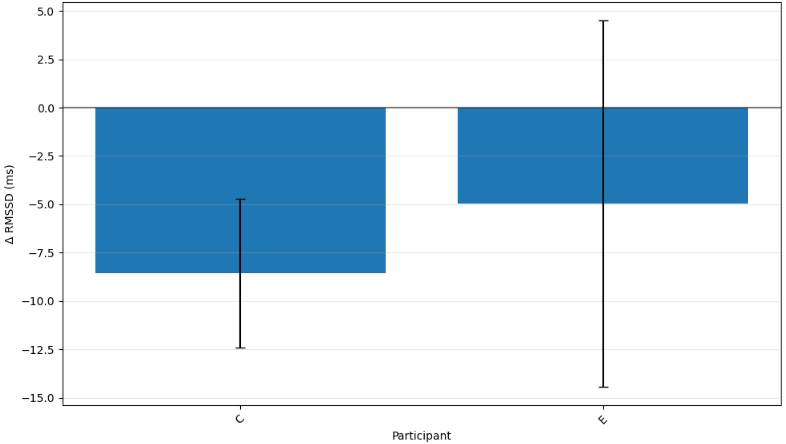}
    \includegraphics[width=0.48\textwidth]{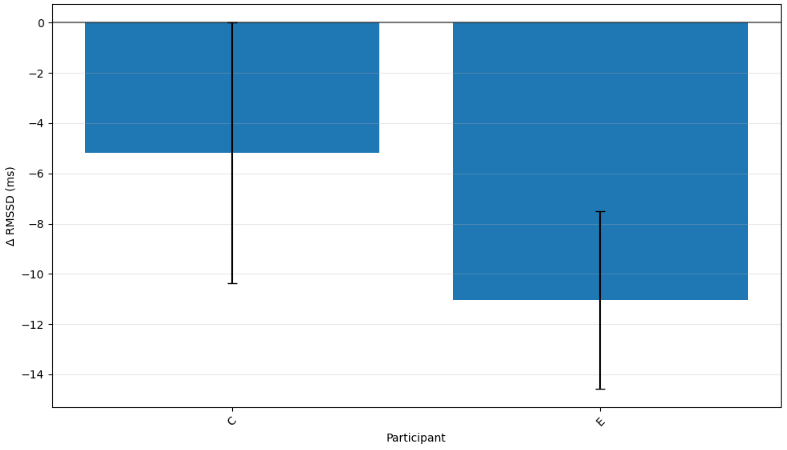}
    \caption{View Annotations}
    \smallskip
    \small \textit{Change in HRV at 15, 30, 60 and 120 minutes after spirituality-related activities.}
    \label{fig:spirituality_hrv}
\end{figure}

\begin{figure}[htbp]
    \centering
    \includegraphics[width=0.48\textwidth]{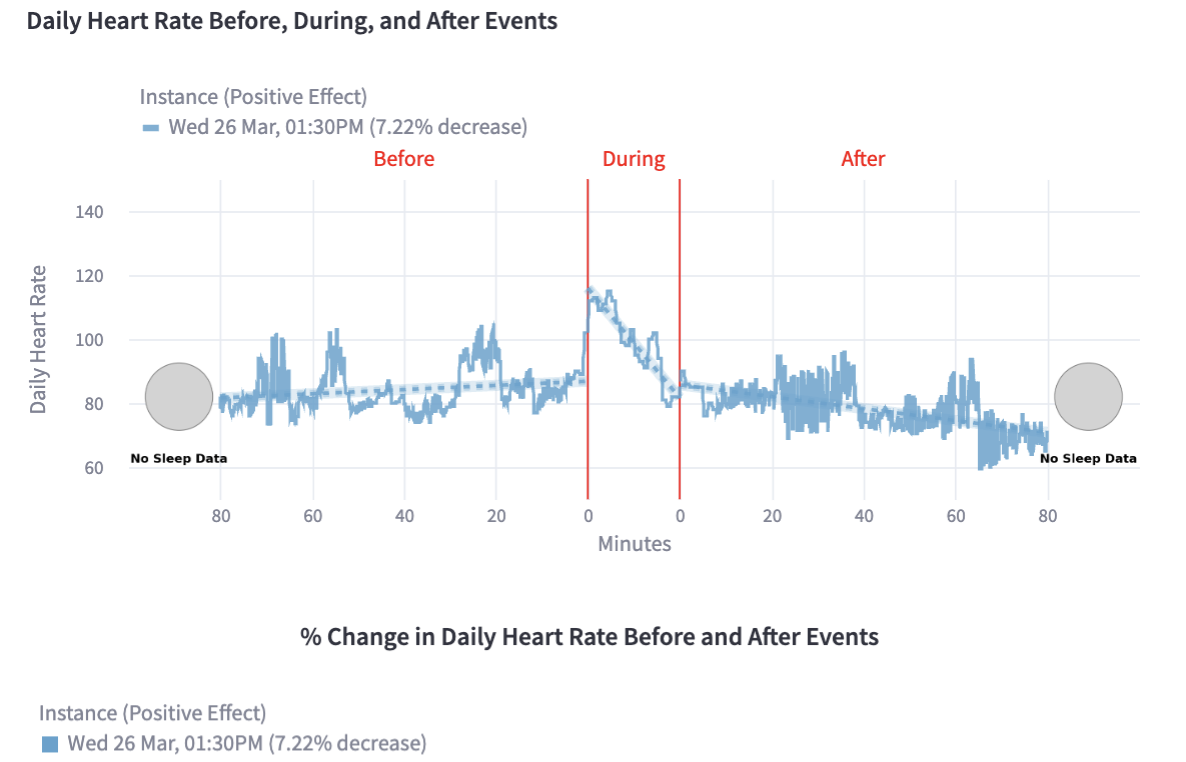}
    \includegraphics[width=0.48\textwidth]{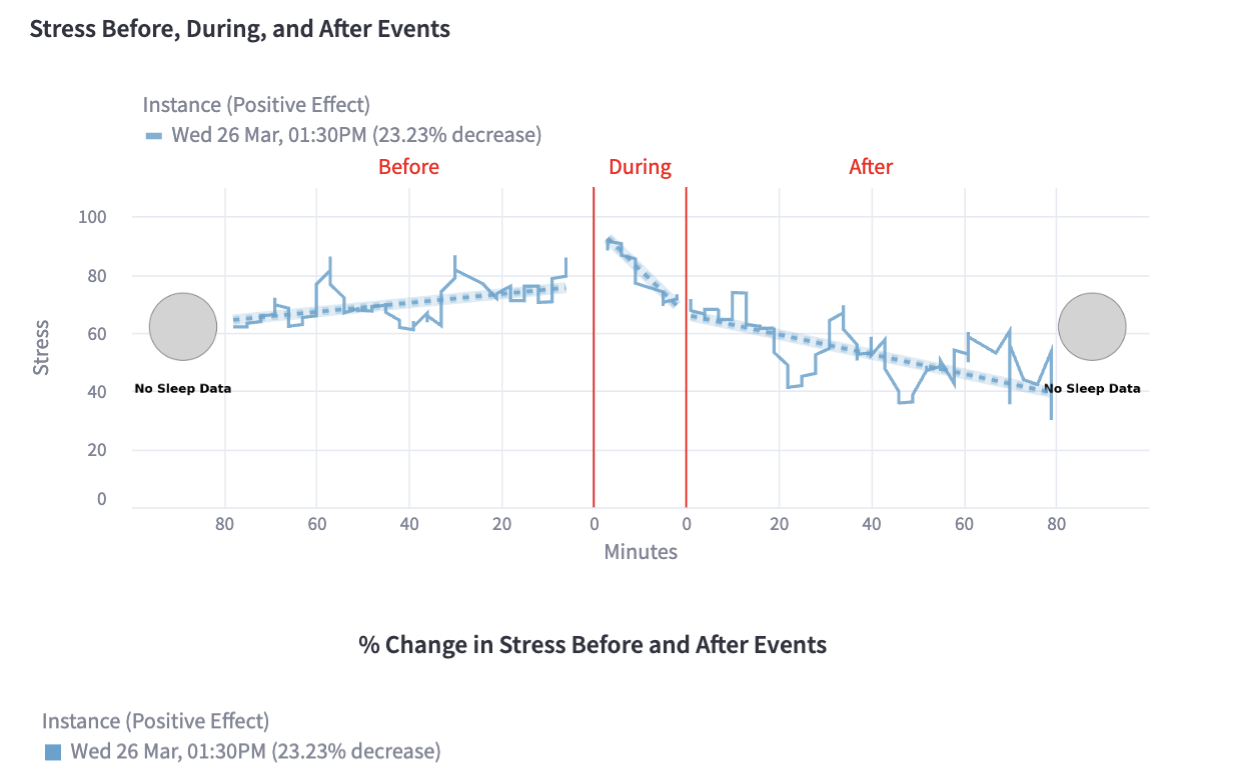}
    \caption{View Annotations}
    \smallskip
    \small \textit{In an exit interview, one participant shared an observation about their physiological response \textbf{patterns} during meetings—they became calmer over the course of the meeting.}
    \label{fig:interview_meeting}
\end{figure}

% %-------------------------------------------------
% \section{Calculation Details}\label{app:calc_details}
% %-------------------------------------------------

% \subsection{BBI Smoothing}\label{app:bbi_smoothing}

% Separately from the RMSSD computation, we apply a smoothing operation to the
% raw beat-to-beat interval (BBI) series to attenuate motion artefacts and other
% high-frequency noise.  A centred rolling mean with a window length of
% \(W = 180\) s (3 min) is evaluated every 30 s:

% \[
% \mathrm{BBI}_{\text{smoothed}}(t) \;=\;
% \frac{1}{W}\;
% \sum_{i = t-W}^{t} \mathrm{BBI}(i).
% \]

% The smoothed trace is used only for visualisation in the dashboard; all heart-rate-variability (HRV) metrics rely on the unsmoothed (raw) BBI values to preserve beat-to-beat detail.

% \subsection{RMSSD (HRV) Calculation}\label{app:rmssd_calc}

% The root-mean-square of successive differences (RMSSD) is computed over
% non-overlapping 5-min windows following the standard guideline \cite{taskforce1996}:

% \[
% \mathrm{RMSSD} \;=\;
% \sqrt{\frac{1}{N-1}\,
% \sum_{k=1}^{N-1}\!
% \bigl(\mathrm{BBI}_{k+1} - \mathrm{BBI}_{k}\bigr)^2},
% \]

% where